\documentclass[aps,pre,epsf,superscriptaddress,amsmath,amssymb,amsfonts,twocolumn,showpacs]{revtex4-1}

\usepackage{graphicx}
\usepackage{epsfig}
\usepackage{dcolumn}
\usepackage{bm}
\usepackage{braket}
\usepackage{amsmath}
\usepackage{color}
\newcommand{\abs}[1]{\left| #1 \right|} 
\usepackage{hyperref}
\usepackage{multirow}
\DeclareMathOperator{\sech}{sech} 

\begin{document}
\title{Spontaneous generation of dark-bright and dark-antidark solitons\\ upon quenching a particle-imbalanced bosonic mixture}

\author{H. Kiehn}
\affiliation{Center for Optical Quantum Technologies, Department of Physics, University of Hamburg, 
Luruper Chaussee 149, 22761 Hamburg Germany}  
\author{S. I. Mistakidis}
\affiliation{Center for Optical Quantum Technologies, Department of Physics, University of Hamburg, 
Luruper Chaussee 149, 22761 Hamburg Germany}
\author{G. C. Katsimiga}
\affiliation{Center for Optical Quantum Technologies, Department of Physics, University of Hamburg, 
Luruper Chaussee 149, 22761 Hamburg Germany}  
\author{P. Schmelcher}
\affiliation{Center for Optical Quantum Technologies, Department of Physics, University of Hamburg, 
Luruper Chaussee 149, 22761 Hamburg Germany}\affiliation{The Hamburg Centre for Ultrafast Imaging,
University of Hamburg, Luruper Chaussee 149, 22761 Hamburg,
Germany}

\date{\today}

\begin{abstract} 

We unveil the dynamical formation of multiple localized structures in the form of dark-bright
and dark-antidark solitary waves that emerge upon quenching a one-dimensional particle-imbalanced Bose-Bose mixture. 
Interspecies interaction quenches drive the system out-of-equilibrium while the so-called miscible/immiscible 
threshold is crossed in a two directional manner. 
Dark-bright entities are spontaneously generated for quenches towards the phase separated
regime and dark-antidark states are formed in the reverse process. 
The distinct mechanisms of creation of the aforementioned states are discussed in detail
and their controlled generation is showcased. 
In both processes, it is found that the number of solitary waves generated is larger for larger particle
imbalances, a result that is enhanced for stronger postquench interspecies interactions. 
Additionally the confining geometry highly affects the production of both types of states
with a decaying solitary wave formation occurring for tighter traps. 
Furthermore, in both of the aforementioned transitions, 
the breathing frequencies measured for the species differ   
significantly for highly imbalanced mixtures. 
Finally, the robustness of the dynamical formation of dark-bright and dark-antidark solitons is also demonstrated 
in quasi one-dimensional setups. 

\end{abstract}

\maketitle

\section{Introduction}
Over the past two decades, nonlinear wave phenomena~\cite{kevrekidisEmergentNonlinearPhenomena} 
in Bose-Einstein condensates (BECs)~\cite{StringariBEC,PethickSmith} have attracted considerable interest. 
Early experiments on single component BECs~\cite{AndersonEarlyBEC,DavisEarlyBEC,BradleyEarlyBEC} triggered 
a vast amount of studies devoted in investigating the properties of nonlinear excitations that arise in them,
such as dark~\cite{burgerDark,frantzeskakisDark,becker2008oscillations,WellerDark,AndersonDark,shomroniDarkExperiment} 
and bright~\cite{streckerBright,khaykovichBright,CornishBright,Dabrowska-WusterBright} solitons.
These structures exist as stable configurations in highly elongated quasi one-dimensional (1D) BECs 
featuring repulsive and attractive interatomic interactions respectively.
Besides single component BECs, the experimental realization of two-component 
mixtures~\cite{myatt1997SympatheticCooling,HallBinaryMixurePhaseSeparation,StamperKurnSpinorBEC,ModugnoRbKMixture,BlochRbMixture}
enabled the investigation of more complex compounds, the so-called vector solitons. 
These include multiple dark and/or bright states such as dark-dark~\cite{HoeferDarkDarkSolitonsBEC,YanDarkDarkSolitonsBEC},
and dark-bright (DB)~\cite{RajendranDarkBrightInteraction,YinDarkBrightInteraction,YanDarkBrightInteraction,
KatsimigaDarkBrightInteraction,BuschDarkBrightBECTheory,NistazakisDarkBrightBECTheory,
VijayajayanthiDarkBrightBECTheory,ValeriyDarkBrightSolitonStipesBECTheory,
AlvarezDarkBrightBECTheory,AchilleosDarkBrightBECTheory,
AlvarezDarkBrightBECImpurities,AchilleosDarkBrightBECTheory2,MiddlekampDarkBrightBECTheory,KatsimigaDarkBrightDynamics} 
(see also the review \cite{KevrekidisMultiComponentSolitonsReview}), as well as more peculiar entities 
like dark-antidark (DAD) 
solitons~\cite{DanailaDarkAntidarkBEC,KevrekidisFamiliesOfMatterWaves,MistakidisQuenchInducedPhaseSeparation,KatsimigaDarkBrightDynamics}. 
The latter antidark state consists of a density hump on top of the matter-wave background.
Initially, such solitons were discovered and broadly studied 
in the field of nonlinear optics~\cite{ChenDarkBrightNonlinearOpticsExperiment,OstrovskayaDarkBrightNonlinearOpticsExperiment,
ChristodoulidesDarkBrightNonlinearOpticsTheory,AfanasyevDarkBrightNonlinearOpticsTheory,
KivsharDarkAntidarkNonlinearOpticsTheory,RadhakrishnanDarkBrightNonlinearOpticsTheory,
BuryakDarkBrightNonlinearOpticsTheory,SheppardDarkBrightNonlinearOpticsTheory,ParkDarkBrightNonlinearOpticsTheory}, 
but they were later on shown to appear also in BECs \cite{becker2008oscillations,HamnerDarkBrightBECCounterflow}. 
Focusing on DB solitons multiples of these types of entities have been experimentally 
realized~\cite{HamnerDarkBrightBECCounterflow}. 
Moreover, considerable attention has been paid in understanding their properties 
and interactions not only in the so-called integrable Manakov limit~\cite{manakov1974theory,prinari2015dark} but also upon breaking
integrability. The latter breaking can be achieved e.g. by the presence of external 
trapping~\cite{yan2015dark,KaramatskosDarkBrightPairs}, 
or by considering different interactions between the same or different atomic 
species~\cite{yan2015dark,KatsimigaDarkBrightInteraction,
KaramatskosDarkBrightPairs,KatsimigaDarkBrightPairs,KatsimigaDarkBrightBifurcationCollision}. 

The generation of DB states is a topic of ongoing research activity for a multitude of reasons.
Remarkably these states appear in binary BECs with repulsive interparticle interactions,
an environment into which bright solitons, being localized density peaks, would disperse 
if dark solitons were absent. This symbiotic nature of DB solitons leads to their remarkably prolonged 
lifetimes compared to their single component analogues. 
Thus DB states constitute ideal candidates for applications  
as e.g. coherent storage as well as processing of optical fields, multi-channel signals and their 
switching~\cite{RajendranDarkBrightInteraction,folman2000controlling,petrov2009trapping,rajendran2011matter}. 

DB solitons exist both in weakly miscible \cite{HamnerDarkBrightBECCounterflow} and immiscible 
\cite{HallBinaryMixurePhaseSeparation} BEC mixtures. 
The miscibility condition within mean-field theory is determined by 
$a_{AB}^2 < a_{AA} a_{BB}$~\cite{HoMiscibilityCondition,TimmermansMiscibilityCondition,
AoMiscibilityCondition,esryMisibilityCondition} 
where $a_{jj}$ is the intra- and $a_{ij}$ denotes the interspecies scattering lengths. 
In turn these scattering lengths can be experimentally tuned via Feshbach resonances 
\cite{InouyeFeshbachResonance,VogelsFeshbachResonancePrediction,ThalhammerFeshbachResonance} 
allowing the realization of a transition between the miscible and immiscible regime of 
interactions~\cite{PappTunableMiscibility,WangTunableMiscibility,TojoControlledPhaseSeparation}. 
The tunability of the scattering lengths opened a new path for studying, in a controllable manner, 
the properties of binary BECs and the solitonic structures that form in them. 
In this way, vector solitons have been identified upon considering e.g. 
temporal~\cite{kanna2013non,manikandan2016manipulating,yu2016nonautonomous} 
and spatial~\cite{cheng2009effective,belmonte2011solitary,mareeswaran2014vector}
variations of the scattering lengths. 
In the same context control over the system's parameters has been examined by performing temporal modulations of the confinement
frequencies~\cite{kanna2013non,manikandan2016manipulating,yu2016nonautonomous} 
and also by adding appropriate time-dependent gain and loss terms~\cite{RajendranDarkBrightInteraction,rajendran2011matter}.  
Moreover, control over the system's dynamical evolution and pattern formation has been reported in~\cite{EtoPhaseSeparation},
upon quenching the binary BEC towards miscibility. 
Also very recently the spontaneous generation of vector solitons upon quenching the interspecies interactions
crossing the miscibility/immiscibility threshold both within mean-field but also considering the many-body 
treatment of this problem has been reported~\cite{MistakidisQuenchInducedPhaseSeparation}. Here, robustly propagating DAD states   
were shown to dynamically emerge.  The latter observation rises a natural question on how different variations of the systems' 
parameters can lead to the controllable creation of these states at least within mean-field theory.

Motivated by the aforementioned studies, in the present work 
we take a closer look on the multi-soliton formation upon considering the interspecies interaction-quenched dynamics of a 
particle imbalanced BEC mixture while crossing 
the miscibility/immiscibility threshold in both directions.
In this way we aim at deepening our understanding regarding vector soliton generation 
in an integrability broken scenario. In this context dynamical 
instabilities lead to peculiar evolution of the resulting states including mass redistribution between the 
solitons~\cite{KatsimigaDarkBrightPairs,KatsimigaDarkBrightBifurcationCollision}, 
and thus shape changing collisions~\cite{rajendran2011matter,manikandan2014generalized}.
In particular, we investigate how the number of vector solitons formed can 
be controlled under different variations of the binary systems' parameters as e.g. the trapping frequency and the 
associated with each species particle number. 

In the miscible-to-immiscible quench scenario, 
the unstable dynamics leads to the filamentation of the density of both species
in line with our earlier findings~\cite{MistakidisQuenchInducedPhaseSeparation}.
This filamentation process entails the formation of DB solitary waves. 
The formation of the latter is verified upon fitting the waveforms corresponding 
to a dark and a bright soliton stemming from the exact 
(in the Manakov limit) single DB state~\cite{YanDarkBrightInteraction,
KatsimigaDarkBrightInteraction,KatsimigaDarkBrightBifurcationCollision}. 
Focusing on the initial stages of the nonequilibrium dynamics, an increase of the number of the DB states generated is 
found upon varying the particle number of the minority species until the particle balanced limit is reached. 
A result that is more enhanced for transitions that enter deeper in the immiscible phase~\cite{MistakidisQuenchInducedPhaseSeparation}. 
Lastly, a decaying formation of DB states is observed as the trapping frequency is increased. 
In all cases both clouds undergo a collective breathing motion.
Moreover, the generated DB states are seen to oscillate within the parabolic trap, 
featuring multiple collisions with one another.
However, due to the non-integrable nature of the system, these collisions are also accompanied by 
a transfer of mass between the soliton constituents~\cite{KatsimigaDarkBrightPairs,KatsimigaDarkBrightBifurcationCollision}.
The latter effect reduces, during evolution, the number of states generated with significantly fewer DB solitary waves remaining 
and robustly propagating at large evolution times. 
Finally, inspecting the corresponding breathing frequency of each bosonic cloud we demonstrate 
that significant differences occur for highly particle imbalanced mixtures~\cite{sartori2013dynamics}.

Following the reverse process, namely an immiscible-to-miscible transition, robust  
DAD solitons are formed~\cite{MistakidisQuenchInducedPhaseSeparation}. 
The binary system, is the 1D analogue of the so-called ``ball and shell" 
configuration~\cite{BigelowBallShell,MertesBallShell}, with 
one species occupying the center of the trap and the other species residing on the two sides of the harmonic potential. 
After the quench the outside species is allowed to ``fall" towards the central region inducing a counterflow. 
The latter leads in turn to the appearance of interference 
fringes~\cite{AndrewsBECInterference,NaraschewskiBECInterference,HostonBECInterference} and dark solitons emerge 
in the regions of destructive interference~\cite{HamnerDarkBrightBECCounterflow}.  
It is the effective potential created by the emergent dark states that causes the other species 
to split and fill the newly formed density dips, leading in turn to the formation of DAD structures.
An almost linear increase of the number of DAD states formed occurs either as the particle number of the inner 
species is increased or entering deeper in the miscible phase. 
Also, tighter trapping results to fewer DAD solitons 
occurring during the nonequilibrium dynamics for larger postquench 
interspecies interactions. 
Finally, different breathing frequencies are measured for the 
two species independently of the postquench interaction 
strength and the particle imbalance. 
A remark is appropriate here. 
Note that the quench-induced dynamics studied in the present work is restricted to a 1D geometry. 
As such, the dynamical formation of the DB and the DAD solitons identified herein,
can be tested in current quasi-1D experiments as it will be demonstrated later on. 
However, we must stress at this point that alterations of the observed interaction quench dynamics 
can take place in higher dimensional settings, e.g. due to the presence of transversal instabilities of dark solitons. 

Our presentation is structured as follows. 
In section \ref{theory} we provide the theoretical framework of 
our mean-field approach in the form of a set of two coupled Gross-Pitaevskii equations (GPEs). 
We also briefly comment on the numerical methods utilized herein. 
In section \ref{MiscibleToImmiscibleDynamics} we present our results for miscible-to-immiscible quenches while section 
\ref{ImmiscibleToMiscibleDynamics} contains our results for the reverse process. 
The robustness of our results in a quasi 1D harmonic trap is demonstrated in section \ref{quasi1D_simulations}. 
Finally section \ref{conclusions} summarizes our findings and also provides future perspectives.

\section{Theoretical Framework}\label{theory} 

\subsection{Mean-Field ansatz for a binary Bose-mixture}  

We consider a binary mixture of repulsively interacting BECs
composed of two different hyperfine states, namely $\ket{F=1, m_F=-1}$ and $\ket{F=2, m_F=1}$, 
of $^{87}$Rb~\cite{EgorovRbScatteringLengths} being 
confined in a one-dimensional (1D) harmonic oscillator potential.
Such a cigar-shaped geometry can be realized 
experimentally~\cite{becker2008oscillations,HoeferDarkDarkSolitonsBEC,MiddlekampDarkBrightBECTheory} 
upon considering a highly anisotropic trap with the longitudinal 
and transverse trapping frequencies obeying  
$\omega_x \ll \omega_{\perp}$. 
Within mean-field theory the dynamics of this binary mixture can be described 
by the following system of coupled GPEs~\cite{kevrekidisEmergentNonlinearPhenomena,StringariBEC}:
\begin{eqnarray}
i\hbar \partial_{\tilde{t}} \tilde{\Psi}_{\sigma} =
\left( -\frac{\hbar^2}{2m} \partial_{\tilde{x}}^2 +\tilde{V}(\tilde{x}) -\mu_{\sigma} + \sum_{k=A,B} 
\tilde{g}_{\sigma k} |\tilde{\Psi}_k|^2\right)\!\tilde{\Psi}_{\sigma}.\nonumber\\
\label{model}
\end{eqnarray}
In the above expression $\tilde{\Psi}_{\sigma}(\tilde{x},\tilde{t})$ ($\sigma=A,B$) denotes the wavefunction 
for the $A$- and $B$-species respectively.
Each $\tilde{\Psi}_{\sigma}(\tilde{x},\tilde{t})$ is normalized to the
corresponding number of atoms i.e. $N_{\sigma} = \int_{-\infty}^{+\infty} |\tilde{\Psi}_{\sigma}|^2 d\tilde{x}$.  
Also $\tilde{V}(\tilde{x})$ represents the external trapping
potential, while $m_A=m_B=m$ and $\mu_{\sigma}$ refer to the atomic mass and chemical potential for each of the species respectively.  
The effective 1D coupling constants are given by $\tilde{g}_{\sigma k}=2\hbar\omega_{\perp} a_{\sigma k}$, 
where $a_{\sigma k}$ are the three $s$-wave scattering lengths 
(with $a_{AB}=a_{BA}$) 
accounting for collisions between atoms that belong to  
the same ($a_{\sigma \sigma}$) or different ($a_{\sigma k}, \sigma \ne k$) species.
We note that both the intra and interspecies scattering lengths can 
be manipulated experimentally by means of Feshbach~\cite{kohler2006production,chin2010feshbach} 
or confinement induced resonances~\cite{olshanii1998atomic,kim2006suppression}.

For the numerical findings to be presented below, we choose to express the system of Eqs.~(\ref{model})
in the following dimensionless form:
\begin{align}\label{eqn::GPE_A}
    i \frac{\partial \Psi_A}{\partial t} &= \left[ -\frac{1}{2}{\partial_x^2} + V + g_{AA} |\Psi_A|^2 + g_{AB} |\Psi_B|^2 - \mu_A\right] \Psi_A,\\\label{eqn::GPE_B}
    i \frac{\partial \Psi_B}{\partial t} &= \left[ -\frac{1}{2}{\partial_x^2} + V + g_{BB} |\Psi_B|^2 + g_{AB} |\Psi_A|^2 - \mu_B\right] \Psi_B.
\end{align}
In the above Eqs.~(\ref{eqn::GPE_A})-(\ref{eqn::GPE_B}), 
the interaction coefficients are normalized to the intraspecies scattering 
length, $a_{AA}$, of the $A$-species, namely $g_{AB}\equiv \tilde{g}_{AB}/\tilde{g}_{AA}$, 
and $g_{BB}\equiv \tilde{g}_{BB}/\tilde{g}_{AA}$. Moreover,
densities $|\Psi_{\sigma}|^2$, length, energy and time are measured 
in units of $2a_{AA}$, $\sqrt{\frac{\hbar}{m \omega_\perp}}$, $\hbar \omega_\perp$ 
and $\omega_\perp^{-1}$ respectively.
Finally, $V(x)=\frac{1}{2} \omega^2 x^2$ is the dimensionless trapping potential with 
$\omega_A=\omega_B \equiv\omega = \omega_x/\omega_\perp$.

In the present work we aim at revealing the out-of-equilibrium dynamics of the aforementioned binary system 
[see Eqs.~(\ref{eqn::GPE_A})-(\ref{eqn::GPE_B})] upon performing an interspecies interaction quench from the miscible to the 
immiscible regime of interactions and vice versa. To this end we fix the intraspecies coefficients to the experimentally relevant 
values for a  $^{87}$Rb mixture, i.e.  $g_{AA} = 1$, $g_{BB} = 0.95$~\cite{EgorovRbScatteringLengths}.
To induce the dynamics an interspecies interaction quench is performed with the system being initially relaxed to its ground state 
configuration lying within the miscible regime of interactions. The latter is characterized by an initial (prequench) 
interaction $g^i_{AB}$  subject to the miscibility condition $g^{i}_{AB} < \sqrt{g_{AA} g_{BB}}$~\cite{AoMiscibilityCondition}.
Consecutively the system is quenched towards the immiscible (alias phase separated) regime having the desired postquench amplitude
$g^f_{AB}$ [see Sec.\ref{MiscibleToImmiscibleDynamics}]. 
The same procedure is followed for the inverse process, namely considering a 
relaxed stated for immiscible components ($g^i_{AB} > \sqrt{g_{AA} g_{BB}}$)
and quenching the system towards the miscible domain [see Sec.\ref{ImmiscibleToMiscibleDynamics}]. 

In both quench scenarios under investigation in order to obtain the ground state 
configuration a fixed-point numerical iteration scheme is employed~\cite{KelleyNewtonKrylov}. 
To simulate the nonequilibrium dynamics of the binary mixture governed by Eqs. 
(\ref{eqn::GPE_A})-(\ref{eqn::GPE_B}) a fourth-order Runge-Kutta integrator is employed 
and a second-order finite differences method is used for the spatial derivatives. 
The spatial and time discretization are $dx=0.1$ and $dt=0.005$ respectively. 
Moreover our numerical computations are restricted to a finite region by employing hard-wall 
boundary conditions. The latter are chosen wide enough to avoid boundary effects. 
In particular, in the dimensionless units adopted herein and also for the majority of our simulations, the hard-walls 
are located at $x_{\pm}=\pm 80$ and we do not observe any appreciable density for $|x| > 60$.

\section{Quench Dynamics from the Miscible to the Immiscible Regime}\label{MiscibleToImmiscibleDynamics}

In the following we discuss the vector soliton formation considering quenches from the miscible to the immiscible regime,
namely for $g_{AB}^f > \sqrt{g_{AA} g_{BB}}$. Before proceeding, it is worth mentioning at this point that the latter relation 
is valid for the homogeneous case (V=0). Nonetheless, corrections due to the presence of the trap
are only significant for strong trapping frequencies~\cite{NavarroModifiedMiscibilityCondition} 
that are not considered in the present work. 
In what follows the trapping frequency 
is fixed to $\omega=0.05$ unless it is stated otherwise. 
Additionally, since the miscibility depends on the interaction strength between the species, for a 
$^{87}$Rb mixture considered herein the transition between the two regimes occurs at $g^{th}_{AB}=\sqrt{0.95}\approx0.974$.
Thus for quenches with $g^{f}_{AB}< g^{th}_{AB}$ it is expected that 
the two species prefer overlapping thereby residing around the trap center.  
However, for postquench interspecies interactions having $g^{f}_{AB}>g^{th}_{AB}$ component separation takes place.
\begin{figure}[htb]
\centering \includegraphics[width=0.5\textwidth]{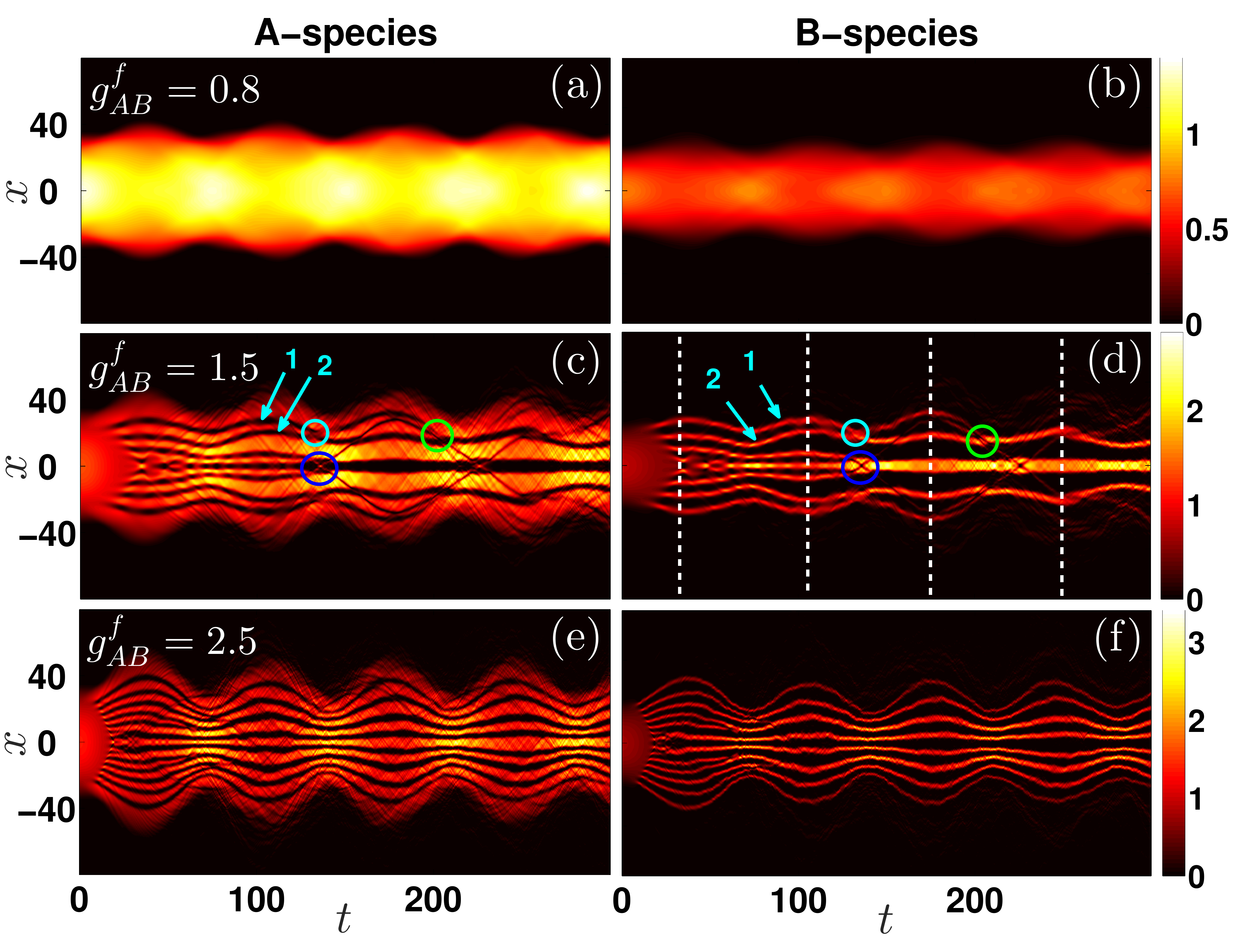}
\caption{\label{fig::MiscImmGabComparison} Spatio-temporal evolution of the one-body density $|\Psi_{\sigma}|^2$ of species $\sigma=A$ 
(left panels) and species $\sigma=B$ (right panels). 
An interspecies interaction quench from $g_{AB}^i=0.2$ to different postquench values (a), 
(b) $g_{AB}^f=0.8$, (c), (d) $g_{AB}^f=1.5$ and (e), (f) $g_{AB}^f=2.5$ is performed. 
In all cases the particle number of species is $N_A=65$ and $N_B=25$ respectively, while both species are trapped in a 
harmonic oscillator of frequency $\omega=0.05$. The system is prepared in its ground state with $g_{AA}=1.0$, $g_{BB}=0.95$ and 
$g_{AB}^i=0.2$. 
The dashed vertical lines mark time instants before and after specific collision events indicated by the circles. 
All quantities shown are given in dimensionless units. }
\end{figure}

\subsection{Density evolution within the miscible domain and breathing mode}

Figures~\ref{fig::MiscImmGabComparison} (a)-(f) illustrate the quenched density evolution upon increasing $g_{AB}^f$.
In all cases the system is initially prepared in the miscible regime with $g_{AB}^i=0.2$, while $N_A=65$ and $N_B=25$.  
As per our discussion above when the interaction is quenched e.g to $g_{AB}^f=0.8$ 
depicted in Figs.~\ref{fig::MiscImmGabComparison} (a)-(b) for the $A$-species and the $B$-species respectively, 
component separation is absent. 
Instead the two species are overlapping while residing around the trap center for all evolution times. 
Notice also that since $g_{BB}<g_{AA}$, the $B$-species is localized closer to the trap center when compared to the $A$-species.
Moreover, a collective breathing motion consisting of a periodic expansion and contraction
of each bosonic cloud is observed during evolution. 

\begin{figure}[htb]
\centering
\includegraphics[width=0.5\textwidth]{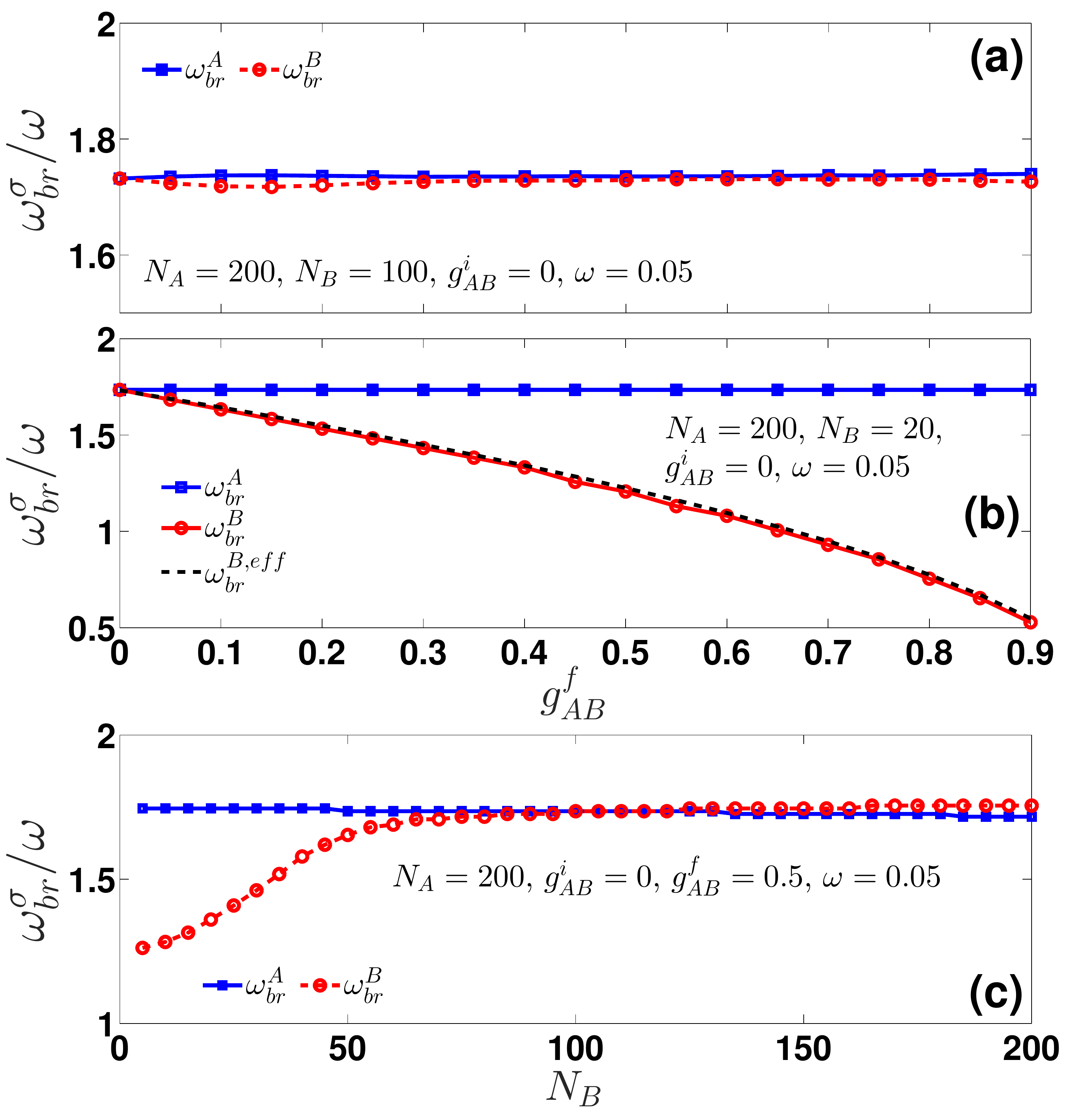}
\caption{\label{fig::breathing_miscible} (a), (b) Breathing mode frequency, $\omega_{br}^{\sigma}$, of each component 
($\sigma=A,B$) of 
the binary bosonic mixture for increasing the postquench interspecies interaction strength $g_{AB}^f$. 
In (a) the system consists of $N_A=200$ and $N_B=100$ particles, while in (b) $N_A=200$ and $N_B=20$ (see legends).  
(c) $\omega_{br}^{\sigma}$ for varying particle number of the $B$-species. 
The remaining system parameters are fixed (see legend). 
In all cases the bosonic mixture is harmonically trapped with $\omega=0.05$ and it is initialized in its ground state 
characterized by $g_{AB}^i=0$. At $t=0$ a quench is performed to $g_{AB}^f$. 
All quantities are given in dimensionless units.}
\end{figure}

Let us now investigate the effect that different variations of the binary systems' parameters 
have on the breathing frequency. 
In all cases illustrated in Figs.~\ref{fig::breathing_miscible} (a)-(c) in order to obtain the 
breathing frequency, $\omega^{\sigma}_{br}$, we start from the non-interacting limit ($g^i_{AB}=0$)
and we measure the Fourier transform of $\braket{x^2}=\int^{+\infty}_{-\infty}x^2|
\Psi_{\sigma}|^2dx$ for each species $\sigma=A,B$. 
In this way we deduce $\omega^{\sigma}_{br}$ upon varying either $g^f_{AB}$ [see Figs.~\ref{fig::breathing_miscible} (a) and (b)]
or the particle number, $N_B$, of the $B$-species [see Fig.~\ref{fig::breathing_miscible} (c)]. 
It is found that for particle number imbalances $N_B/N_A=0.5$, $\omega_{br}^{\sigma}$
is independent of the $g^f_{AB}$ variation acquiring 
the constant value of $\omega_{br}\approx\sqrt{3} \omega$ \cite{sartori2013dynamics}, see Fig.~\ref{fig::breathing_miscible} (a). 
However, for larger particle imbalances as e.g. $N_B/N_A=0.1$ depicted in Fig.~\ref{fig::breathing_miscible} (b)
a completely different behavior is observed, with the breathing frequency of the minority $B$-species
being drastically affected by the postquench variation. In particular as $g^f_{AB}$ increases towards the miscibility/immiscibility 
threshold, a monotonic decrease of $\omega^B_{br}$ is observed \cite{sartori2013dynamics}, e.g. with $\omega^B_{br}=0.5$ for 
$g^f_{AB}=0.9$.  
This decrease of $\omega^B_{br}$ suggests that instead of the harmonic trapping a modified frequency comes into play.  
Since the $A$-species is the majority component, 
we assume that it creates an effective potential \cite{mistakidis2019quench,ferrier2014mixture} of 
the form $V_{eff}(x)=V(x)+g_{AB}^f|\Psi_A(x,0)|^2$ [see Eqs. (\ref{eqn::GPE_A})-(\ref{eqn::GPE_B})] 
into which the $B$-species is trapped and that $|\Psi_A(x,0)|^2$ has roughly the form of a Thomas-Fermi (TF) profile. 
As a consequence we obtain the following 
effective trapping frequency $\omega_{eff} = \omega \sqrt{1-\frac{g_{AB}^f}{g_{AA}}}$. Thus the corresponding effective breathing 
frequency is $\omega^{B,eff}_{br}=\sqrt{3}\omega_{eff}$. A remarkable agreement between $\omega^{B,eff}_{br}$ and 
$\omega^{B}_{br}$ is also shown in Fig.~\ref{fig::breathing_miscible} (b) (see dashed black 
line).  Finally, in line with the above discussion, 
deep in the miscible regime $\omega^{B}_{br}$ shows an increasing tendency 
for larger $N_B$ and acquires the constant 
value of $\omega_{br}\approx\sqrt{3} \omega$ 
for imbalances $N_B/N_A\gtrsim0.3$ [see Fig.~\ref{fig::breathing_miscible} (c)].

\subsection{Density evolution towards the immiscible regime and identification of DB states}

\begin{figure}[htb]
\centering
\includegraphics[width=0.47\textwidth]{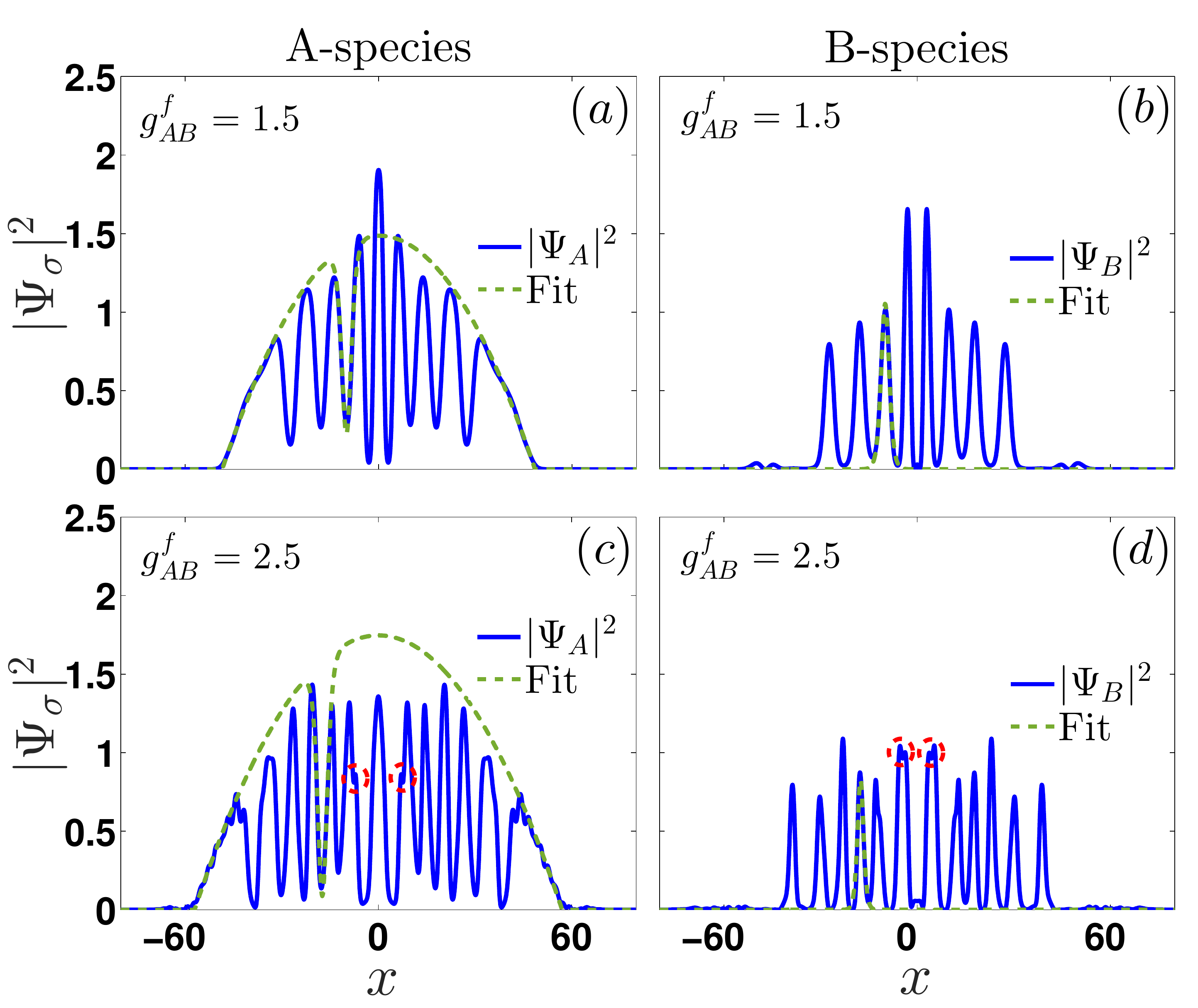}
\caption{\label{fig::MiscImmProfileFit} Snapshots of the one-body densities (solid blue lines) of the $A$-species (left panels) and 
$B$-species (right panels) at $t=35$. The postquench interspecies interaction strength is (a), (b) $g_{AB}^f=1.5$ and (c), 
(d) $g_{AB}^f=2.5$. Each dashed green line illustrates a fitting of a single (a), (c) dark and (b), (d) bright 
soliton. A TF density profile is used as an envelope of the background density for the fitting. 
The red dashed circles indicate the collision of the most inner DB pairs.  
The remaining system parameters are the same as in Fig. \ref{fig::MiscImmGabComparison}. 
All quantities shown are given in dimensionless units.}
\end{figure}

In contrast to the above-observed dynamics within the miscible domain, for values of $g^f_{AB}$ that are above 
a certain threshold the miscible-to-immiscible phase transition takes place. 
As it is evident in Figs.~\ref{fig::MiscImmGabComparison} (c)-(f) the
dynamics in this regime is unstable and the initially localized, for each species, configuration    
breaks into several filaments~\cite{MistakidisQuenchInducedPhaseSeparation}. 
The filaments refer to the individual density branches appearing in $\abs{\Psi_{\sigma}}^2$ during evolution. 
For further details regarding this dynamical instability we refer the reader 
e.g. to Ref.~\cite{TommasiniBogoliubov}  
and references therein where the homogeneous case is investigated or the recent work of 
Ref.~\cite{MistakidisQuenchInducedPhaseSeparation} that includes the presence of the trap. 
By comparing the filamentation of each density for two different postquench interactions, we observe that for values of $g_{AB}^f$
that enter deeper into the immiscible phase more filaments are formed [compare Figs.~\ref{fig::MiscImmGabComparison} (c) and (d) 
to Figs.~\ref{fig::MiscImmGabComparison} (e) and (f) respectively].
Yet another important observation is that in all cases the number of filaments formed at the initial stages of the out-of-equilibrium 
dynamics is significantly larger compared to the remaining number of filaments robustly propagating at larger evolution times. In 
particular e.g. in Figs.~\ref{fig::MiscImmGabComparison} (c) and (d) around $t=35$ eight filaments are found to be spontaneously 
generated in both species [see also Figs. \ref{fig::MiscImmProfileFit} (a), (b)] while only four and three of them are present in the 
$A$ and $B$-species respectively for $t\geqslant 200$. 

Focusing on the initial stages of the dynamics, let us closer inspect the filaments formed. 
Figures~\ref{fig::MiscImmProfileFit} (a) and (b) present profile snapshots of 
the densities $|\Psi_A(x)|^2$ and $|\Psi_B(x)|^2$ of the majority and the minority species respectively at $t=35$ 
and for $g_{AB}^f=1.5$.    
Indeed, eight filaments are formed in each of the species with the density dips of the majority 
one being filled by density peaks of the minority species. 
Since such a filling mechanism resembles the formation of DB solitons in defocusing media 
below we attempt an identification of the structures formed to these symbiotic states.

As a first step towards confirmation we invoke the exact, at the integrable Manakov limit, 
single DB soliton solution~\cite{YanDarkBrightInteraction,
KatsimigaDarkBrightInteraction,KatsimigaDarkBrightBifurcationCollision}. 
The latter reads $\Psi_A(x,t)=A \left(\cos\phi \tanh \left[D \left(x-x_0(t)\right)\right]+i \sin \phi\right)$ 
and $\Psi_B(x,t)=B \sech \left[D \left(x-x_0(t)\right) \right] e^{ikx +i\theta(t)+i(\mu_B-\mu_A)t}$. 
Here, $\phi$ denotes the soliton's phase angle, while $A\cos\phi$, and $B$ refer to the amplitude of the dark and the bright 
soliton respectively. Furthermore, $D$ and $x_0(t)$ correspond to the common inverse width and the soliton's center respectively.
Finally, $k=D\tan\phi$ is the constant wave-number of the bright soliton and $\theta(t)$ its phase.
The above expressions are used for the density profile fits shown with dashed lines in Figs.~\ref{fig::MiscImmProfileFit} (a) and (b).
Evidently a remarkably good agreement between our numerical findings and the fitted profiles is seen [see also Table \ref{table}]. 
The latter verifies that indeed the filamentation process leads to structures that possess a DB solitary wave character. 
We note that these emergent dark states are characterized by a phase jump being a 
multiple of $\pi$ (results not shown here for brevity). 
The same overall phenomenology is observed for even higher values of $g^f_{AB}$ such as the one illustrated in the profile snapshots 
of Figs.~\ref{fig::MiscImmProfileFit} (c) and (d) for $g^f_{AB}=2.5$. 
Here, the number of DB states is significantly increased 
compared to the $g^f_{AB}=1.5$ scenario but again an excellent agreement 
between the numerical data and the fitted DB waveforms can be inferred [see also Table \ref{table}].    
Notice also that since in this case the number of solitary waves generated is large, already from these early stages of the dynamics
collisions between the DB states are clearly captured by this snapshot. 
Indeed, the collisions leave their fingerprints in the deformed densities, see the dashed circles which mark the two most inner pairs 
symmetrically placed around $x=0$ depicted in Figs.~\ref{fig::MiscImmProfileFit} (c) and (d).  
%
\begin{table}\centering
\begin{tabular}{|p{0.3cm}|p{1.1cm}||p{1.48cm}|p{1.48cm}|p{1.48cm}||p{1.48cm}|}
 \hline
 \multicolumn{6}{|c|}{{\bf DB characteristics between collisions}} \\
 \hline \hline 
  
 &~~~~~&~~~$t=35$~~~ & ~~~ $t=106$~~~&~~~$t=175$~~~&~~~$t=250$~~~~\\
 \hline \hline
 \parbox[t]{2mm}{\multirow{6}{*}{\rotatebox[origin=c]{90}{DB soliton 1}}}
 &$A\cos\phi$  & $1.1683$ & $1.1434$ & $1.2908$ &$1.2301$\\
 &$B$ & $0.9087$ & $1.0299$ & $0.7435$ & $1.1371$\\
 & $D$  & $0.5684$ & $0.5372$  & $0.4484$ & $0.4195$\\
 &$x_{0_{B_{\bf{1}}}}$  & $27.3425$ & $27.3687$  & $30.8764$ & $22.6316$\\
 &$x_{0_{D_{\bf{1}}}}$  & $27.3715$ & $27.3613$  & $31.1916$ & $22.4180$\\
 &$N_b$  & $2.9053$ & $\textbf{3.9491}$ & $\textbf{2.4656}$ & $\textbf{6.1651}$\\
 \hline 
 \hline
 \parbox[t]{2mm}{\multirow{6}{*}{\rotatebox[origin=c]{90}{DB soliton 2}}}
&$A\cos\phi$  & $1.1023$ & $1.2351$ & $1.2136$  & $-$\\
 &$B$ & $0.9788$ & $1.0675$ & $1.1102$ & $-$\\
 & $D$  & $0.5604$ & $0.7378$  & $0.5444$ & $-$\\
 &$x_{0_{B_{\bf{2}}}}$  & $17.8409$ & $20.4372$  & $19.7581$ & $-$\\
 &$x_{0_{D_{\bf{2}}}}$  & $17.8281$ & $20.5576$  & $19.7446$ & $-$\\
 &$N_b$  & $3.4191$ & $\textbf{3.0893}$ & $\textbf{4.5278}$ & $-$\\
 \hline
\end{tabular}
\caption{Fitted DB soliton characteristics referring to the different collisional events that take place during 
the spatio-temporal evolution of both species shown in Figs.~\ref{fig::MiscImmGabComparison} (c) and (d). 
Recall that $A\cos\phi$ [B] refers to the amplitude of the dark [bright] solitons generated in the $A$ [$B$] species. 
The selected time instants, i.e. $t=35$, $t=106$, $t=175$, and $t=250$, for which we have performed the fits in the density profiles 
are indicated in Figs.~\ref{fig::MiscImmGabComparison} (c) and (d) by vertical 
dashed lines, and the collision events during propagation are marked with coloured circles. 
The number of particles of each bright solitary wave, $N_b$, prior and after each collision point is 
depicted in boldface. Note also that in all cases the accuracy of the fitted values is of about $0.98$.}
\label{table}
\end{table}

After the initial stages of formation the DB solitary waves are left to dynamically evolve within the parabolic trap 
[see Figs.~\ref{fig::MiscImmGabComparison} (c)-(d) for $t>50$].
A closer inspection of the evolution of the density of each species reveals a collective breathing motion with 
four contraction or focus points where the DB states are closest. 
It is this breathing motion that leads in turn to the collision events that occur during evolution. 
Several collision events take place both around the center of the trap as well as at the edges of each bosonic cloud. 
These events are much more pronounced for smaller $g^f_{AB}$  
[compare Figs.~\ref{fig::MiscImmGabComparison} (c)-(d) with Figs.~\ref{fig::MiscImmGabComparison} (e)-(f)] with the DB states 
that are less mobile, i.e. those closer to the trap center, featuring more dramatic collisions. 
Recalling now that we operate in the nonintegrable limit these collisions can be rather 
asymmetric~\cite{KatsimigaDarkBrightBifurcationCollision} leading to a transfer of mass 
between the solitary wave constituents and thus 
to shape changing collisions~\cite{rajendran2011matter,manikandan2014generalized,KatsimigaDarkBrightBifurcationCollision}. 
It is this mass redistribution that results in a decreasing number of DB solitons but renders them wider and robustly propagating 
for large evolution times. 

In order to monitor such events and shed light on the observed dynamics we consider several collision points at different time 
instants during evolution [see the circles in Figs. \ref{fig::MiscImmGabComparison} (c), (d)]. 
For these selected events we perform fittings prior and after 
its collision vertex. In doing so we immediately have access to the relevant amplitudes and widths of the states  
and we can further compare the number of particles, $N_b$, contained in each bright solitary wave prior and after a collision. 
In order to perform the latter calculation we make use of the analytical expression for the single DB 
state~\cite{YanDarkBrightInteraction,KatsimigaDarkBrightInteraction,KatsimigaDarkBrightBifurcationCollision}, 
i.e. $N_b=\int~dx~|\Psi_B|^2=2B^2/D$, that is exact at the integrable limit but approximately holds also in our setup 
\footnote{For the numerical evaluation of $N_b^{num}$ for a selected DB pair we used $N_b^{num}=(1/N_B)\int_{c_1}^{c_2} dx 
\abs{\Psi_B}^2$. 
Here, $N_B$ is the total number of particles in the $B$-species and the integration interval $[ c_1$, $c_2 ]$ refers to the spatial 
area around the center of the selected bright soliton. 
The observed deviation between $N_b^{num}$ and $N_b$ is smaller than $1\%$.}. 
The corresponding outcomes of such a process are summarized in Table~\ref{table} for e.g. postquench amplitude $g^f_{AB}=1.5$.
In particular, in this Table we present the fitted solitary wave parameters when monitoring the two most outer DB states 
labeled as ``1" and ``2" in Figs.~\ref{fig::MiscImmGabComparison} (c)-(d). 
These DB waves participate in two distinct collision events that take place during evolution 
one at around $t\approx 140$ and a second one around $t\approx 210$ 
[see the light blue and green circles in Figs.~\ref{fig::MiscImmGabComparison} (c)-(d)]. 
By measuring the number of particles, $N_b$, contained in each bright solitary wave prior and after the first collision namely at 
$t=106$ and $t=175$ respectively (see Table~\ref{table}) it is found that indeed a mass redistribution between the solitary waves 
occurs (see the corresponding boldfaced $N_b$ values in Table~\ref{table}). Notice that after the first collision 
the number of particles contained in ``2" is almost two times that of state ``1''.
Remarkably, at later propagation times, i.e. after the second collision takes place,   
an almost complete transfer of mass results in a single but wider bright solitary wave (see Table~\ref{table} at $t=250$)
that remains and robustly propagates in the BEC medium. 
Notice that its corresponding dark counterpart is also wider [see Fig.~\ref{fig::MiscImmGabComparison} (c) at $t=250$]. 
We remark that the symmetric, with respect to $x=0$, DB states have exactly the same dynamical evolution. 

Finally, we perform the relevant fittings for the huge merger that appears around the trap center after the collision of at least 
three DB pairs that takes place at around $t\approx 140$ [see the blue circle in Fig.~\ref{fig::MiscImmGabComparison} (d)]. 
From this fitting we can verify via the particle number conservation prior and after the collision point, 
that indeed a total transfer of mass of the two outer DB pairs to the central 
one occurs. 
In particular, $N_{b_I}=4$, $N_{b_{II}}=5$ and $N_{b_{III}}=4$ prior the collision 
leading to the single DB solitary wave with $N_{b}=13$. 

\begin{figure*}[htb]
\centering \includegraphics[width=0.8\textwidth]{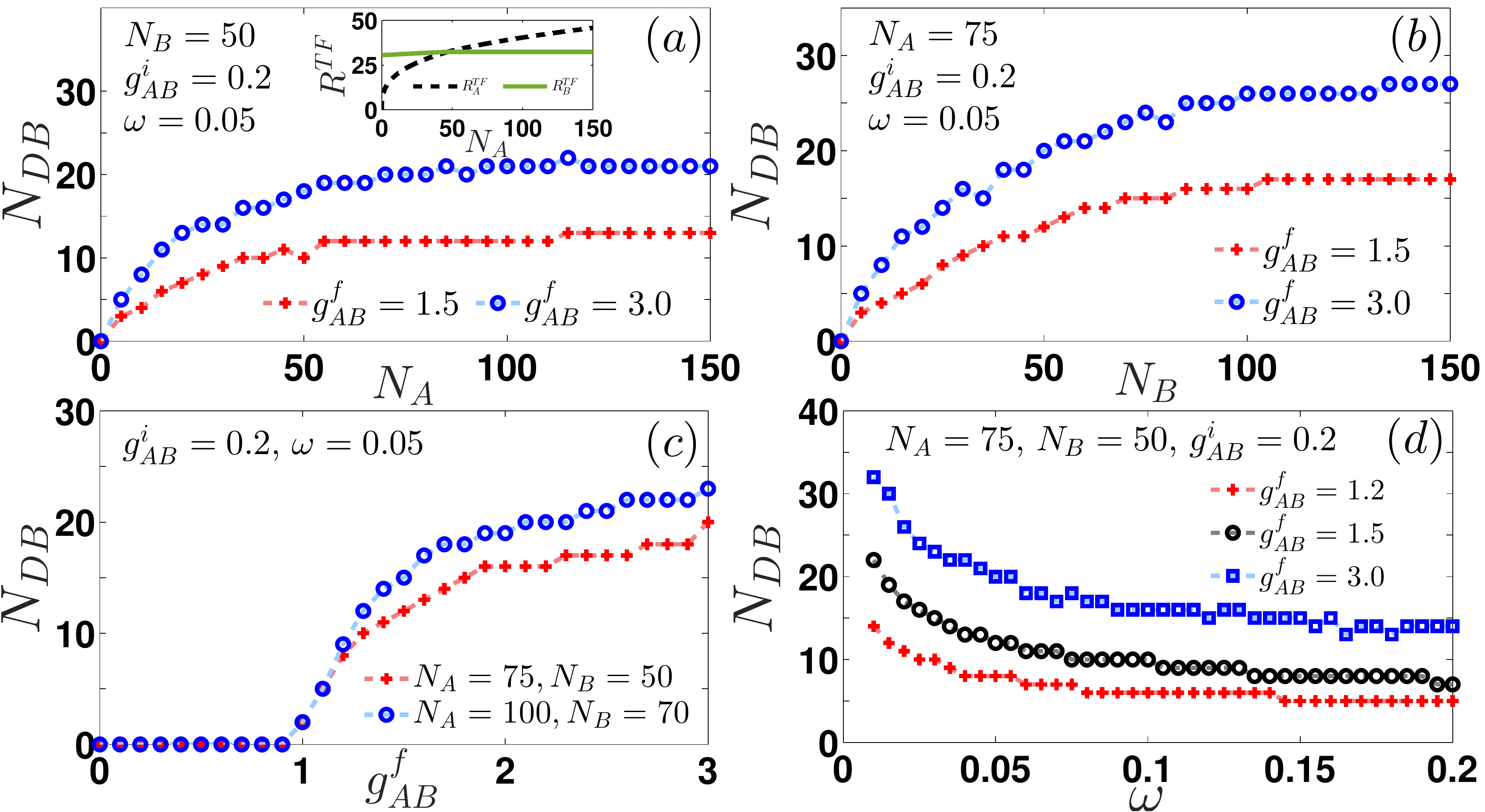}
\caption{\label{fig::MiscImmOverview} Number of DB solitons, $N_{DB}$, formed when considering a quench from the miscible to the 
immiscible phase for varying system parameters. 
(a) $N_{DB}$ upon increasing the particle number, $N_A$, of the $A$-species. 
Inset illustrates the TF radii, $R^{TF}_{\sigma \bar{\sigma}}$ with $\sigma=A,B$ and $\bar{\sigma}=B,A$, for the same variation. 
(b) The same as (a) but varying the particle number, $N_B$, of the $B$-species. 
$N_{DB}$ as a function of (c) the final postquench interaction $g_{AB}^f$ and (d) the trapping frequency $\omega$. 
In all cases the different parameters used are indicated in the respective legends.
Additionally, $g_{AA}=1.0$, $g_{BB}=0.95$ and the system is initialized in its ground state characterized by the corresponding 
$g_{AB}^i$. 
All quantities are given in dimensionless units.}
\end{figure*}

\subsection{Controlling the number of DB states}

Having identified the DB solitary waves generated via the filamentation process, in what follows we aim at controlling their 
formation. 
Thus we investigate how the number of DB states, $N_{DB}$, changes 
upon considering different variations of the binary systems' 
parameters. 

In particular, Fig.~\ref{fig::MiscImmOverview} $(a)$ illustrates $N_{DB}$ upon increasing the particle number, $N_A$, 
of the $A$-species for $N_B=50$. As it can be seen, solitary wave formation occurs already from small particle numbers ($N_A>10$) in 
the $A$-species. 
As $N_A$ increases $N_{DB}$ increases in a square root fashion with more DB states being generated for larger postquench 
interactions $g^f_{AB}$.
This behaviour can be intuitively understood by considering the size of the two BECs. 
To estimate the system's size, we start from Eqs.~(\ref{eqn::GPE_A})-(\ref{eqn::GPE_B}) and by making use of the TF 
approximation we derive, within the miscible regime, the following expressions for the corresponding radii of the two 
clouds~\cite{PethickSmith} 
\begin{align}
    R_{\sigma}^{TF} &= \omega^{-2/3} \left(\frac{3}{2}\left(g_{\sigma \sigma} N_{\sigma} 
    + g_{\sigma \bar{\sigma}} N_{\bar{\sigma}} \right)\right)^{1/3}, \label{radiusa}\\
    R_{\bar{\sigma}}^{TF} &= \omega^{-2/3} \left(\frac{3}{2}\frac{g_{\sigma\sigma} g_{\bar{\sigma} \bar{\sigma}} 
    - g_{\sigma \bar{\sigma}}^2}{\left(g_{\sigma \sigma}-g_{\sigma \bar{\sigma}}\right)}N_{\bar{\sigma}}\right)^{1/3}.
    \label{radiusb}
\end{align}
In the above expressions, $\sigma=A,B$ ($\bar{\sigma}=B,A$) denotes the majority (minority) component. 
It is now important to note that DB solitary waves will only form in regions where both species are present.
The latter sets the scale for solitary wave formation as the region defined by the smaller of the two aforementioned radii.    
In the inset of Fig.~\ref{fig::MiscImmOverview} $(a)$ $R^{TF}_A$, and $R^{TF}_B$ 
are shown as a function of $N_A$ for $g^f_{AB}=1.5$. 
For $N_A < N_B$ the TF radius of the $A$-species is the smaller one and as such is provided by Eq.~(\ref{radiusb}).
Indeed, $R^{TF}_A$ increases until the particle balanced limit is reached ($N_A=N_B=50$) where $R^{TF}_A=R^{TF}_B$~\footnote{Note that 
these equalities are exact at the integrable limit,
but approximately hold also here with a deviation of about $2\%$}. 
It is in this region, i.e. $N_A<N_B$, that the growth rate of $N_{DB}$ is significant.
On the other hand when $N_A > N_B$,  $R^{TF}_B$ becomes the smaller radius. 
In this case, unlike $R_A^{TF}$,  $R^{TF}_B$ remains almost constant for increasing $N_A$. 
Thus, since $R^{TF}_B$ remains constant as $N_A$ is increased, 
a plateau of almost constant DB soliton production appears. 

The same qualitative picture is expected to hold also upon varying the number of particles, 
$N_B$, of the second species, since fixing $N_A$ and varying $N_B$ simply interchanges the role of the two components.
Indeed as illustrated in Fig.~\ref{fig::MiscImmOverview} $(b)$, solely focusing on the case where $N_B<N_A$,
again a square-root-like increase of $N_{DB}$ is observed leading to more solitary waves being generated as we enter deeper  
into the immiscible regime of interactions. Notice that for $N_B \geq N_A$ again an almost constant production of DB states is 
observed. 
However since in this case $N_A$ is fixed to a larger value compared to the previous variation, $N_{DB}$ acquires larger values before 
the particle balanced limit is reached. 

The influence of the postquench interspecies interaction $g_{AB}^f$ on the DB formation 
is shown in Fig.~\ref{fig::MiscImmOverview} $(c)$. 
Here and as per our discussion above, for $g^f_{AB}<g^{th}_{AB}\equiv0.974$ (i.e. within the miscible domain) $N_{DB}=0$. 
Solitary wave formation takes place only for values of $g^f_{AB}>g^{th}_{AB}$, with $N_{DB}$ increasing in an almost square root 
manner for increasing postquench interactions. 
Additionally here, $N_{DB}$ is larger even for slightly larger particle number imbalances. 
Recall that the observed dynamics shown in Figs.~\ref{fig::MiscImmGabComparison} (c)-(f) dictates also an increase of $N_{DB}$ being 
generated faster for larger $g_{AB}^f$. 
The latter is consistent with earlier predictions \cite{MistakidisQuenchInducedPhaseSeparation} regarding  
particle balanced mixtures. 

Furthermore, in order to examine the effect of the trapping geometry on the solitary wave formation, we next calculate
$N_{DB}$ upon varying the trapping frequency $\omega$. The outcome of this parametric variation is depicted in
Fig.~\ref{fig::MiscImmOverview} $(d)$. $N_{DB}$ decreases rapidly as $\omega$ increases but acquires 
larger values for larger $g^f_{AB}$. 
Note also that as we approach the homogeneous limit, $\omega \rightarrow 0$, 
$N_{DB}$ tends to infinity a result that is explained by the 
increasing size of the condensates. Indeed the radii of the condensates [see Eqs. (\ref{radiusa})-(\ref{radiusb}) depend on 
$\omega^{-2/3}$ which is in accordance to what we observe. 

Finally, let us stress at this point that the observed dynamical evolution of the binary mixture can be experimentally 
monitored by performing in-situ single-shot absorption 
measurements~\cite{MistakidisQuenchInducedPhaseSeparation,bloch2008many,katsimiga2017many}.
Indeed, the DB solitary waves building upon the densities of the different species can be captured by such images.
For this purpose we further consider mixtures with larger particle numbers in both species showcasing
the generalization of our findings.
Evidently, as can be seen in Table~\ref{tableii} 
the same overall controlled generation of DB states persists upon varying distinct parameters of the system. 
\begin{table}\centering
\begin{tabular}{|p{0.01cm}p{1.1cm}|p{1.48cm}|p{1.48cm}|p{1.48cm}|p{1.48cm}|}
 \hline
 \multicolumn{6}{|c|}{{\bf Miscible to Immiscible transition}} \\
 \hline \hline 
  
 &~~$g^f_{AB}$~~~&~~~~~~$\omega$~~~ & ~~~~~$N_A$~~~& ~~~~~$N_B$~~~&~~~~$N_{DB}$~~~~\\
 \hline \hline
 &~~$1.2$  &~~~~~$0.05$ &~~~~ $10^4$ &~~~~ ${\bf 10^3}$ &~~~~~$47$\\
 \hline
 &~~$1.2$  &~~~~~$0.05$ &~~~~ $10^4$ &~~ ${\bf 5\times 10^3}$ &~~~~$114$\\
 \hline
 &~~$1.2$  &~~~~~$0.05$ &~~~~ $10^4$ &~~ ${\bf 8\times 10^3}$ &~~~~$136$\\
 \hline \hline
 &~~${\bf 1.5}$  &~~~~~$0.05$ &~~~~ $10^4$ &~~ ${5\times 10^3}$ &~~~~$160$\\
 \hline 
 &~~${\bf 2.0}$  &~~~~~$0.05$ &~~~~ $10^4$ &~~ ${5\times 10^3}$ &~~~~$237$\\
 \hline \hline
 &~~${1.2}$  &~~~~~${\bf 0.10}$ &~~~~ $10^4$ &~~ ${5\times 10^3}$ &~~~~~$96$\\
 \hline 
 &~~${1.2}$  &~~~~~${\bf 0.15}$ &~~~~ $10^4$ &~~ ${5\times 10^3}$ &~~~~~$85$\\ 
 \hline
\end{tabular}
\caption{Controlled DB solitary wave formation for a miscible-to-immiscible transition with $g^i_{AB}=0.2$, 
upon significantly enlarging the system size. In all cases $N_A=10^4$ is fixed
while the bold-faced quantities are the ones that are varied in each distinct simulation.    
Notice the significantly larger DB soliton count as the system size increases. }
\label{tableii}
\end{table}

\section{Quench Dynamics from the Immiscible to the Miscible Regime}\label{ImmiscibleToMiscibleDynamics} 

Up to now we only considered quenches from the miscible to the immiscible regime of interactions.
Next, immiscible-to-miscible transitions will be investigated. For this reverse process, the binary system 
is initially prepared in its ground state having a fixed $g_{AB}^i=1.5$. 
Then, the $A$-species resides at the edges of the harmonic trap since it possesses a larger intraspecies coefficient ($g_{AA}$)
while the $B$-species occupies the central region. Such an initial state configuration  
represents the 1D analogue of the so-called ``ball and shell" structure found in higher dimensional 
mixtures~\cite{BigelowBallShell,MertesBallShell}.  
After this initial state preparation the system is abruptly quenched to a lower $g_{AB}^f$ value lying within the miscible regime 
of interactions. 
\begin{figure}[htb]
\centering \includegraphics[width=0.5\textwidth]{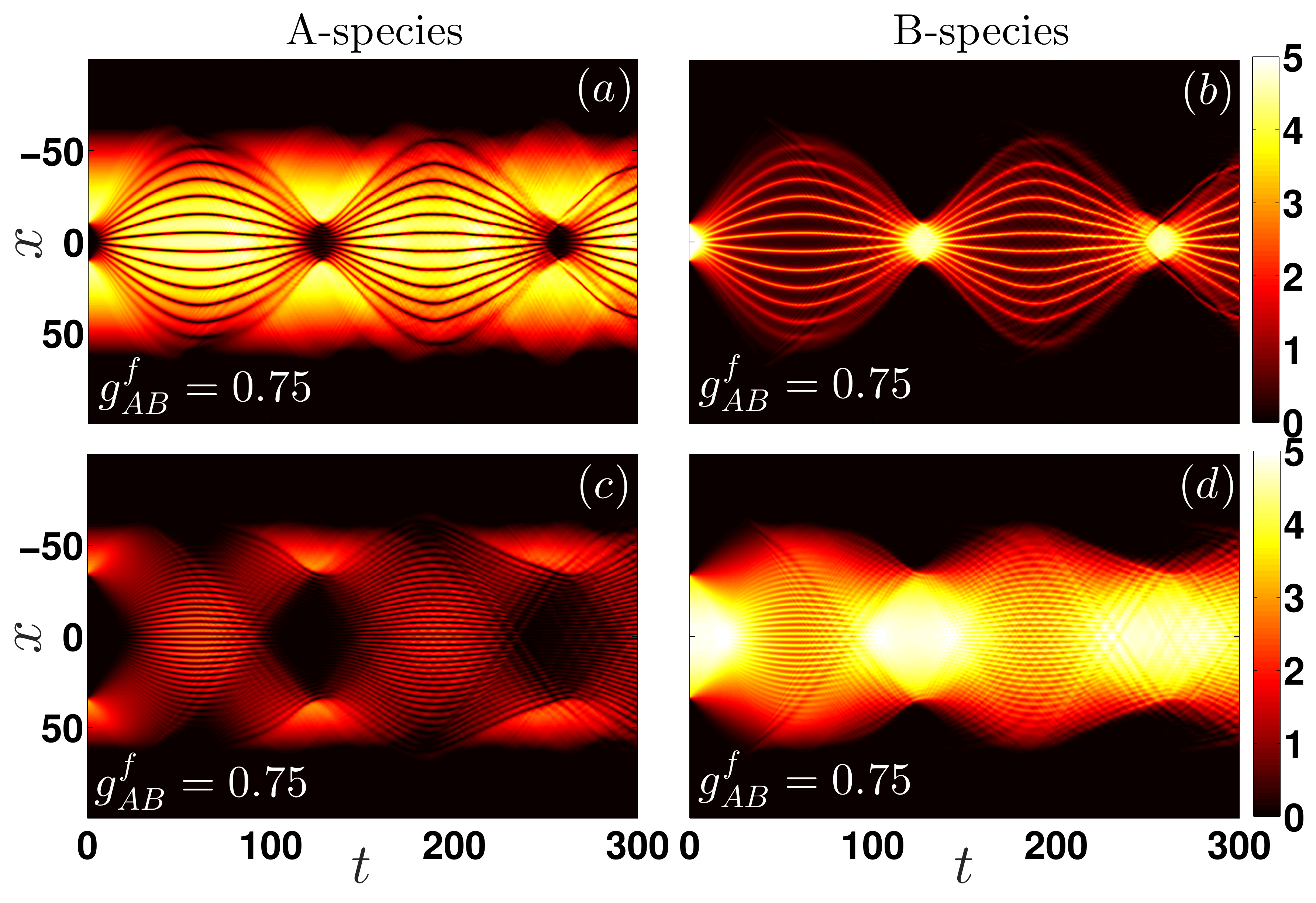} 
\caption{\label{fig::ImmMiscGabComparison} One-body density evolution $|\Psi_{\sigma}|^2$ of $\sigma=A$  (left panels) and 
$\sigma=B$-species (right panels) respectively, performing an interspecies interaction quench from $g_{AB}^i=1.5$ to $g_{AB}^f=0.75$. 
The system consists of (a), (b) $N_A=300$, $N_B=100$ and (c), (d) $N_A=100$, $N_B=300$ bosons. 
In all cases both species are trapped in a harmonic oscillator of frequency $\omega=0.05$. 
The system is initialized in its ground state with $g_{AA}=1.0$, $g_{BB}=0.95$ and $g_{AB}^i=1.5$. 
All quantities are given in dimensionless units. }
\end{figure}

\subsection{Effective counterflow dynamics and emergence of DAD states}

\begin{figure}[htb]
\centering
\includegraphics[width=0.5\textwidth]{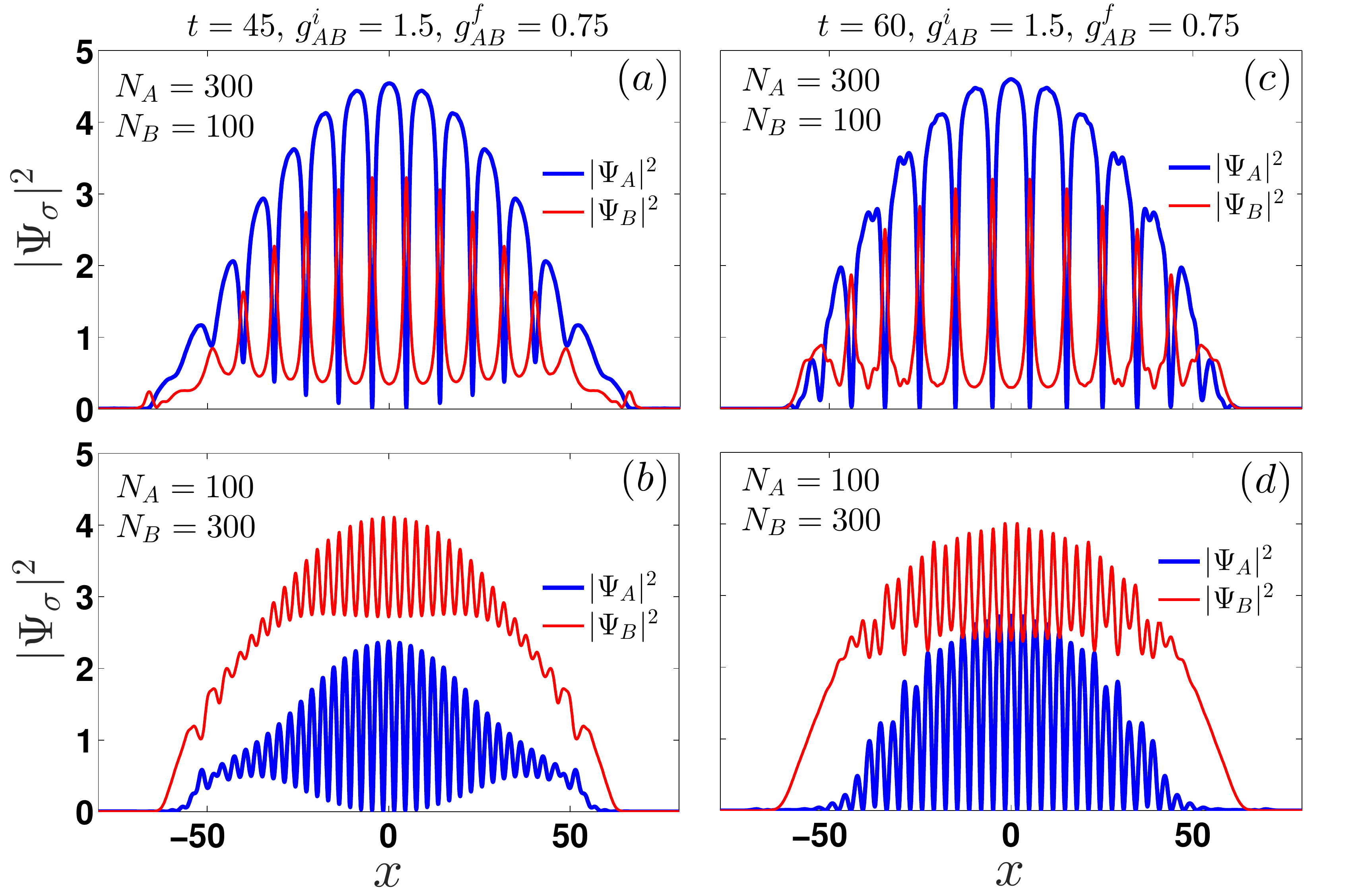}
\caption{\label{fig::ImmMiscFit} Profiles of the one-body densities of each species (see legend) at distinct time instants of the 
evolution. 
In (a), (c) $N_A=300$, $N_B=100$ while in (b), (d) $N_A=100$ and $N_B=300$. 
The remaining system parameters are the same as in Fig. \ref{fig::ImmMiscGabComparison}. 
All quantities shown are in dimensionless units. }
\end{figure}

Case examples of the spatio-temporal evolution of the densities of both species for two different particle 
number imbalances, namely for $N_B/N_A=1/3$ and $N_B/N_A=3$, but for the same postquench interspecies repulsion are shown in  
Figs.~\ref{fig::ImmMiscGabComparison} (a), (b) and (c), (d) respectively.
In particular, Figs.~\ref{fig::ImmMiscGabComparison} (a) and (c)  
depict the evolution of the density of the $A$-species and Figs.~\ref{fig::ImmMiscGabComparison} (b) and (d)  
the corresponding evolution of the $B$-species. In both cases pattern formation takes place from the early 
stages of the nonequilibrium dynamics but with a dramatically larger number of states formed when the inner $B$-species
is the majority component of the bosonic mixture. 
The latter outcome can be inferred by inspecting the corresponding profile snapshots illustrated in Figs.~\ref{fig::ImmMiscFit} 
(a)-(d). Here, Figs.~\ref{fig::ImmMiscFit} (a), (b) and (c), (d) 
show the density profiles of both species at the initial stages of 
formation, i.e. at $t=45$ and $t=60$ respectively, for $N_B/N_A=1/3$ and $N_B/N_A=3$. 
Evidently, at least twelve localized entities appear in the former snapshots, namely when the outer $A$-species is the majority 
component, while at least thirty occur when the inner $B$-species is the majority component. 
In this latter case the localized structures closely resemble interference fringes. 
Notice that in both cases, since component mixing is favored in this regime, the $B$-species tend to fill the quench induced minima 
of the $A$-species. 
However, when the $B$-species is the majority component an overfill of the corresponding $A$-species minima occurs 
[see the relevant profiles in Figs. \ref{fig::ImmMiscFit} (b), (d)] and thus no robust localized configuration can be realized. 
Before delving in the associated dynamics, we should emphasize that in contrast to the previous quench scenario,
the structures formed in this case are not only different in nature when compared to the DB solitary waves previously generated,
but appear also to be more robust during evolution [compare Figs.~\ref{fig::ImmMiscGabComparison} (a)-(d) to 
Figs.~\ref{fig::MiscImmGabComparison} (c)-(f)]. 
Thus the breaking of integrability within the miscible domain doesn't
seem to affect the evolution of the states generated, a result that is in line with earlier 
predictions~\cite{MistakidisQuenchInducedPhaseSeparation}. 

In particular, a closer inspection of the spatio-temporal evolution of the densities 
[see Figs.~\ref{fig::ImmMiscGabComparison} (a)-(d) and Figs.~\ref{fig::ImmMiscFit} (a)-(d)]
reveals that dark soliton entities emerge during evolution in the $A$-species, 
while antidark solitons develop in the corresponding $B$-species. 
Note that the antidark solitons are states consisting of a density peak on top of the BEC 
background~\cite{DanailaDarkAntidarkBEC,KevrekidisFamiliesOfMatterWaves,
MistakidisQuenchInducedPhaseSeparation,KatsimigaDarkBrightDynamics}.  
Thus in contrast to the DB solitary waves
formed in the previous quench scenario, here multiple DAD solitons are spontaneously generated. 
The formation of these DAD states has been reported also in the particle balanced case~\cite{MistakidisQuenchInducedPhaseSeparation}. 
Below, and in contrast to earlier findings we will explore in detail the mechanism of the formation of these DAD
states and we will also attempt to control their generation upon considering different variations of the binary systems' 
parameters. 

To unravel the underlying mechanism of formation of the aforementioned DAD states 
we must first recall that the initial state configuration
consists of two spatially separated components. 
As the system is dynamically quenched towards the species miscible regime,
the interspecies interaction energy, $\Delta E_{int}=2\abs{\Psi_A}^2\abs{\Psi_B}^2(g_{AB}^f-g_{AB}^i)$, 
decreases the more the smaller the $g_{AB}^f$ becomes. 
Then, the $B$-species is no longer deemed to reside at the central region of the parabolic potential but instead it can expand. 
More importantly the $A$-species residing previously symmetrically around the edges of the trap is now allowed to ``fall" towards the 
center. 
It is this counterflow of the $A$-species that causes the formation of the dark solitons right at the trap center where   
the two parts of the gas destructively interfere~\cite{AndrewsBECInterference}. 
Note also that in our initial state preparation we start with a constant zero phase in the 
$A$-species and thus we expect and indeed 
observe an even number of dark solitons to emerge~\cite{Zakharov1973} 
[see also here the profiles shown in Figs.~\ref{fig::ImmMiscFit} (a), (c)].   
This dark soliton formation leads in turn, due to the effective potential that 
these states create, to the breaking of the $B$-species into several 
density peaks each of which filling the newly formed density dips.
Essentially it is energetically beneficial for the $B$-species to avoid regions where the $A$-species 
has a high density, since the interaction between the species is repulsive ($g^f_{AB}>0$). 
Thus the $B$-species tends to fill the $A$-species minima. 
This effect is more severe for weaker repulsions that enter deeper in the miscible regime, with the number of the 
emergent DAD states found to increase for smaller $g_{AB}^f$ (see our discussion below).
However since the binary system is in the miscible regime where a mixed state is favorable the generated density
peaks are formed on top of their BEC background. 
The above-discussed mechanism of formation of the observed DAD states 
is fundamentally different in nature, and is to be contrasted with the unstable dynamics that leads to the 
formation of the DB structures identified in the miscible-to-immiscible transition.

To testify the formation of dark solitons in our setup, we next rely on predictions stemming 
from counterflow experiments~\cite{WellerDark}. 
It is known that for single component BECs a critical distance
between the two initially separated BECs exists below which soliton formation takes place~\cite{Scott1998}.  
This critical distance in the dimensionless units adopted herein reads 
$D_c = \pi \left( \frac{6 \pi N_A g_{AA}}{\omega_{eff}^A} \right)^{1/3}$. 
Here, $\omega_{eff}^{A} = \omega \sqrt{1-\frac{g_{AB}^f}{g_{BB}}}$ is the modified trapping frequency that 
incorporates the postquench cross interaction term $g_{AB}^f$. 
This effective potential picture \cite{mistakidis2019quench,ferrier2014mixture}, $V_{eff}=V(x) + g_{AB}^f|\Psi_B(x,0)|^2$, is 
analogous to the one introduced in section \ref{MiscibleToImmiscibleDynamics}. 
\begin{figure}[htb]
\centering
\includegraphics[width=0.5\textwidth]{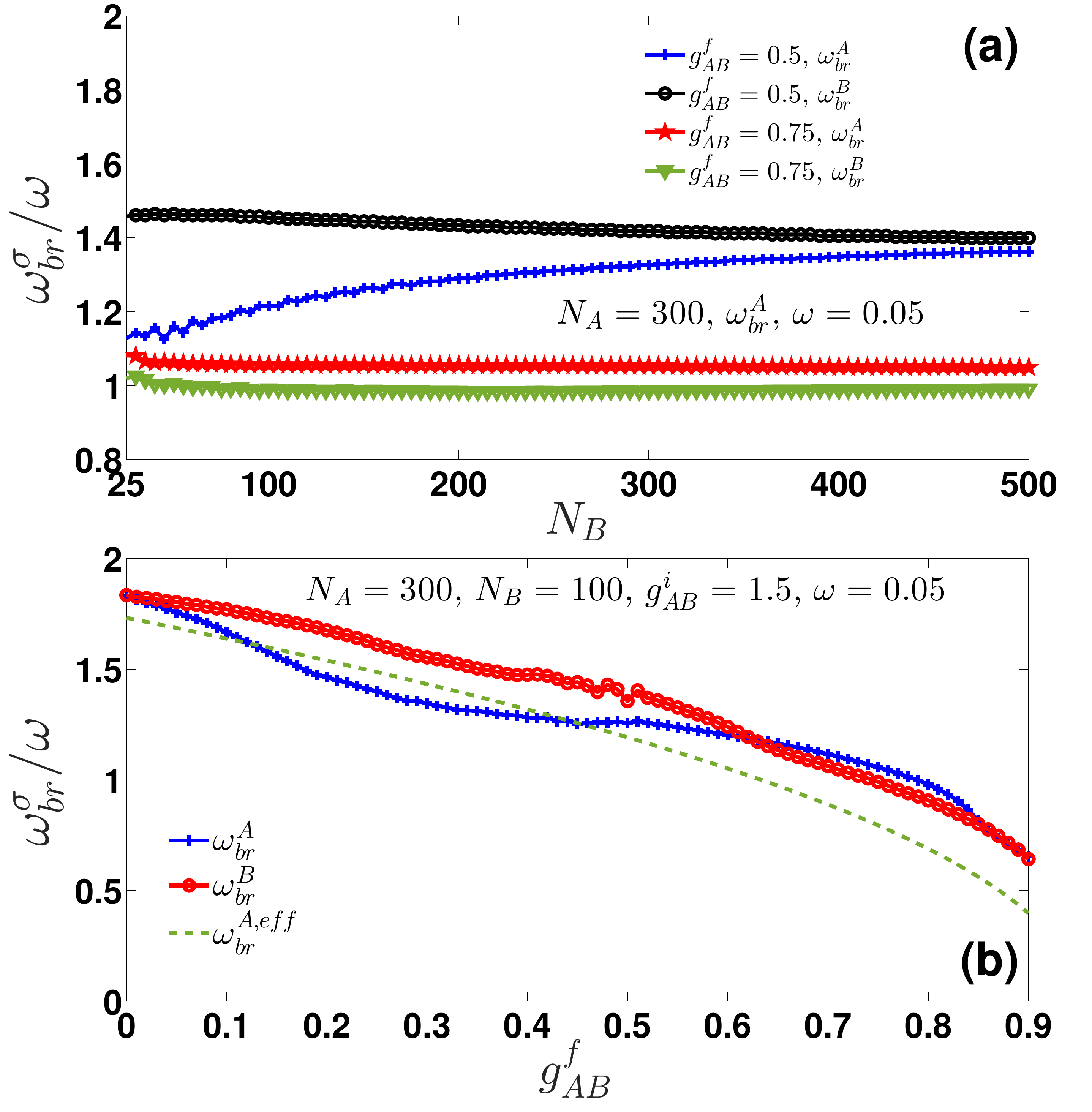}
\caption{\label{Breathimm} (a) Normalized breathing frequency, $\omega_{br}^{\sigma}/\omega$, 
of each component ($\sigma=A,B$) of the binary bosonic mixture upon varying the particle number, $N_{B}$, of the $B$-species. 
(b) $\omega_{br}^{\sigma}/\omega$ for different postquench interspecies interaction strengths $g_{AB}^f$. 
The remaining system parameters are fixed (see legends). 
In all cases the bosonic mixture is harmonically trapped with $\omega=0.05$ and it is initialized in its ground state 
characterized by $g_{AB}^i=1.5$. At $t=0$ a quench is performed to $g_{AB}^f$. 
All quantities are given in dimensionless units. }
\end{figure}
We remark here that in contrast to the previous quench scenario [section \ref{MiscibleToImmiscibleDynamics}] 
the $A$-species performs the counterflow dynamics with the $B$-species residing at the trap center. 
Thus, we consider the dynamics of the $A$-species into the effective potential formed by the $B$-species and the external trap. 
For the case example of $N_B/N_A=1/3$ presented in Fig. \ref{fig::ImmMiscGabComparison} (a) $D_c =197$ while 
$D=20.8$, verifying that the observed structures building upon the majority $A$-species are indeed dark solitons.  

\begin{figure*}[htb]
\centering
\includegraphics[width=0.8\textwidth]{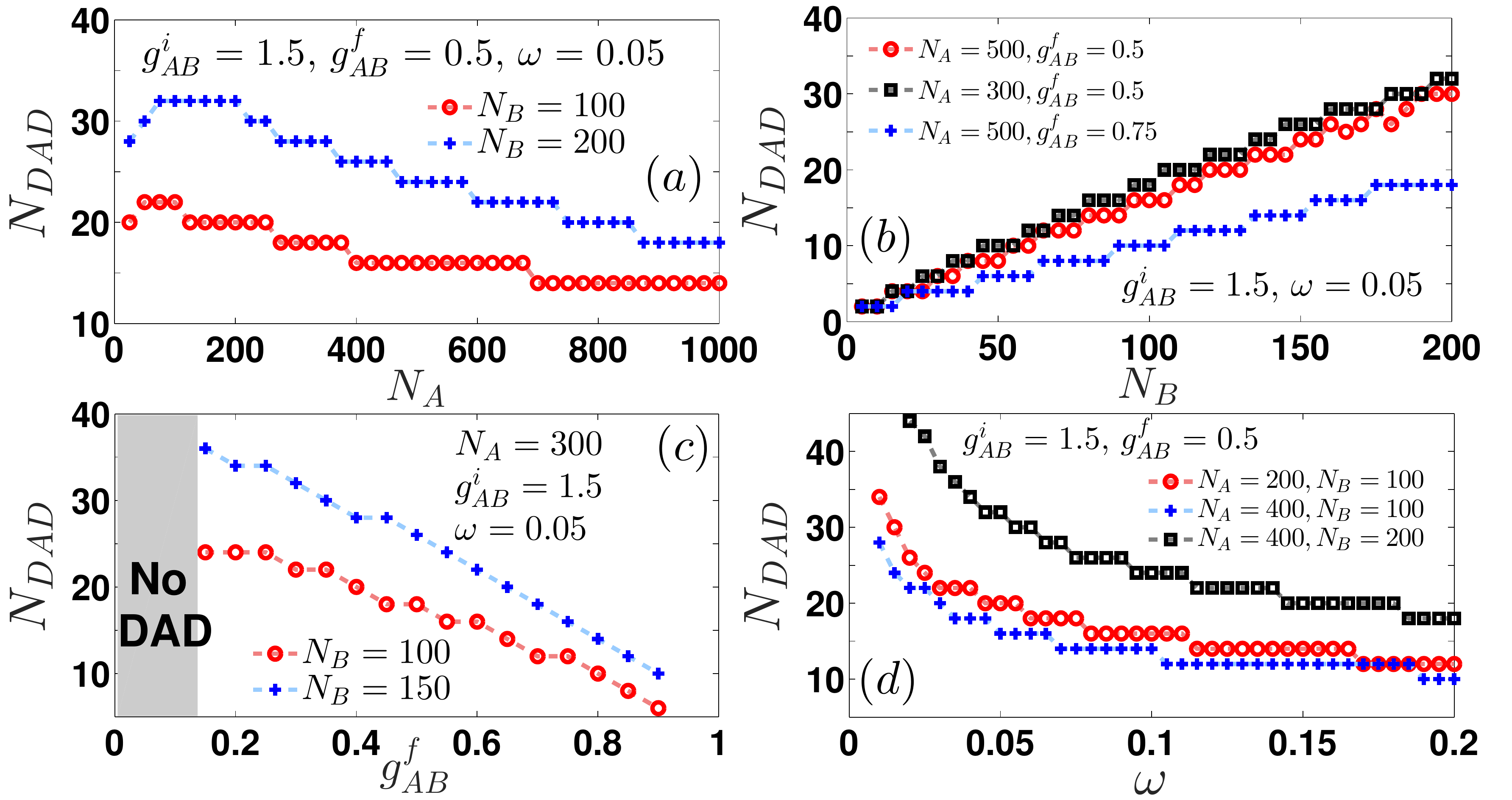}
\caption{\label{fig::ImmMiscOverview} 
Number of DAD solitons, $N_{DAD}$, created upon an interspecies interaction quench from the immiscible to the 
miscible phase for varying system parameters. 
(a) $N_{DAD}$ upon increasing the particle number, $N_A$, of the $A$-species. 
(b) $N_{DAD}$ but varying the particle number, $N_B$, of the $B$-species. 
$N_{DAD}$ as a function of (c) the final postquench interaction $g_{AB}^f$, (d) the trapping frequency $\omega$. 
In all cases the different parameters used are indicated in the respective legends.
Additionally, $g_{AA}=1.0$, $g_{BB}=0.95$ and the system is initialized in its ground state characterized by the corresponding 
$g_{AB}^i$. 
All quantities shown are in dimensionless units. }
\end{figure*}

\subsection{Breathing mode}

Besides the formation of the DAD structures the two species also undergo a collective breathing motion 
as it can be readily seen in Fig.~\ref{fig::ImmMiscGabComparison}. 
Results for the corresponding normalized breathing frequency, $\omega_{br}^{\sigma}/\omega$, 
are shown in Figs.~\ref{Breathimm} (a), (b),  
upon considering the effect of different 
particle number imbalances and postquench interspecies interaction strengths respectively. 
Referring to a fixed postquench interaction strength $\omega_{br}^A$ and $\omega_{br}^B$ are significantly different 
especially in the region where $N_A>N_B$, see Fig.~\ref{Breathimm} (a). 
This deviation is much more pronounced for smaller $g_{AB}^f$. 
Within the region $N_A<N_B$, the frequencies $\omega_{br}^A$ and $\omega_{br}^B$ tend to approach 
each other but they never coincide. 
For fixed particle number imbalances, i.e. $N_B/N_A=1/3$, an increase of $\omega_{br}^{\sigma}/\omega$ is observed as the system is 
quenched to lower values of $g^f_{AB}$ that enter deeper in the miscible regime of interactions [see Fig.~\ref{Breathimm} (b)]. 
More specifically, both species appear to have almost the same breathing frequency for quenches 
near the miscibility/immiscibility threshold but deviate significantly for values of $g^f_{AB}$ within the interval
$[0.6,0.1]$. In this later case, $\omega_{br}^{B}/\omega$ appears to be higher for all values of the postquench interaction 
$g^f_{AB}$ until the noninteracting limit is reached, with both species acquiring the value 
$\omega_{br}^{\sigma}/\omega\approx \sqrt3$. 
Moreover, the effective breathing frequency, $\omega_{br}^{A,eff}$ (see also section \ref{MiscibleToImmiscibleDynamics}), seems to 
underestimate $\omega_{br}^{\sigma}/\omega$ the more, the closest we are to the 
miscibility/immiscibility threshold. The observed deviations are mainly attributed to the fact that the effective model
assumes a TF profile for the minority $B$-species which differs considerably from the actual density of this species 
due to the dynamical generation of antidark solitons. 

\subsection{Controlled creation of DADs}

Having discussed the mechanism of formation of DAD states, we next investigate whether their number can be tuned via 
parametric variations. Figures~\ref{fig::ImmMiscOverview} (a)-(d) summarize our findings.
We find that the number of DAD states, $N_{DAD}$, strongly depends 
on variations of the particle number either of the $A$-  
or the $B$-species [see Figs.~\ref{fig::ImmMiscOverview} (a), (b)]. 
In particular, in both cases illustrated in Fig.~\ref{fig::ImmMiscOverview} (a)  
a linear increase of $N_{DAD}$ for increasing $N_A$ is observed. 
This increase is followed by a plateau of constant soliton count until the particle balanced limit is reached, 
with $N_{DAD}$ being larger for larger $N_B$.
Further increase of $N_A$, namely for $N_B/N_A<1$ leads to a gradual decrease of $N_{DAD}$ 
with less entities formed for smaller ratios. 
This behavior can be intuitively explained by recalling that increasing $N_A$ also increases the 
size of the $A$-species. 
The latter initially reside at the edges of the trap squeezing more and more the central $B$-species 
and thus reducing the initial 
separation of the outer BECs. Smaller initial spatial separation leads in turn to less kinetic energy for the 
counterflowing parts of the $A$-species and thus to less solitons being spontaneously generated. 
To inspect more carefully the case of $N_B/N_A<1$, Fig. \ref{fig::ImmMiscOverview} (b) shows $N_{DAD}$ for 
varying $N_B$ and for different values of $g_{AB}^f$ but fixed $N_A$. 
We observe that increasing $N_B$ leads to a linear increase of $N_{DAD}$ with the latter being larger 
for smaller $g_{AB}^f$. 
Also $N_{DAD}$ is larger for smaller ratios of $N_B/N_A$ and fixed $g_{AB}^f$. 

The impact of $g_{AB}^f$ on $N_{DAD}$ is also examined in Fig. \ref{fig::ImmMiscOverview} (c) for 
constant particle imbalances. 
As it can be seen, the number of solitons increases in an almost linear manner for smaller $g_{AB}^f$, namely 
deeper in the miscible domain. 
The above-mentioned tendency can be understood as follows. 
For smaller values of $g_{AB}^f$ the two species featuring less repulsion tend to be closer 
to one another favoring a phase mixed state. 
As such the $A$-species acquires more kinetic energy during the dynamics caused by a quench 
to weaker $g_{AB}^f$ and thus resulting in a larger $N_{DAD}$. 
However, it is important to note that as the noninteracting limit is approached the DAD states become gradually less robust.  

Focusing on the antidarks formed in the $B$-species it is found that as $g^f_{AB}$ decreases 
the amplitude of these states also decreases.   
To appreciate this behaviour we introduce the visibility
\begin{equation}\label{eqn:visibility}
    \nu = \frac{n_{max}-n_{min}}{n_{max}+n_{min}}.
\end{equation}
Here, $n_{max}$ is the density at the peak of an antidark and $n_{min}$ is the corresponding background density in the minimum. 
In the case that the minimum densities are notably different on the left/right side of the peak, we define $n_{min}$ as the average of 
the two minima. 
As an example the maximum visibility of the central antidark soliton, is $\nu=80.5\%$ in Fig.~\ref{fig::ImmMiscFit} (a) and grows to 
$\nu=82.8\%$ in Fig.~\ref{fig::ImmMiscFit} (c), while is found to be $\nu=20.4\%$ and $\nu=25.7\%$ in Figs.~\ref{fig::ImmMiscFit} (b), 
(d) respectively.  
Moreover, the maximum visibility measured for the corresponding central antidark state for $g_{AB}^f=0.1$ is $\nu =10\%$, while it 
approaches zero for $g_{AB}^f\rightarrow 0$. 
Thus, in the following, we define a visibility threshold for DAD formation. 
In particular, we consider as DAD states the ones whose central antidark structure 
exhibits a visibility $\nu\geq 10\%$. 
For instance in the case of $N_B/N_A=1/3$, according to our simulations this condition is fulfilled 
only for quenches with $g_{AB}^f> 0.1$ [see also Fig. \ref{fig::ImmMiscOverview} (c)]. 

The role of the trapping geometry is examined in Fig.~\ref{fig::ImmMiscOverview} (d). 
Here, a hyperbolic-like increase of $N_{DAD}$ can be seen for a decreasing trapping frequency $\omega$, 
leading to more DAD structures being spontaneously generated for larger $N_B/N_A$ ratios. 
The observed increase in $N_{DAD}$ can be easily explained by the overall size of the system 
which for lower $\omega$'s increases providing in this way more space for solitons to form. 

Finally, Table~\ref{tableiii} illustrates that the same overall phenomenology regarding the controlled formation of 
multiple DAD solitons holds for even larger bosonic ensembles. 

\begin{table}\centering
\begin{tabular}{|p{0.01cm}p{1.1cm}|p{1.48cm}|p{1.48cm}|p{1.48cm}|p{1.48cm}|}
 \hline
 \multicolumn{6}{|c|}{{\bf Immiscible to Miscible transition}} \\
 \hline \hline 
  
 &~~$g^f_{AB}$~~~&~~~~~~$\omega$~~~ & ~~~~~$N_A$~~~& ~~~~~$N_B$~~~&~~~~$N_{DAD}$~~~~\\
 \hline \hline
 &~~$0.75$  &~~~~~$0.05$ &~~~~ $10^4$ &~~~~ ${\bf 10^3}$ &~~~~~$40$\\
 \hline
 &~~$0.75$  &~~~~~$0.05$ &~~~~ $10^4$ &~~ ${\bf 5\times 10^3}$ &~~~~$160$\\
 \hline
 &~~$0.75$  &~~~~~$0.05$ &~~~~ $10^4$ &~~ ${\bf 8\times 10^3}$ &~~~~$234$\\
 \hline \hline
 &~~${\bf 0.80}$  &~~~~~$0.05$ &~~~~ $10^4$ &~~ ${5\times 10^3}$ &~~~~$142$\\
 \hline 
 &~~${\bf 0.40}$  &~~~~~$0.05$ &~~~~ $10^4$ &~~ ${5\times 10^3}$ &~~~~$334$\\
 \hline \hline
 &~~$0.75$  &~~~~~${\bf 0.10}$ &~~~~ $10^4$ &~~ ${5\times 10^3}$ &~~~~$130$\\
 \hline 
 &~~$0.75$  &~~~~~${\bf 0.15}$ &~~~~ $10^4$ &~~ ${5\times 10^3}$ &~~~~$120$\\ 
 \hline
\end{tabular}
\caption{Controlled DAD formation for an immiscible-to-miscible transition with $g^i_{AB}=1.5$, 
upon significantly enlarging the system size. In all cases $N_A=10^4$ is fixed
while the bold-faced quantities are the ones that are varied in each distinct simulation.    
Notice the significantly larger DAD soliton generation as the system size increases. }
\label{tableiii}
\end{table}

\section{Dynamics in a Quasi One-Dimensional Harmonic Trap}\label{quasi1D_simulations}

To further showcase that our above-described results regarding the spontaneous generation of solitonic structures can be detected in contemporary ultracold atom 
experiments we simulate the corresponding interaction dynamics in a quasi 1D geometry. 
In particular, we consider a binary BEC consisting of $^{87}$Rb atoms prepared in its hyperfine states $\ket{F=1,m_F= -1}$ (species A) and 
$\ket{F=2, m_F=1}$ (species B) and being confined in a quasi 1D harmonic trapping potential \cite{EgorovRbScatteringLengths,YanDarkDarkSolitonsBEC}. 
The underlying dynamics of the system is governed by the following set of three-dimensional coupled GPE equations
\begin{equation}\label{3D_GPE}
\begin{split}
&i\hbar \frac{\partial  \Psi_{\sigma} (\textbf{r}, t) }{\partial t} \\& = \left[-\frac{\hbar^2\nabla^2}{2m_{\sigma}}+ V_{\sigma}(\textbf{r})+  \sum_{\sigma'=A,B} g_{\sigma \sigma'}|\Psi_{\sigma'}
(\textbf{r}, t)|^2 \right ] 
\Psi_{\sigma}(\textbf{r}, t)  .
\end{split}
\end{equation}
In this expression, the spatial coordinates are denoted by $\textbf{r}=(x,y,z)$, the species index $\sigma = A, B$ while $m_{\sigma}$ and $\Psi_{\sigma}(\textbf{r}, t)$ 
denote the mass and the wavefunction of the $\sigma$-species respectively. 
Also, the intraspecies interaction strenghts $ g_{\sigma\sigma}=2\pi \hbar a_{\sigma \sigma}/m_{\sigma}$ involve the scattering lengths 
$a_{AA}$ and $a_{BB}$, whereas the interspecies coupling is $g_{AB} = 2 \pi \hbar a_{AB}/m_{AB}$ where $m_{AB} = m_A m_B/(m_A+m_B)$ is the reduced mass. 
Furthermore, we assume a three-dimensional harmonic oscillator potential, namely 
$V_{\sigma} (\textbf{r}) = V_{\sigma}(x,y,z) = m_{\sigma} \omega^2_{\sigma} (x^2 +\alpha_{\sigma}^2 y^2 + \lambda_{\sigma}^2 z^2 )/2$ with 
$\omega _{\sigma} $ being the trapping frequency along the $x$ direction, while $\alpha_{\sigma}$ and $\lambda_{\sigma}$ are two anisotropy 
parameters. 
For our purposes, the atoms of both components possess the same mass (two different hyperfine states of the
same atomic species) and are confined in the harmonic potential $V_{\sigma} (\textbf{r})$. 
As a consequence $m_A = m_B = m$, $\omega_A = \omega_B = \omega_x$, $\alpha_A = \alpha_B = \omega_y/\omega_x$, and $\lambda_A = \lambda_B = \omega_z/\omega_x$. 
The length and the energy scales of the system are expressed in terms of the harmonic oscillator length $a_{\rm osc} = \sqrt{\hbar/m \omega_x}$ and 
the energy quanta $\hbar \omega_x$ respectively. 
For the convenience in our numerical simulations the above set of coupled GPE equations [see Eq. (\ref{3D_GPE})] is expressed into a dimensionless 
form by rescaling the spatial and time coordinates as $x' = x/a_{\rm osc}$, $y' =y/a_{\rm osc}$, $z' = z/a_{\rm osc}$  
and $t'= \omega_x t$ repsectively. 
Additionally, the species wavefunction is $\Psi'_{\sigma}(x', y', z') = \sqrt{a_{\rm osc}^3/N_{\sigma}} \Psi_{\sigma}(x, y, z, t)$ and 
we also set $\hbar = 1$. 
For simplicity below we omit the primes. 

\begin{figure}[htb]
\centering \includegraphics[width=0.5\textwidth]{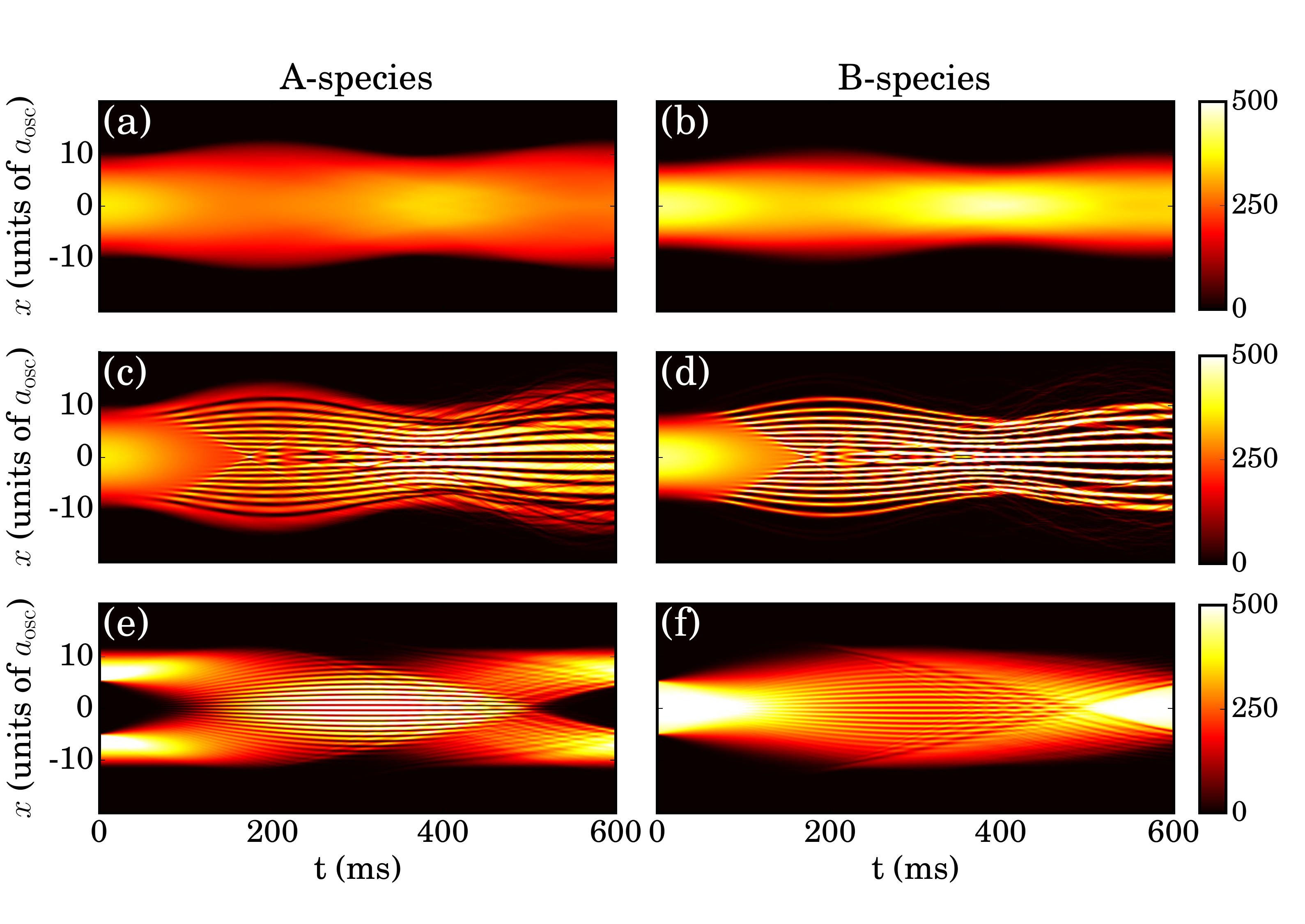} 
\caption{\label{fig:quasi-1D} One-body density evolution $|\Psi_{\sigma}|^2$ of the $\sigma=A$ (left panels) and the $\sigma=B$-species (right panels) 
respectively, performing an interspecies interaction quench from (a), (b) $a_{AB}^i=20a_0$ to $a_{AB}^f=80a_0$, (c), (d) $a_{AB}^i=20a_0$ to $a_{AB}^f=150a_0$, 
(e), (f) $a_{AB}^i=150a_0$ to $a_{AB}^f=75a_0$. 
In all cases the system consists of $N_A=5000$, $N_B=3000$ bosons and both species are trapped in a quasi 1D 
harmonic oscillator with frequencies $\omega_x=2\pi \times1.5Hz$, $\omega_y=2\pi \times140Hz$ and $\omega_z=2\pi \times178Hz$. 
The system is initialized in its ground state with $a_{AA}=100.4a_0$ and $a_{BB}=95.44a_0$. }
\end{figure}

Moreover, due to the quasi 1D geometry of the external trap $\omega_z, \omega_y  \gg \omega_x$, the wavefunction 
of each species is decomposed as follows  
\begin{equation}\label{ansatz_quasi1D}
\Psi_{\sigma} (x, y, z, t) = \psi_{\sigma} (x,t) \phi_{\sigma} (y) \phi_{\sigma} (z).
\end{equation} 
Here,  $\phi_{\sigma} (y) $, $\phi_{\sigma} (z) $ refer to the normalized ground state wavefunctions in the $y$ and $z$ spatial direction respectively. 
Note also that for all the simulations to be presented below we use $N_{A}=5000$ and $N_{B}=3000$ atoms in the corresponding species and 
the values $a_{AA}=100.4a_0$, $a_{BB}=95.44a_0$ for the intraspecies scattering lengths \cite{EgorovRbScatteringLengths}.
The latter correspond to the experimentally relevant 
scattering lengths of $^{87}$Rb, where $a_0$ is the Bohr radius. 
Additionally, for the harmonic oscillator potential we use the trapping frequencies $(\omega_x,\omega_y,\omega_z) = 2\pi \times (1.5, 140, 178)Hz$
as per the experiment of \cite{YanDarkBrightInteraction}. 
Notice that in this section we chose to present our findings in dimensional units in order to provide a direct connection with current state-of-the-art experimental settings. 
Our system is initially prepared in its ground state configuration being characterized by an initial interspecies scattering length $a_{AB}^i$. 
To trigger the dynamics an 
interaction quench is performed (as in Sections \ref{MiscibleToImmiscibleDynamics}, \ref{ImmiscibleToMiscibleDynamics}) towards $a_{AB}^f$ and we monitor the time-evolution 
of the binary mixture up to $600$ ms utilizing Eq. (\ref{3D_GPE}), see Fig. \ref{fig:quasi-1D}. 
Evidently, following an interspecies interaction quench within the miscible phase e.g. from $a_{AB}^i=20a_0$ towards $a_{AB}^f=80a_0$ both species undergo a breathing 
motion manifested by the expansion and contraction of their one-body densities [Figs. \ref{fig:quasi-1D} (a), (b)]. 
However, for a quench from the miscible towards the immiscible regime of interactions the spontaneous generation of DB soliton structures is expected and indeed observed [Figs. \ref{fig:quasi-1D} (c), (d)].  
Notice that dark solitons build upon the majority component and the bright ones develop in the minority species following a quench from $a_{AB}^i=20a_0$ to $a_{AB}^f=150a_0$. 
Finally, upon considering an interspecies interaction quench from the immiscible to the miscible phase DAD solitonic structures appear in the 
corresponding one-body density evolution as a result of the quench-induced counterflow dynamics as shown in Figs. \ref{fig:quasi-1D} (e), (f) for 
a quench from $a_{AB}^i=150a_0$ to $a_{AB}^f=75a_0$. 
Indeed, we observe that dark states build upon the majority species and antidark structures emerge in the minority species. 
From the above discussion we can conclude that the dynamical formation of the DB and the DAD solitons discussed in the preceding
Sections \ref{MiscibleToImmiscibleDynamics}, \ref{ImmiscibleToMiscibleDynamics} appears also in quasi 1D setups 
and as such can be potentially observed in current state-of-the-art ultracold atom experiments. 
Indeed, the waveforms identified herein are known to be robust in quasi 1D 
geometries~\cite{YanDarkBrightInteraction}. 
Exposure of these structures to higher-dimensional setups may lead to transversal instabilities and thus the formation of nonlinear 
excitations of a different kind such as vortex-bright ones in 
two spatial dimensions~\cite{LawVortexBright,PolaVortexBright,mukherjee2019quench} 
and vortex-line-bright solitons or vortex-ring-bright solitons in three dimensions respectively~\cite{charalampidis2016so}.

\section{Conclusions}\label{conclusions} 

In the present work the interaction quench dynamics of a 1D harmonically confined particle imbalanced BEC mixture
has been investigated within a mean-field theoretical approach. In particular upon considering quenches crossing the 
miscibility/immiscibility threshold in both directions the dynamical formation of 
multiple solitary waves in this non-integrable model has been unraveled.
For miscible-to-immiscible transitions the unstable dynamics leads to 
the filamentation of the density of both species in line with our recent findings regarding particle balanced 
mixtures~\cite{MistakidisQuenchInducedPhaseSeparation}.
However for the particle imbalanced scenario studied herein we were able to 
identify that via the filamentation process DB solitary waves emerge. 
The number of the latter is significantly reduced during the dynamics. 
Indeed, due to the non-integrable 
nature of the system, the emergent DB structures
feature highly asymmetric collisions with mass 
redistribution~\cite{KatsimigaDarkBrightBifurcationCollision}  
between the solitary wave constituents. 
A controlled generation of DB solitons is achieved by considering distinct variations
of the experimentally tractable system parameters such as the particle numbers, the interspecies repulsion coefficient, and 
the trapping frequency.
Finally, and since both clouds undergo a collective breathing motion,
we further evaluate the relevant breathing frequency of each species. 
We find that the two breathing frequencies are substantially different 
for strong particle imbalances. 
A deviation that is more dramatic for larger postquench repulsions~\cite{sartori2013dynamics}. 

On the other hand, for immiscible-to-miscible transitions the spontaneous creation of multiple DAD solitons
is observed linking to previous studies~\cite{MistakidisQuenchInducedPhaseSeparation}.
Here, we extend our investigations shedding light on the underlying mechanism that leads to the formation of these states.
It is showcased that these states emerge due to a counterflow process that takes 
place as the system is quenched towards miscibility. 
Additionally, the generated DAD solitons are found to be robust throughout the evolution.
This result is to be contrasted to the observed decrease of the DB states formed in the previous transition, 
indicating that the breaking of integrability in this regime does not seem to affect the dynamics.  
Measuring the breathing frequency of each bosonic cloud for such a transition, 
we find that it differs from one component to the other 
independently of the postquench interaction strength or the particle number imbalance.
The above behavior of the breathing frequency for this immiscible-to-miscible transition,
is to the best of our knowledge, reported herein for the first time. 
Also here, the controlled creation of multiple DAD states is revealed
by manipulating the particle numbers, the interspecies interaction, and the trapping frequency. 

Finally, we should emphasize that in both quench processes the observed dynamics persists for large 
bosonic ensembles with particle numbers being of the order of $10^4$ suggesting a possible observation of our findings
in current state-of-the-art experiments. 

There are several possible extensions of the present work that can be considered in future endeavours. 
Of particular interest would be to examine the 
quench-induced dynamics of 1D spinor BECs~\cite{bersano2018three}. 
In such settings dark-dark-bright and dark-bright-bright soliton complexes 
are known to form~\cite{bersano2018three,nistazakis2008bright}. 
Therefore, attempting to control the generation of multiple such states upon quenching 
would be a natural generalization of our current effort.
Also for this spinorial BEC environment one could further go beyond the 
mean-field approximation \cite{KatsimigaDarkBrightDynamics,katsimiga2017many,katsimiga2018many},
addressing the fate of multiple solitonic entities in the presence of quantum fluctuations. 
Yet another fruitful direction would be to investigate the controlled dynamical formation
of other nonlinear excitations such as the beating dark-dark solitons.
It is known that these states can be experimentally created in miscible binary mixtures upon considering 
the fast counterflow of both components~\cite{HoeferDarkDarkSolitonsBEC}. 
Finally, a generalization of the quench-induced dynamics of binary BEC mixtures to higher
dimensions represents an intriguing perspective. 
It is well-known that e.g. in two-dimensions there exist 
vortex-bright soliton states~\cite{LawVortexBright,PolaVortexBright,mukherjee2019quench}, while in three-dimensions structures 
consisting of vortex-line-bright and vortex-ring-bright solitons have been identified~\cite{charalampidis2016so}.
It would be interesting to examine if such structures can be dynamically 
produced under the influence of an interspecies interaction quench.

\section*{Acknowledgements} 
S. I. M and G. C. K would like to thank P. G. Kevrekidis for fruitful discussions. 
S. I. M and G. C. K gratefully acknowledge K. Mukherjee for insightful discussions and assistance regarding the quasi one-dimensional 
simulations.

\bibliography{sample}

\begin{thebibliography}{107}%
\makeatletter
\providecommand \@ifxundefined [1]{%
 \@ifx{#1\undefined}
}%
\providecommand \@ifnum [1]{%
 \ifnum #1\expandafter \@firstoftwo
 \else \expandafter \@secondoftwo
 \fi
}%
\providecommand \@ifx [1]{%
 \ifx #1\expandafter \@firstoftwo
 \else \expandafter \@secondoftwo
 \fi
}%
\providecommand \natexlab [1]{#1}%
\providecommand \enquote  [1]{``#1''}%
\providecommand \bibnamefont  [1]{#1}%
\providecommand \bibfnamefont [1]{#1}%
\providecommand \citenamefont [1]{#1}%
\providecommand \href@noop [0]{\@secondoftwo}%
\providecommand \href [0]{\begingroup \@sanitize@url \@href}%
\providecommand \@href[1]{\@@startlink{#1}\@@href}%
\providecommand \@@href[1]{\endgroup#1\@@endlink}%
\providecommand \@sanitize@url [0]{\catcode `\\12\catcode `\$12\catcode
  `\&12\catcode `\#12\catcode `\^12\catcode `\_12\catcode `\%12\relax}%
\providecommand \@@startlink[1]{}%
\providecommand \@@endlink[0]{}%
\providecommand \url  [0]{\begingroup\@sanitize@url \@url }%
\providecommand \@url [1]{\endgroup\@href {#1}{\urlprefix }}%
\providecommand \urlprefix  [0]{URL }%
\providecommand \Eprint [0]{\href }%
\providecommand \doibase [0]{http://dx.doi.org/}%
\providecommand \selectlanguage [0]{\@gobble}%
\providecommand \bibinfo  [0]{\@secondoftwo}%
\providecommand \bibfield  [0]{\@secondoftwo}%
\providecommand \translation [1]{[#1]}%
\providecommand \BibitemOpen [0]{}%
\providecommand \bibitemStop [0]{}%
\providecommand \bibitemNoStop [0]{.\EOS\space}%
\providecommand \EOS [0]{\spacefactor3000\relax}%
\providecommand \BibitemShut  [1]{\csname bibitem#1\endcsname}%
\let\auto@bib@innerbib\@empty
\bibitem [{\citenamefont {Kevrekidis}\ \emph {et~al.}(2007)\citenamefont
  {Kevrekidis}, \citenamefont {Frantzeskakis},\ and\ \citenamefont
  {Carretero-Gonz{\'a}lez}}]{kevrekidisEmergentNonlinearPhenomena}%
  \BibitemOpen
  \bibfield  {author} {\bibinfo {author} {\bibfnamefont {P.~G.}\ \bibnamefont
  {Kevrekidis}}, \bibinfo {author} {\bibfnamefont {D.~J.}\ \bibnamefont
  {Frantzeskakis}}, \ and\ \bibinfo {author} {\bibfnamefont {R.}~\bibnamefont
  {Carretero-Gonz{\'a}lez}},\ }\href@noop {} {\emph {\bibinfo {title} {Emergent
  nonlinear phenomena in Bose-Einstein condensates: theory and experiment}}},\
  Vol.~\bibinfo {volume} {45}\ (\bibinfo  {publisher} {Springer Science \&
  Business Media},\ \bibinfo {year} {2007})\BibitemShut {NoStop}%
\bibitem [{\citenamefont {Pitaevskii}\ and\ \citenamefont
  {Stringari}(2016)}]{StringariBEC}%
  \BibitemOpen
  \bibfield  {author} {\bibinfo {author} {\bibfnamefont {L.}~\bibnamefont
  {Pitaevskii}}\ and\ \bibinfo {author} {\bibfnamefont {S.}~\bibnamefont
  {Stringari}},\ }\href@noop {} {\emph {\bibinfo {title} {Bose-Einstein
  condensation and superfluidity}}},\ Vol.\ \bibinfo {volume} {164}\ (\bibinfo
  {publisher} {Oxford University Press},\ \bibinfo {year} {2016})\BibitemShut
  {NoStop}%
\bibitem [{\citenamefont {Pethick}\ and\ \citenamefont
  {Smith}(2002)}]{PethickSmith}%
  \BibitemOpen
  \bibfield  {author} {\bibinfo {author} {\bibfnamefont {C.~J.}\ \bibnamefont
  {Pethick}}\ and\ \bibinfo {author} {\bibfnamefont {H.}~\bibnamefont
  {Smith}},\ }\href@noop {} {\emph {\bibinfo {title} {Bose-Einstein
  condensation in dilute gases}}}\ (\bibinfo  {publisher} {Cambridge University
  Press},\ \bibinfo {year} {2002})\BibitemShut {NoStop}%
\bibitem [{\citenamefont {Anderson}\ \emph {et~al.}(1995)\citenamefont
  {Anderson}, \citenamefont {Ensher}, \citenamefont {Matthews}, \citenamefont
  {Wieman},\ and\ \citenamefont {Cornell}}]{AndersonEarlyBEC}%
  \BibitemOpen
  \bibfield  {author} {\bibinfo {author} {\bibfnamefont {M.~H.}\ \bibnamefont
  {Anderson}}, \bibinfo {author} {\bibfnamefont {J.~R.}\ \bibnamefont
  {Ensher}}, \bibinfo {author} {\bibfnamefont {M.~R.}\ \bibnamefont
  {Matthews}}, \bibinfo {author} {\bibfnamefont {C.~E.}\ \bibnamefont
  {Wieman}}, \ and\ \bibinfo {author} {\bibfnamefont {E.~A.}\ \bibnamefont
  {Cornell}},\ }\href {\doibase 10.1126/science.269.5221.198} {\ \textbf
  {\bibinfo {volume} {269}},\ \bibinfo {pages} {198} (\bibinfo {year}
  {1995})}\BibitemShut {NoStop}%
\bibitem [{\citenamefont {Davis}\ \emph {et~al.}(1995)\citenamefont {Davis},
  \citenamefont {Mewes}, \citenamefont {Andrews}, \citenamefont {van Druten},
  \citenamefont {Durfee}, \citenamefont {Kurn},\ and\ \citenamefont
  {Ketterle}}]{DavisEarlyBEC}%
  \BibitemOpen
  \bibfield  {author} {\bibinfo {author} {\bibfnamefont {K.~B.}\ \bibnamefont
  {Davis}}, \bibinfo {author} {\bibfnamefont {M.~O.}\ \bibnamefont {Mewes}},
  \bibinfo {author} {\bibfnamefont {M.~R.}\ \bibnamefont {Andrews}}, \bibinfo
  {author} {\bibfnamefont {N.~J.}\ \bibnamefont {van Druten}}, \bibinfo
  {author} {\bibfnamefont {D.~S.}\ \bibnamefont {Durfee}}, \bibinfo {author}
  {\bibfnamefont {D.~M.}\ \bibnamefont {Kurn}}, \ and\ \bibinfo {author}
  {\bibfnamefont {W.}~\bibnamefont {Ketterle}},\ }\href {\doibase
  10.1103/PhysRevLett.75.3969} {\bibfield  {journal} {\bibinfo  {journal}
  {Phys. Rev. Lett.}\ }\textbf {\bibinfo {volume} {75}},\ \bibinfo {pages}
  {3969} (\bibinfo {year} {1995})}\BibitemShut {NoStop}%
\bibitem [{\citenamefont {Bradley}\ \emph {et~al.}(1997)\citenamefont
  {Bradley}, \citenamefont {Sackett},\ and\ \citenamefont
  {Hulet}}]{BradleyEarlyBEC}%
  \BibitemOpen
  \bibfield  {author} {\bibinfo {author} {\bibfnamefont {C.~C.}\ \bibnamefont
  {Bradley}}, \bibinfo {author} {\bibfnamefont {C.~A.}\ \bibnamefont
  {Sackett}}, \ and\ \bibinfo {author} {\bibfnamefont {R.~G.}\ \bibnamefont
  {Hulet}},\ }\href {\doibase 10.1103/PhysRevLett.78.985} {\bibfield  {journal}
  {\bibinfo  {journal} {Phys. Rev. Lett.}\ }\textbf {\bibinfo {volume} {78}},\
  \bibinfo {pages} {985} (\bibinfo {year} {1997})}\BibitemShut {NoStop}%
\bibitem [{\citenamefont {Burger}\ \emph {et~al.}(1999)\citenamefont {Burger},
  \citenamefont {Bongs}, \citenamefont {Dettmer}, \citenamefont {Ertmer},
  \citenamefont {Sengstock}, \citenamefont {Sanpera}, \citenamefont
  {Shlyapnikov},\ and\ \citenamefont {Lewenstein}}]{burgerDark}%
  \BibitemOpen
  \bibfield  {author} {\bibinfo {author} {\bibfnamefont {S.}~\bibnamefont
  {Burger}}, \bibinfo {author} {\bibfnamefont {K.}~\bibnamefont {Bongs}},
  \bibinfo {author} {\bibfnamefont {S.}~\bibnamefont {Dettmer}}, \bibinfo
  {author} {\bibfnamefont {W.}~\bibnamefont {Ertmer}}, \bibinfo {author}
  {\bibfnamefont {K.}~\bibnamefont {Sengstock}}, \bibinfo {author}
  {\bibfnamefont {A.}~\bibnamefont {Sanpera}}, \bibinfo {author} {\bibfnamefont
  {G.~V.}\ \bibnamefont {Shlyapnikov}}, \ and\ \bibinfo {author} {\bibfnamefont
  {M.}~\bibnamefont {Lewenstein}},\ }\href@noop {} {\bibfield  {journal}
  {\bibinfo  {journal} {Phys. Rev. Lett.}\ }\textbf {\bibinfo {volume} {83}},\
  \bibinfo {pages} {5198} (\bibinfo {year} {1999})}\BibitemShut {NoStop}%
\bibitem [{\citenamefont {Frantzeskakis}(2010)}]{frantzeskakisDark}%
  \BibitemOpen
  \bibfield  {author} {\bibinfo {author} {\bibfnamefont {D.}~\bibnamefont
  {Frantzeskakis}},\ }\href@noop {} {\bibfield  {journal} {\bibinfo  {journal}
  {J. Phys. A: Math. and Theor.}\ }\textbf {\bibinfo {volume} {43}},\ \bibinfo
  {pages} {213001} (\bibinfo {year} {2010})}\BibitemShut {NoStop}%
\bibitem [{\citenamefont {Becker}\ \emph {et~al.}(2008)\citenamefont {Becker},
  \citenamefont {Stellmer}, \citenamefont {Soltan-Panahi}, \citenamefont
  {D{\"o}rscher}, \citenamefont {Baumert}, \citenamefont {Richter},
  \citenamefont {Kronj{\"a}ger}, \citenamefont {Bongs},\ and\ \citenamefont
  {Sengstock}}]{becker2008oscillations}%
  \BibitemOpen
  \bibfield  {author} {\bibinfo {author} {\bibfnamefont {C.}~\bibnamefont
  {Becker}}, \bibinfo {author} {\bibfnamefont {S.}~\bibnamefont {Stellmer}},
  \bibinfo {author} {\bibfnamefont {P.}~\bibnamefont {Soltan-Panahi}}, \bibinfo
  {author} {\bibfnamefont {S.}~\bibnamefont {D{\"o}rscher}}, \bibinfo {author}
  {\bibfnamefont {M.}~\bibnamefont {Baumert}}, \bibinfo {author} {\bibfnamefont
  {E.-M.}\ \bibnamefont {Richter}}, \bibinfo {author} {\bibfnamefont
  {J.}~\bibnamefont {Kronj{\"a}ger}}, \bibinfo {author} {\bibfnamefont
  {K.}~\bibnamefont {Bongs}}, \ and\ \bibinfo {author} {\bibfnamefont
  {K.}~\bibnamefont {Sengstock}},\ }\href@noop {} {\bibfield  {journal}
  {\bibinfo  {journal} {Nat. Phys.}\ }\textbf {\bibinfo {volume} {4}},\
  \bibinfo {pages} {496} (\bibinfo {year} {2008})}\BibitemShut {NoStop}%
\bibitem [{\citenamefont {Weller}\ \emph {et~al.}(2008)\citenamefont {Weller},
  \citenamefont {Ronzheimer}, \citenamefont {Gross}, \citenamefont {Esteve},
  \citenamefont {Oberthaler}, \citenamefont {Frantzeskakis}, \citenamefont
  {Theocharis},\ and\ \citenamefont {Kevrekidis}}]{WellerDark}%
  \BibitemOpen
  \bibfield  {author} {\bibinfo {author} {\bibfnamefont {A.}~\bibnamefont
  {Weller}}, \bibinfo {author} {\bibfnamefont {J.~P.}\ \bibnamefont
  {Ronzheimer}}, \bibinfo {author} {\bibfnamefont {C.}~\bibnamefont {Gross}},
  \bibinfo {author} {\bibfnamefont {J.}~\bibnamefont {Esteve}}, \bibinfo
  {author} {\bibfnamefont {M.~K.}\ \bibnamefont {Oberthaler}}, \bibinfo
  {author} {\bibfnamefont {D.~J.}\ \bibnamefont {Frantzeskakis}}, \bibinfo
  {author} {\bibfnamefont {G.}~\bibnamefont {Theocharis}}, \ and\ \bibinfo
  {author} {\bibfnamefont {P.~G.}\ \bibnamefont {Kevrekidis}},\ }\href
  {\doibase 10.1103/PhysRevLett.101.130401} {\bibfield  {journal} {\bibinfo
  {journal} {Phys. Rev. Lett.}\ }\textbf {\bibinfo {volume} {101}},\ \bibinfo
  {pages} {130401} (\bibinfo {year} {2008})}\BibitemShut {NoStop}%
\bibitem [{\citenamefont {Anderson}\ \emph {et~al.}(2001)\citenamefont
  {Anderson}, \citenamefont {Haljan}, \citenamefont {Regal}, \citenamefont
  {Feder}, \citenamefont {Collins}, \citenamefont {Clark},\ and\ \citenamefont
  {Cornell}}]{AndersonDark}%
  \BibitemOpen
  \bibfield  {author} {\bibinfo {author} {\bibfnamefont {B.~P.}\ \bibnamefont
  {Anderson}}, \bibinfo {author} {\bibfnamefont {P.~C.}\ \bibnamefont
  {Haljan}}, \bibinfo {author} {\bibfnamefont {C.~A.}\ \bibnamefont {Regal}},
  \bibinfo {author} {\bibfnamefont {D.~L.}\ \bibnamefont {Feder}}, \bibinfo
  {author} {\bibfnamefont {L.~A.}\ \bibnamefont {Collins}}, \bibinfo {author}
  {\bibfnamefont {C.~W.}\ \bibnamefont {Clark}}, \ and\ \bibinfo {author}
  {\bibfnamefont {E.~A.}\ \bibnamefont {Cornell}},\ }\href {\doibase
  10.1103/PhysRevLett.86.2926} {\bibfield  {journal} {\bibinfo  {journal}
  {Phys. Rev. Lett.}\ }\textbf {\bibinfo {volume} {86}},\ \bibinfo {pages}
  {2926} (\bibinfo {year} {2001})}\BibitemShut {NoStop}%
\bibitem [{\citenamefont {Shomroni}(2009)}]{shomroniDarkExperiment}%
  \BibitemOpen
  \bibfield  {author} {\bibinfo {author} {\bibfnamefont {I.}~\bibnamefont
  {Shomroni}},\ }\href@noop {} {\bibfield  {journal} {\bibinfo  {journal} {Nat.
  Phys.}\ }\textbf {\bibinfo {volume} {5}},\ \bibinfo {pages} {193} (\bibinfo
  {year} {2009})}\BibitemShut {NoStop}%
\bibitem [{\citenamefont {Strecker}\ \emph {et~al.}(2002)\citenamefont
  {Strecker}, \citenamefont {Partridge}, \citenamefont {Truscott},\ and\
  \citenamefont {Hulet}}]{streckerBright}%
  \BibitemOpen
  \bibfield  {author} {\bibinfo {author} {\bibfnamefont {K.~E.}\ \bibnamefont
  {Strecker}}, \bibinfo {author} {\bibfnamefont {G.~B.}\ \bibnamefont
  {Partridge}}, \bibinfo {author} {\bibfnamefont {A.~G.}\ \bibnamefont
  {Truscott}}, \ and\ \bibinfo {author} {\bibfnamefont {R.~G.}\ \bibnamefont
  {Hulet}},\ }\href@noop {} {\bibfield  {journal} {\bibinfo  {journal}
  {Nature}\ }\textbf {\bibinfo {volume} {417}},\ \bibinfo {pages} {150}
  (\bibinfo {year} {2002})}\BibitemShut {NoStop}%
\bibitem [{\citenamefont {Khaykovich}\ \emph {et~al.}(2002)\citenamefont
  {Khaykovich}, \citenamefont {Schreck}, \citenamefont {Ferrari}, \citenamefont
  {Bourdel}, \citenamefont {Cubizolles}, \citenamefont {Carr}, \citenamefont
  {Castin},\ and\ \citenamefont {Salomon}}]{khaykovichBright}%
  \BibitemOpen
  \bibfield  {author} {\bibinfo {author} {\bibfnamefont {L.}~\bibnamefont
  {Khaykovich}}, \bibinfo {author} {\bibfnamefont {F.}~\bibnamefont {Schreck}},
  \bibinfo {author} {\bibfnamefont {G.}~\bibnamefont {Ferrari}}, \bibinfo
  {author} {\bibfnamefont {T.}~\bibnamefont {Bourdel}}, \bibinfo {author}
  {\bibfnamefont {J.}~\bibnamefont {Cubizolles}}, \bibinfo {author}
  {\bibfnamefont {L.}~\bibnamefont {Carr}}, \bibinfo {author} {\bibfnamefont
  {Y.}~\bibnamefont {Castin}}, \ and\ \bibinfo {author} {\bibfnamefont
  {C.}~\bibnamefont {Salomon}},\ }\href@noop {} {\bibfield  {journal} {\bibinfo
   {journal} {Science}\ }\textbf {\bibinfo {volume} {296}},\ \bibinfo {pages}
  {1290} (\bibinfo {year} {2002})}\BibitemShut {NoStop}%
\bibitem [{\citenamefont {Cornish}\ \emph {et~al.}(2006)\citenamefont
  {Cornish}, \citenamefont {Thompson},\ and\ \citenamefont
  {Wieman}}]{CornishBright}%
  \BibitemOpen
  \bibfield  {author} {\bibinfo {author} {\bibfnamefont {S.~L.}\ \bibnamefont
  {Cornish}}, \bibinfo {author} {\bibfnamefont {S.~T.}\ \bibnamefont
  {Thompson}}, \ and\ \bibinfo {author} {\bibfnamefont {C.~E.}\ \bibnamefont
  {Wieman}},\ }\href {\doibase 10.1103/PhysRevLett.96.170401} {\bibfield
  {journal} {\bibinfo  {journal} {Phys. Rev. Lett.}\ }\textbf {\bibinfo
  {volume} {96}},\ \bibinfo {pages} {170401} (\bibinfo {year}
  {2006})}\BibitemShut {NoStop}%
\bibitem [{\citenamefont {Dabrowska-W\"{u}ster}\ \emph
  {et~al.}(2007)\citenamefont {Dabrowska-W\"{u}ster}, \citenamefont
  {W\"{u}ster},\ and\ \citenamefont {Davis}}]{Dabrowska-WusterBright}%
  \BibitemOpen
  \bibfield  {author} {\bibinfo {author} {\bibfnamefont {B.~J.}\ \bibnamefont
  {Dabrowska-W\"{u}ster}}, \bibinfo {author} {\bibfnamefont {S.}~\bibnamefont
  {W\"{u}ster}}, \ and\ \bibinfo {author} {\bibfnamefont {M.~J.}\ \bibnamefont
  {Davis}},\ }in\ \href {http://www.osapublishing.org/abstract.cfm?URI=Q
  -2007-QWE28} {\emph {\bibinfo {booktitle} {Quantum-Atom Optics Downunder}}}\
  (\bibinfo  {publisher} {Optical Society of America},\ \bibinfo {year}
  {2007})\ p.\ \bibinfo {pages} {QWE28}\BibitemShut {NoStop}%
\bibitem [{\citenamefont {Myatt}\ \emph {et~al.}(1997)\citenamefont {Myatt},
  \citenamefont {Burt}, \citenamefont {Ghrist}, \citenamefont {Cornell},\ and\
  \citenamefont {Wieman}}]{myatt1997SympatheticCooling}%
  \BibitemOpen
  \bibfield  {author} {\bibinfo {author} {\bibfnamefont {C.~J.}\ \bibnamefont
  {Myatt}}, \bibinfo {author} {\bibfnamefont {E.~A.}\ \bibnamefont {Burt}},
  \bibinfo {author} {\bibfnamefont {R.~W.}\ \bibnamefont {Ghrist}}, \bibinfo
  {author} {\bibfnamefont {E.~A.}\ \bibnamefont {Cornell}}, \ and\ \bibinfo
  {author} {\bibfnamefont {C.~E.}\ \bibnamefont {Wieman}},\ }\href@noop {}
  {\bibfield  {journal} {\bibinfo  {journal} {Phys. Rev. Lett.}\ }\textbf
  {\bibinfo {volume} {78}},\ \bibinfo {pages} {586} (\bibinfo {year}
  {1997})}\BibitemShut {NoStop}%
\bibitem [{\citenamefont {Hall}\ \emph {et~al.}(1998)\citenamefont {Hall},
  \citenamefont {Matthews}, \citenamefont {Ensher}, \citenamefont {Wieman},\
  and\ \citenamefont {Cornell}}]{HallBinaryMixurePhaseSeparation}%
  \BibitemOpen
  \bibfield  {author} {\bibinfo {author} {\bibfnamefont {D.~S.}\ \bibnamefont
  {Hall}}, \bibinfo {author} {\bibfnamefont {M.~R.}\ \bibnamefont {Matthews}},
  \bibinfo {author} {\bibfnamefont {J.~R.}\ \bibnamefont {Ensher}}, \bibinfo
  {author} {\bibfnamefont {C.~E.}\ \bibnamefont {Wieman}}, \ and\ \bibinfo
  {author} {\bibfnamefont {E.~A.}\ \bibnamefont {Cornell}},\ }\href {\doibase
  10.1103/PhysRevLett.81.1539} {\bibfield  {journal} {\bibinfo  {journal}
  {Phys. Rev. Lett.}\ }\textbf {\bibinfo {volume} {81}},\ \bibinfo {pages}
  {1539} (\bibinfo {year} {1998})}\BibitemShut {NoStop}%
\bibitem [{\citenamefont {Stamper-Kurn}\ \emph {et~al.}(1998)\citenamefont
  {Stamper-Kurn}, \citenamefont {Andrews}, \citenamefont {Chikkatur},
  \citenamefont {Inouye}, \citenamefont {Miesner}, \citenamefont {Stenger},\
  and\ \citenamefont {Ketterle}}]{StamperKurnSpinorBEC}%
  \BibitemOpen
  \bibfield  {author} {\bibinfo {author} {\bibfnamefont {D.~M.}\ \bibnamefont
  {Stamper-Kurn}}, \bibinfo {author} {\bibfnamefont {M.~R.}\ \bibnamefont
  {Andrews}}, \bibinfo {author} {\bibfnamefont {A.~P.}\ \bibnamefont
  {Chikkatur}}, \bibinfo {author} {\bibfnamefont {S.}~\bibnamefont {Inouye}},
  \bibinfo {author} {\bibfnamefont {H.-J.}\ \bibnamefont {Miesner}}, \bibinfo
  {author} {\bibfnamefont {J.}~\bibnamefont {Stenger}}, \ and\ \bibinfo
  {author} {\bibfnamefont {W.}~\bibnamefont {Ketterle}},\ }\href {\doibase
  10.1103/PhysRevLett.80.2027} {\bibfield  {journal} {\bibinfo  {journal}
  {Phys. Rev. Lett.}\ }\textbf {\bibinfo {volume} {80}},\ \bibinfo {pages}
  {2027} (\bibinfo {year} {1998})}\BibitemShut {NoStop}%
\bibitem [{\citenamefont {Modugno}\ \emph {et~al.}(2001)\citenamefont
  {Modugno}, \citenamefont {Ferrari}, \citenamefont {Roati}, \citenamefont
  {Brecha}, \citenamefont {Simoni},\ and\ \citenamefont
  {Inguscio}}]{ModugnoRbKMixture}%
  \BibitemOpen
  \bibfield  {author} {\bibinfo {author} {\bibfnamefont {G.}~\bibnamefont
  {Modugno}}, \bibinfo {author} {\bibfnamefont {G.}~\bibnamefont {Ferrari}},
  \bibinfo {author} {\bibfnamefont {G.}~\bibnamefont {Roati}}, \bibinfo
  {author} {\bibfnamefont {R.~J.}\ \bibnamefont {Brecha}}, \bibinfo {author}
  {\bibfnamefont {A.}~\bibnamefont {Simoni}}, \ and\ \bibinfo {author}
  {\bibfnamefont {M.}~\bibnamefont {Inguscio}},\ }\href {\doibase
  10.1126/science.1066687} {\ \textbf {\bibinfo {volume} {294}},\ \bibinfo
  {pages} {1320} (\bibinfo {year} {2001})}\BibitemShut {NoStop}%
\bibitem [{\citenamefont {Bloch}\ \emph {et~al.}(2001)\citenamefont {Bloch},
  \citenamefont {Greiner}, \citenamefont {Mandel}, \citenamefont {H\"ansch},\
  and\ \citenamefont {Esslinger}}]{BlochRbMixture}%
  \BibitemOpen
  \bibfield  {author} {\bibinfo {author} {\bibfnamefont {I.}~\bibnamefont
  {Bloch}}, \bibinfo {author} {\bibfnamefont {M.}~\bibnamefont {Greiner}},
  \bibinfo {author} {\bibfnamefont {O.}~\bibnamefont {Mandel}}, \bibinfo
  {author} {\bibfnamefont {T.~W.}\ \bibnamefont {H\"ansch}}, \ and\ \bibinfo
  {author} {\bibfnamefont {T.}~\bibnamefont {Esslinger}},\ }\href {\doibase
  10.1103/PhysRevA.64.021402} {\bibfield  {journal} {\bibinfo  {journal} {Phys.
  Rev. A}\ }\textbf {\bibinfo {volume} {64}},\ \bibinfo {pages} {021402}
  (\bibinfo {year} {2001})}\BibitemShut {NoStop}%
\bibitem [{\citenamefont {Hoefer}\ \emph {et~al.}(2011)\citenamefont {Hoefer},
  \citenamefont {Chang}, \citenamefont {Hamner},\ and\ \citenamefont
  {Engels}}]{HoeferDarkDarkSolitonsBEC}%
  \BibitemOpen
  \bibfield  {author} {\bibinfo {author} {\bibfnamefont {M.~A.}\ \bibnamefont
  {Hoefer}}, \bibinfo {author} {\bibfnamefont {J.~J.}\ \bibnamefont {Chang}},
  \bibinfo {author} {\bibfnamefont {C.}~\bibnamefont {Hamner}}, \ and\ \bibinfo
  {author} {\bibfnamefont {P.}~\bibnamefont {Engels}},\ }\href {\doibase
  10.1103/PhysRevA.84.041605} {\bibfield  {journal} {\bibinfo  {journal} {Phys.
  Rev. A}\ }\textbf {\bibinfo {volume} {84}},\ \bibinfo {pages} {041605}
  (\bibinfo {year} {2011})}\BibitemShut {NoStop}%
\bibitem [{\citenamefont {Yan}\ \emph {et~al.}(2012)\citenamefont {Yan},
  \citenamefont {Chang}, \citenamefont {Hamner}, \citenamefont {Hoefer},
  \citenamefont {Kevrekidis}, \citenamefont {Engels}, \citenamefont
  {Achilleos}, \citenamefont {Frantzeskakis},\ and\ \citenamefont
  {Cuevas}}]{YanDarkDarkSolitonsBEC}%
  \BibitemOpen
  \bibfield  {author} {\bibinfo {author} {\bibfnamefont {D.}~\bibnamefont
  {Yan}}, \bibinfo {author} {\bibfnamefont {J.~J.}\ \bibnamefont {Chang}},
  \bibinfo {author} {\bibfnamefont {C.}~\bibnamefont {Hamner}}, \bibinfo
  {author} {\bibfnamefont {M.}~\bibnamefont {Hoefer}}, \bibinfo {author}
  {\bibfnamefont {P.~G.}\ \bibnamefont {Kevrekidis}}, \bibinfo {author}
  {\bibfnamefont {P.}~\bibnamefont {Engels}}, \bibinfo {author} {\bibfnamefont
  {V.}~\bibnamefont {Achilleos}}, \bibinfo {author} {\bibfnamefont {D.~J.}\
  \bibnamefont {Frantzeskakis}}, \ and\ \bibinfo {author} {\bibfnamefont
  {J.}~\bibnamefont {Cuevas}},\ }\href
  {http://stacks.iop.org/0953-4075/45/i=11/a=115301} {\bibfield  {journal}
  {\bibinfo  {journal} {J. Phys. B: At. Mol. Opt. Phys.}\ }\textbf {\bibinfo
  {volume} {45}},\ \bibinfo {pages} {115301} (\bibinfo {year}
  {2012})}\BibitemShut {NoStop}%
\bibitem [{\citenamefont {Rajendran}\ \emph {et~al.}(2009)\citenamefont
  {Rajendran}, \citenamefont {Muruganandam},\ and\ \citenamefont
  {Lakshmanan}}]{RajendranDarkBrightInteraction}%
  \BibitemOpen
  \bibfield  {author} {\bibinfo {author} {\bibfnamefont {S.}~\bibnamefont
  {Rajendran}}, \bibinfo {author} {\bibfnamefont {P.}~\bibnamefont
  {Muruganandam}}, \ and\ \bibinfo {author} {\bibfnamefont {M.}~\bibnamefont
  {Lakshmanan}},\ }\href {http://stacks.iop.org/0953-4075/42/i=14/a=145307}
  {\bibfield  {journal} {\bibinfo  {journal} {J. Phys. B: At. Mol. Opt. Phys.}\
  }\textbf {\bibinfo {volume} {42}},\ \bibinfo {pages} {145307} (\bibinfo
  {year} {2009})}\BibitemShut {NoStop}%
\bibitem [{\citenamefont {Yin}\ \emph {et~al.}(2011)\citenamefont {Yin},
  \citenamefont {Berloff}, \citenamefont {P\'erez-Garc\'{\i}a}, \citenamefont
  {Novoa}, \citenamefont {Carpentier},\ and\ \citenamefont
  {Michinel}}]{YinDarkBrightInteraction}%
  \BibitemOpen
  \bibfield  {author} {\bibinfo {author} {\bibfnamefont {C.}~\bibnamefont
  {Yin}}, \bibinfo {author} {\bibfnamefont {N.~G.}\ \bibnamefont {Berloff}},
  \bibinfo {author} {\bibfnamefont {V.~M.}\ \bibnamefont
  {P\'erez-Garc\'{\i}a}}, \bibinfo {author} {\bibfnamefont {D.}~\bibnamefont
  {Novoa}}, \bibinfo {author} {\bibfnamefont {A.~V.}\ \bibnamefont
  {Carpentier}}, \ and\ \bibinfo {author} {\bibfnamefont {H.}~\bibnamefont
  {Michinel}},\ }\href {\doibase 10.1103/PhysRevA.83.051605} {\bibfield
  {journal} {\bibinfo  {journal} {Phys. Rev. A}\ }\textbf {\bibinfo {volume}
  {83}},\ \bibinfo {pages} {051605} (\bibinfo {year} {2011})}\BibitemShut
  {NoStop}%
\bibitem [{\citenamefont {Yan}\ \emph {et~al.}(2011)\citenamefont {Yan},
  \citenamefont {Chang}, \citenamefont {Hamner}, \citenamefont {Kevrekidis},
  \citenamefont {Engels}, \citenamefont {Achilleos}, \citenamefont
  {Frantzeskakis}, \citenamefont {Carretero-Gonz\'alez},\ and\ \citenamefont
  {Schmelcher}}]{YanDarkBrightInteraction}%
  \BibitemOpen
  \bibfield  {author} {\bibinfo {author} {\bibfnamefont {D.}~\bibnamefont
  {Yan}}, \bibinfo {author} {\bibfnamefont {J.~J.}\ \bibnamefont {Chang}},
  \bibinfo {author} {\bibfnamefont {C.}~\bibnamefont {Hamner}}, \bibinfo
  {author} {\bibfnamefont {P.~G.}\ \bibnamefont {Kevrekidis}}, \bibinfo
  {author} {\bibfnamefont {P.}~\bibnamefont {Engels}}, \bibinfo {author}
  {\bibfnamefont {V.}~\bibnamefont {Achilleos}}, \bibinfo {author}
  {\bibfnamefont {D.~J.}\ \bibnamefont {Frantzeskakis}}, \bibinfo {author}
  {\bibfnamefont {R.}~\bibnamefont {Carretero-Gonz\'alez}}, \ and\ \bibinfo
  {author} {\bibfnamefont {P.}~\bibnamefont {Schmelcher}},\ }\href {\doibase
  10.1103/PhysRevA.84.053630} {\bibfield  {journal} {\bibinfo  {journal} {Phys.
  Rev. A}\ }\textbf {\bibinfo {volume} {84}},\ \bibinfo {pages} {053630}
  (\bibinfo {year} {2011})}\BibitemShut {NoStop}%
\bibitem [{\citenamefont {Katsimiga}\ \emph
  {et~al.}(2017{\natexlab{a}})\citenamefont {Katsimiga}, \citenamefont
  {Stockhofe}, \citenamefont {Kevrekidis},\ and\ \citenamefont
  {Schmelcher}}]{KatsimigaDarkBrightInteraction}%
  \BibitemOpen
  \bibfield  {author} {\bibinfo {author} {\bibfnamefont {G.~C.}\ \bibnamefont
  {Katsimiga}}, \bibinfo {author} {\bibfnamefont {J.}~\bibnamefont
  {Stockhofe}}, \bibinfo {author} {\bibfnamefont {P.~G.}\ \bibnamefont
  {Kevrekidis}}, \ and\ \bibinfo {author} {\bibfnamefont {P.}~\bibnamefont
  {Schmelcher}},\ }\href {\doibase 10.1103/PhysRevA.95.013621} {\bibfield
  {journal} {\bibinfo  {journal} {Phys. Rev. A}\ }\textbf {\bibinfo {volume}
  {95}},\ \bibinfo {pages} {013621} (\bibinfo {year}
  {2017}{\natexlab{a}})}\BibitemShut {NoStop}%
\bibitem [{\citenamefont {Busch}\ and\ \citenamefont
  {Anglin}(2001)}]{BuschDarkBrightBECTheory}%
  \BibitemOpen
  \bibfield  {author} {\bibinfo {author} {\bibfnamefont {T.}~\bibnamefont
  {Busch}}\ and\ \bibinfo {author} {\bibfnamefont {J.~R.}\ \bibnamefont
  {Anglin}},\ }\href {\doibase 10.1103/PhysRevLett.87.010401} {\bibfield
  {journal} {\bibinfo  {journal} {Phys. Rev. Lett.}\ }\textbf {\bibinfo
  {volume} {87}},\ \bibinfo {pages} {010401} (\bibinfo {year}
  {2001})}\BibitemShut {NoStop}%
\bibitem [{\citenamefont {Nistazakis}\ \emph
  {et~al.}(2008{\natexlab{a}})\citenamefont {Nistazakis}, \citenamefont
  {Frantzeskakis}, \citenamefont {Kevrekidis}, \citenamefont {Malomed},\ and\
  \citenamefont {Carretero-Gonz\'alez}}]{NistazakisDarkBrightBECTheory}%
  \BibitemOpen
  \bibfield  {author} {\bibinfo {author} {\bibfnamefont {H.~E.}\ \bibnamefont
  {Nistazakis}}, \bibinfo {author} {\bibfnamefont {D.~J.}\ \bibnamefont
  {Frantzeskakis}}, \bibinfo {author} {\bibfnamefont {P.~G.}\ \bibnamefont
  {Kevrekidis}}, \bibinfo {author} {\bibfnamefont {B.~A.}\ \bibnamefont
  {Malomed}}, \ and\ \bibinfo {author} {\bibfnamefont {R.}~\bibnamefont
  {Carretero-Gonz\'alez}},\ }\href {\doibase 10.1103/PhysRevA.77.033612}
  {\bibfield  {journal} {\bibinfo  {journal} {Phys. Rev. A}\ }\textbf {\bibinfo
  {volume} {77}},\ \bibinfo {pages} {033612} (\bibinfo {year}
  {2008}{\natexlab{a}})}\BibitemShut {NoStop}%
\bibitem [{\citenamefont {Vijayajayanthi}\ \emph {et~al.}(2008)\citenamefont
  {Vijayajayanthi}, \citenamefont {Kanna},\ and\ \citenamefont
  {Lakshmanan}}]{VijayajayanthiDarkBrightBECTheory}%
  \BibitemOpen
  \bibfield  {author} {\bibinfo {author} {\bibfnamefont {M.}~\bibnamefont
  {Vijayajayanthi}}, \bibinfo {author} {\bibfnamefont {T.}~\bibnamefont
  {Kanna}}, \ and\ \bibinfo {author} {\bibfnamefont {M.}~\bibnamefont
  {Lakshmanan}},\ }\href {\doibase 10.1103/PhysRevA.77.013820} {\bibfield
  {journal} {\bibinfo  {journal} {Phys. Rev. A}\ }\textbf {\bibinfo {volume}
  {77}},\ \bibinfo {pages} {013820} (\bibinfo {year} {2008})}\BibitemShut
  {NoStop}%
\bibitem [{\citenamefont {Brazhnyi}\ and\ \citenamefont
  {Pérez-García}(2011)}]{ValeriyDarkBrightSolitonStipesBECTheory}%
  \BibitemOpen
  \bibfield  {author} {\bibinfo {author} {\bibfnamefont {V.~A.}\ \bibnamefont
  {Brazhnyi}}\ and\ \bibinfo {author} {\bibfnamefont {V.~M.}\ \bibnamefont
  {Pérez-García}},\ }\href {\doibase
  https://doi.org/10.1016/j.chaos.2010.12.012} {\bibfield  {journal} {\bibinfo
  {journal} {Chaos, Solitons \& Fractals}\ }\textbf {\bibinfo {volume} {44}},\
  \bibinfo {pages} {381 } (\bibinfo {year} {2011})}\BibitemShut {NoStop}%
\bibitem [{\citenamefont {Álvarez}\ \emph {et~al.}(2011)\citenamefont
  {Álvarez}, \citenamefont {Cuevas}, \citenamefont {Romero},\ and\
  \citenamefont {Kevrekidis}}]{AlvarezDarkBrightBECTheory}%
  \BibitemOpen
  \bibfield  {author} {\bibinfo {author} {\bibfnamefont {A.}~\bibnamefont
  {Álvarez}}, \bibinfo {author} {\bibfnamefont {J.}~\bibnamefont {Cuevas}},
  \bibinfo {author} {\bibfnamefont {F.}~\bibnamefont {Romero}}, \ and\ \bibinfo
  {author} {\bibfnamefont {P.}~\bibnamefont {Kevrekidis}},\ }\href {\doibase
  https://doi.org/10.1016/j.physd.2010.12.006} {\bibfield  {journal} {\bibinfo
  {journal} {Physica D: Nonlinear Phenomena}\ }\textbf {\bibinfo {volume}
  {240}},\ \bibinfo {pages} {767 } (\bibinfo {year} {2011})}\BibitemShut
  {NoStop}%
\bibitem [{\citenamefont {Achilleos}\ \emph {et~al.}(2011)\citenamefont
  {Achilleos}, \citenamefont {Kevrekidis}, \citenamefont {Rothos},\ and\
  \citenamefont {Frantzeskakis}}]{AchilleosDarkBrightBECTheory}%
  \BibitemOpen
  \bibfield  {author} {\bibinfo {author} {\bibfnamefont {V.}~\bibnamefont
  {Achilleos}}, \bibinfo {author} {\bibfnamefont {P.~G.}\ \bibnamefont
  {Kevrekidis}}, \bibinfo {author} {\bibfnamefont {V.~M.}\ \bibnamefont
  {Rothos}}, \ and\ \bibinfo {author} {\bibfnamefont {D.~J.}\ \bibnamefont
  {Frantzeskakis}},\ }\href {\doibase 10.1103/PhysRevA.84.053626} {\bibfield
  {journal} {\bibinfo  {journal} {Phys. Rev. A}\ }\textbf {\bibinfo {volume}
  {84}},\ \bibinfo {pages} {053626} (\bibinfo {year} {2011})}\BibitemShut
  {NoStop}%
\bibitem [{\citenamefont {Álvarez}\ \emph {et~al.}(2013)\citenamefont
  {Álvarez}, \citenamefont {Cuevas}, \citenamefont {Romero}, \citenamefont
  {Hamner}, \citenamefont {Chang}, \citenamefont {Engels}, \citenamefont
  {Kevrekidis},\ and\ \citenamefont
  {Frantzeskakis}}]{AlvarezDarkBrightBECImpurities}%
  \BibitemOpen
  \bibfield  {author} {\bibinfo {author} {\bibfnamefont {A.}~\bibnamefont
  {Álvarez}}, \bibinfo {author} {\bibfnamefont {J.}~\bibnamefont {Cuevas}},
  \bibinfo {author} {\bibfnamefont {F.~R.}\ \bibnamefont {Romero}}, \bibinfo
  {author} {\bibfnamefont {C.}~\bibnamefont {Hamner}}, \bibinfo {author}
  {\bibfnamefont {J.~J.}\ \bibnamefont {Chang}}, \bibinfo {author}
  {\bibfnamefont {P.}~\bibnamefont {Engels}}, \bibinfo {author} {\bibfnamefont
  {P.~G.}\ \bibnamefont {Kevrekidis}}, \ and\ \bibinfo {author} {\bibfnamefont
  {D.~J.}\ \bibnamefont {Frantzeskakis}},\ }\href
  {http://stacks.iop.org/0953-4075/46/i=6/a=065302} {\bibfield  {journal}
  {\bibinfo  {journal} {J. Phys. B: At. Mol. Opt. Phys.}\ }\textbf {\bibinfo
  {volume} {46}},\ \bibinfo {pages} {065302} (\bibinfo {year}
  {2013})}\BibitemShut {NoStop}%
\bibitem [{\citenamefont {Achilleos}\ \emph {et~al.}(2012)\citenamefont
  {Achilleos}, \citenamefont {Yan}, \citenamefont {Kevrekidis},\ and\
  \citenamefont {Frantzeskakis}}]{AchilleosDarkBrightBECTheory2}%
  \BibitemOpen
  \bibfield  {author} {\bibinfo {author} {\bibfnamefont {V.}~\bibnamefont
  {Achilleos}}, \bibinfo {author} {\bibfnamefont {D.}~\bibnamefont {Yan}},
  \bibinfo {author} {\bibfnamefont {P.~G.}\ \bibnamefont {Kevrekidis}}, \ and\
  \bibinfo {author} {\bibfnamefont {D.~J.}\ \bibnamefont {Frantzeskakis}},\
  }\href {http://stacks.iop.org/1367-2630/14/i=5/a=055006} {\bibfield
  {journal} {\bibinfo  {journal} {New J. Phys.}\ }\textbf {\bibinfo {volume}
  {14}},\ \bibinfo {pages} {055006} (\bibinfo {year} {2012})}\BibitemShut
  {NoStop}%
\bibitem [{\citenamefont {Middelkamp}\ \emph {et~al.}(2011)\citenamefont
  {Middelkamp}, \citenamefont {Chang}, \citenamefont {Hamner}, \citenamefont
  {Carretero-González}, \citenamefont {Kevrekidis}, \citenamefont {Achilleos},
  \citenamefont {Frantzeskakis}, \citenamefont {Schmelcher},\ and\
  \citenamefont {Engels}}]{MiddlekampDarkBrightBECTheory}%
  \BibitemOpen
  \bibfield  {author} {\bibinfo {author} {\bibfnamefont {S.}~\bibnamefont
  {Middelkamp}}, \bibinfo {author} {\bibfnamefont {J.}~\bibnamefont {Chang}},
  \bibinfo {author} {\bibfnamefont {C.}~\bibnamefont {Hamner}}, \bibinfo
  {author} {\bibfnamefont {R.}~\bibnamefont {Carretero-González}}, \bibinfo
  {author} {\bibfnamefont {P.}~\bibnamefont {Kevrekidis}}, \bibinfo {author}
  {\bibfnamefont {V.}~\bibnamefont {Achilleos}}, \bibinfo {author}
  {\bibfnamefont {D.}~\bibnamefont {Frantzeskakis}}, \bibinfo {author}
  {\bibfnamefont {P.}~\bibnamefont {Schmelcher}}, \ and\ \bibinfo {author}
  {\bibfnamefont {P.}~\bibnamefont {Engels}},\ }\href {\doibase
  https://doi.org/10.1016/j.physleta.2010.11.025} {\bibfield  {journal}
  {\bibinfo  {journal} {Phys. Lett. A}\ }\textbf {\bibinfo {volume} {375}},\
  \bibinfo {pages} {642 } (\bibinfo {year} {2011})}\BibitemShut {NoStop}%
\bibitem [{\citenamefont {Katsimiga}\ \emph
  {et~al.}(2017{\natexlab{b}})\citenamefont {Katsimiga}, \citenamefont
  {Koutentakis}, \citenamefont {Mistakidis}, \citenamefont {Kevrekidis},\ and\
  \citenamefont {Schmelcher}}]{KatsimigaDarkBrightDynamics}%
  \BibitemOpen
  \bibfield  {author} {\bibinfo {author} {\bibfnamefont {G.~C.}\ \bibnamefont
  {Katsimiga}}, \bibinfo {author} {\bibfnamefont {G.~M.}\ \bibnamefont
  {Koutentakis}}, \bibinfo {author} {\bibfnamefont {S.~I.}\ \bibnamefont
  {Mistakidis}}, \bibinfo {author} {\bibfnamefont {P.~G.}\ \bibnamefont
  {Kevrekidis}}, \ and\ \bibinfo {author} {\bibfnamefont {P.}~\bibnamefont
  {Schmelcher}},\ }\href {http://stacks.iop.org/1367-2630/19/i=7/a=073004}
  {\bibfield  {journal} {\bibinfo  {journal} {New J. Phys.}\ }\textbf {\bibinfo
  {volume} {19}},\ \bibinfo {pages} {073004} (\bibinfo {year}
  {2017}{\natexlab{b}})}\BibitemShut {NoStop}%
\bibitem [{\citenamefont {Kevrekidis}\ and\ \citenamefont
  {Frantzeskakis}(2016)}]{KevrekidisMultiComponentSolitonsReview}%
  \BibitemOpen
  \bibfield  {author} {\bibinfo {author} {\bibfnamefont {P.}~\bibnamefont
  {Kevrekidis}}\ and\ \bibinfo {author} {\bibfnamefont {D.}~\bibnamefont
  {Frantzeskakis}},\ }\href {\doibase
  https://doi.org/10.1016/j.revip.2016.07.002} {\bibfield  {journal} {\bibinfo
  {journal} {Rev. Phys.}\ }\textbf {\bibinfo {volume} {1}},\ \bibinfo {pages}
  {140 } (\bibinfo {year} {2016})}\BibitemShut {NoStop}%
\bibitem [{\citenamefont {Danaila}\ \emph {et~al.}(2016)\citenamefont
  {Danaila}, \citenamefont {Khamehchi}, \citenamefont {Gokhroo}, \citenamefont
  {Engels},\ and\ \citenamefont {Kevrekidis}}]{DanailaDarkAntidarkBEC}%
  \BibitemOpen
  \bibfield  {author} {\bibinfo {author} {\bibfnamefont {I.}~\bibnamefont
  {Danaila}}, \bibinfo {author} {\bibfnamefont {M.~A.}\ \bibnamefont
  {Khamehchi}}, \bibinfo {author} {\bibfnamefont {V.}~\bibnamefont {Gokhroo}},
  \bibinfo {author} {\bibfnamefont {P.}~\bibnamefont {Engels}}, \ and\ \bibinfo
  {author} {\bibfnamefont {P.~G.}\ \bibnamefont {Kevrekidis}},\ }\href
  {\doibase 10.1103/PhysRevA.94.053617} {\bibfield  {journal} {\bibinfo
  {journal} {Phys. Rev. A}\ }\textbf {\bibinfo {volume} {94}},\ \bibinfo
  {pages} {053617} (\bibinfo {year} {2016})}\BibitemShut {NoStop}%
\bibitem [{\citenamefont {Kevrekidis}\ \emph {et~al.}(2004)\citenamefont
  {Kevrekidis}, \citenamefont {Nistazakis}, \citenamefont {Frantzeskakis},
  \citenamefont {Malomed},\ and\ \citenamefont
  {Carretero-Gonz{\'a}lez}}]{KevrekidisFamiliesOfMatterWaves}%
  \BibitemOpen
  \bibfield  {author} {\bibinfo {author} {\bibfnamefont {P.~G.}\ \bibnamefont
  {Kevrekidis}}, \bibinfo {author} {\bibfnamefont {H.~E.}\ \bibnamefont
  {Nistazakis}}, \bibinfo {author} {\bibfnamefont {D.~J.}\ \bibnamefont
  {Frantzeskakis}}, \bibinfo {author} {\bibfnamefont {B.~A.}\ \bibnamefont
  {Malomed}}, \ and\ \bibinfo {author} {\bibfnamefont {R.}~\bibnamefont
  {Carretero-Gonz{\'a}lez}},\ }\href {\doibase 10.1140/epjd/e2003-00311-6}
  {\bibfield  {journal} {\bibinfo  {journal} {EPJ D - At. Mol. Opt. and Plasm.
  Phys.}\ }\textbf {\bibinfo {volume} {28}},\ \bibinfo {pages} {181} (\bibinfo
  {year} {2004})}\BibitemShut {NoStop}%
\bibitem [{\citenamefont {Mistakidis}\ \emph {et~al.}(2018)\citenamefont
  {Mistakidis}, \citenamefont {Katsimiga}, \citenamefont {Kevrekidis},\ and\
  \citenamefont {Schmelcher}}]{MistakidisQuenchInducedPhaseSeparation}%
  \BibitemOpen
  \bibfield  {author} {\bibinfo {author} {\bibfnamefont {S.~I.}\ \bibnamefont
  {Mistakidis}}, \bibinfo {author} {\bibfnamefont {G.~C.}\ \bibnamefont
  {Katsimiga}}, \bibinfo {author} {\bibfnamefont {P.~G.}\ \bibnamefont
  {Kevrekidis}}, \ and\ \bibinfo {author} {\bibfnamefont {P.}~\bibnamefont
  {Schmelcher}},\ }\href {http://stacks.iop.org/1367-2630/20/i=4/a=043052}
  {\bibfield  {journal} {\bibinfo  {journal} {New J. Phys.}\ }\textbf {\bibinfo
  {volume} {20}},\ \bibinfo {pages} {043052} (\bibinfo {year}
  {2018})}\BibitemShut {NoStop}%
\bibitem [{\citenamefont {Chen}\ \emph {et~al.}(1997)\citenamefont {Chen},
  \citenamefont {Segev}, \citenamefont {Coskun}, \citenamefont
  {Christodoulides},\ and\ \citenamefont
  {Kivshar}}]{ChenDarkBrightNonlinearOpticsExperiment}%
  \BibitemOpen
  \bibfield  {author} {\bibinfo {author} {\bibfnamefont {Z.}~\bibnamefont
  {Chen}}, \bibinfo {author} {\bibfnamefont {M.}~\bibnamefont {Segev}},
  \bibinfo {author} {\bibfnamefont {T.~H.}\ \bibnamefont {Coskun}}, \bibinfo
  {author} {\bibfnamefont {D.~N.}\ \bibnamefont {Christodoulides}}, \ and\
  \bibinfo {author} {\bibfnamefont {Y.~S.}\ \bibnamefont {Kivshar}},\ }\href
  {\doibase 10.1364/JOSAB.14.003066} {\bibfield  {journal} {\bibinfo  {journal}
  {J. Opt. Soc. Am. B}\ }\textbf {\bibinfo {volume} {14}},\ \bibinfo {pages}
  {3066} (\bibinfo {year} {1997})}\BibitemShut {NoStop}%
\bibitem [{\citenamefont {Ostrovskaya}\ \emph {et~al.}(1999)\citenamefont
  {Ostrovskaya}, \citenamefont {Kivshar}, \citenamefont {Chen},\ and\
  \citenamefont {Segev}}]{OstrovskayaDarkBrightNonlinearOpticsExperiment}%
  \BibitemOpen
  \bibfield  {author} {\bibinfo {author} {\bibfnamefont {E.~A.}\ \bibnamefont
  {Ostrovskaya}}, \bibinfo {author} {\bibfnamefont {Y.~S.}\ \bibnamefont
  {Kivshar}}, \bibinfo {author} {\bibfnamefont {Z.}~\bibnamefont {Chen}}, \
  and\ \bibinfo {author} {\bibfnamefont {M.}~\bibnamefont {Segev}},\ }\href
  {\doibase 10.1364/OL.24.000327} {\bibfield  {journal} {\bibinfo  {journal}
  {Opt. Lett.}\ }\textbf {\bibinfo {volume} {24}},\ \bibinfo {pages} {327}
  (\bibinfo {year} {1999})}\BibitemShut {NoStop}%
\bibitem [{\citenamefont
  {Christodoulides}(1988)}]{ChristodoulidesDarkBrightNonlinearOpticsTheory}%
  \BibitemOpen
  \bibfield  {author} {\bibinfo {author} {\bibfnamefont {D.}~\bibnamefont
  {Christodoulides}},\ }\href {\doibase
  https://doi.org/10.1016/0375-9601(88)90511-7} {\bibfield  {journal} {\bibinfo
   {journal} {Phys. Lett. A}\ }\textbf {\bibinfo {volume} {132}},\ \bibinfo
  {pages} {451 } (\bibinfo {year} {1988})}\BibitemShut {NoStop}%
\bibitem [{\citenamefont {Afanasyev}\ \emph {et~al.}(1989)\citenamefont
  {Afanasyev}, \citenamefont {Kivshar}, \citenamefont {Konotop},\ and\
  \citenamefont {Serkin}}]{AfanasyevDarkBrightNonlinearOpticsTheory}%
  \BibitemOpen
  \bibfield  {author} {\bibinfo {author} {\bibfnamefont {V.~V.}\ \bibnamefont
  {Afanasyev}}, \bibinfo {author} {\bibfnamefont {Y.~S.}\ \bibnamefont
  {Kivshar}}, \bibinfo {author} {\bibfnamefont {V.~V.}\ \bibnamefont
  {Konotop}}, \ and\ \bibinfo {author} {\bibfnamefont {V.~N.}\ \bibnamefont
  {Serkin}},\ }\href {\doibase 10.1364/OL.14.000805} {\bibfield  {journal}
  {\bibinfo  {journal} {Opt. Lett.}\ }\textbf {\bibinfo {volume} {14}},\
  \bibinfo {pages} {805} (\bibinfo {year} {1989})}\BibitemShut {NoStop}%
\bibitem [{\citenamefont {Kivshar}\ \emph {et~al.}(1996)\citenamefont
  {Kivshar}, \citenamefont {Afansjev},\ and\ \citenamefont
  {Snyder}}]{KivsharDarkAntidarkNonlinearOpticsTheory}%
  \BibitemOpen
  \bibfield  {author} {\bibinfo {author} {\bibfnamefont {Y.~S.}\ \bibnamefont
  {Kivshar}}, \bibinfo {author} {\bibfnamefont {V.~V.}\ \bibnamefont
  {Afansjev}}, \ and\ \bibinfo {author} {\bibfnamefont {A.~W.}\ \bibnamefont
  {Snyder}},\ }\href {\doibase https://doi.org/10.1016/0030-4018(96)00111-3}
  {\bibfield  {journal} {\bibinfo  {journal} {Opt. Commun.}\ }\textbf {\bibinfo
  {volume} {126}},\ \bibinfo {pages} {348 } (\bibinfo {year}
  {1996})}\BibitemShut {NoStop}%
\bibitem [{\citenamefont {Radhakrishnan}\ and\ \citenamefont
  {Lakshmanan}(1995)}]{RadhakrishnanDarkBrightNonlinearOpticsTheory}%
  \BibitemOpen
  \bibfield  {author} {\bibinfo {author} {\bibfnamefont {R.}~\bibnamefont
  {Radhakrishnan}}\ and\ \bibinfo {author} {\bibfnamefont {M.}~\bibnamefont
  {Lakshmanan}},\ }\href {http://stacks.iop.org/0305-4470/28/i=9/a=025}
  {\bibfield  {journal} {\bibinfo  {journal} {J. Phys. A: Mathematical and
  General}\ }\textbf {\bibinfo {volume} {28}},\ \bibinfo {pages} {2683}
  (\bibinfo {year} {1995})}\BibitemShut {NoStop}%
\bibitem [{\citenamefont {Buryak}\ \emph {et~al.}(1996)\citenamefont {Buryak},
  \citenamefont {Kivshar},\ and\ \citenamefont
  {Parker}}]{BuryakDarkBrightNonlinearOpticsTheory}%
  \BibitemOpen
  \bibfield  {author} {\bibinfo {author} {\bibfnamefont {A.~V.}\ \bibnamefont
  {Buryak}}, \bibinfo {author} {\bibfnamefont {Y.~S.}\ \bibnamefont {Kivshar}},
  \ and\ \bibinfo {author} {\bibfnamefont {D.~F.}\ \bibnamefont {Parker}},\
  }\href {\doibase https://doi.org/10.1016/0375-9601(96)00208-3} {\bibfield
  {journal} {\bibinfo  {journal} {Phys. Lett. A}\ }\textbf {\bibinfo {volume}
  {215}},\ \bibinfo {pages} {57 } (\bibinfo {year} {1996})}\BibitemShut
  {NoStop}%
\bibitem [{\citenamefont {Sheppard}\ and\ \citenamefont
  {Kivshar}(1997)}]{SheppardDarkBrightNonlinearOpticsTheory}%
  \BibitemOpen
  \bibfield  {author} {\bibinfo {author} {\bibfnamefont {A.~P.}\ \bibnamefont
  {Sheppard}}\ and\ \bibinfo {author} {\bibfnamefont {Y.~S.}\ \bibnamefont
  {Kivshar}},\ }\href {\doibase 10.1103/PhysRevE.55.4773} {\bibfield  {journal}
  {\bibinfo  {journal} {Phys. Rev. E}\ }\textbf {\bibinfo {volume} {55}},\
  \bibinfo {pages} {4773} (\bibinfo {year} {1997})}\BibitemShut {NoStop}%
\bibitem [{\citenamefont {Park}\ and\ \citenamefont
  {Shin}(2000)}]{ParkDarkBrightNonlinearOpticsTheory}%
  \BibitemOpen
  \bibfield  {author} {\bibinfo {author} {\bibfnamefont {Q.-H.}\ \bibnamefont
  {Park}}\ and\ \bibinfo {author} {\bibfnamefont {H.~J.}\ \bibnamefont
  {Shin}},\ }\href {\doibase 10.1103/PhysRevE.61.3093} {\bibfield  {journal}
  {\bibinfo  {journal} {Phys. Rev. E}\ }\textbf {\bibinfo {volume} {61}},\
  \bibinfo {pages} {3093} (\bibinfo {year} {2000})}\BibitemShut {NoStop}%
\bibitem [{\citenamefont {Hamner}\ \emph {et~al.}(2011)\citenamefont {Hamner},
  \citenamefont {Chang}, \citenamefont {Engels},\ and\ \citenamefont
  {Hoefer}}]{HamnerDarkBrightBECCounterflow}%
  \BibitemOpen
  \bibfield  {author} {\bibinfo {author} {\bibfnamefont {C.}~\bibnamefont
  {Hamner}}, \bibinfo {author} {\bibfnamefont {J.~J.}\ \bibnamefont {Chang}},
  \bibinfo {author} {\bibfnamefont {P.}~\bibnamefont {Engels}}, \ and\ \bibinfo
  {author} {\bibfnamefont {M.~A.}\ \bibnamefont {Hoefer}},\ }\href {\doibase
  10.1103/PhysRevLett.106.065302} {\bibfield  {journal} {\bibinfo  {journal}
  {Phys. Rev. Lett.}\ }\textbf {\bibinfo {volume} {106}},\ \bibinfo {pages}
  {065302} (\bibinfo {year} {2011})}\BibitemShut {NoStop}%
\bibitem [{\citenamefont {Manakov}(1974)}]{manakov1974theory}%
  \BibitemOpen
  \bibfield  {author} {\bibinfo {author} {\bibfnamefont {S.~V.}\ \bibnamefont
  {Manakov}},\ }\href@noop {} {\bibfield  {journal} {\bibinfo  {journal}
  {Soviet Physics-JETP}\ }\textbf {\bibinfo {volume} {38}},\ \bibinfo {pages}
  {248} (\bibinfo {year} {1974})}\BibitemShut {NoStop}%
\bibitem [{\citenamefont {Prinari}\ \emph {et~al.}(2015)\citenamefont
  {Prinari}, \citenamefont {Vitale},\ and\ \citenamefont
  {Biondini}}]{prinari2015dark}%
  \BibitemOpen
  \bibfield  {author} {\bibinfo {author} {\bibfnamefont {B.}~\bibnamefont
  {Prinari}}, \bibinfo {author} {\bibfnamefont {F.}~\bibnamefont {Vitale}}, \
  and\ \bibinfo {author} {\bibfnamefont {G.}~\bibnamefont {Biondini}},\
  }\href@noop {} {\bibfield  {journal} {\bibinfo  {journal} {J. Math. Phys.}\
  }\textbf {\bibinfo {volume} {56}},\ \bibinfo {pages} {071505} (\bibinfo
  {year} {2015})}\BibitemShut {NoStop}%
\bibitem [{\citenamefont {Yan}\ \emph {et~al.}(2015)\citenamefont {Yan},
  \citenamefont {Tsitoura}, \citenamefont {Kevrekidis},\ and\ \citenamefont
  {Frantzeskakis}}]{yan2015dark}%
  \BibitemOpen
  \bibfield  {author} {\bibinfo {author} {\bibfnamefont {D.}~\bibnamefont
  {Yan}}, \bibinfo {author} {\bibfnamefont {F.}~\bibnamefont {Tsitoura}},
  \bibinfo {author} {\bibfnamefont {P.}~\bibnamefont {Kevrekidis}}, \ and\
  \bibinfo {author} {\bibfnamefont {D.}~\bibnamefont {Frantzeskakis}},\
  }\href@noop {} {\bibfield  {journal} {\bibinfo  {journal} {Phys. Rev. A}\
  }\textbf {\bibinfo {volume} {91}},\ \bibinfo {pages} {023619} (\bibinfo
  {year} {2015})}\BibitemShut {NoStop}%
\bibitem [{\citenamefont {Karamatskos}\ \emph {et~al.}(2015)\citenamefont
  {Karamatskos}, \citenamefont {Stockhofe}, \citenamefont {Kevrekidis},\ and\
  \citenamefont {Schmelcher}}]{KaramatskosDarkBrightPairs}%
  \BibitemOpen
  \bibfield  {author} {\bibinfo {author} {\bibfnamefont {E.~T.}\ \bibnamefont
  {Karamatskos}}, \bibinfo {author} {\bibfnamefont {J.}~\bibnamefont
  {Stockhofe}}, \bibinfo {author} {\bibfnamefont {P.~G.}\ \bibnamefont
  {Kevrekidis}}, \ and\ \bibinfo {author} {\bibfnamefont {P.}~\bibnamefont
  {Schmelcher}},\ }\href {\doibase 10.1103/PhysRevA.91.043637} {\bibfield
  {journal} {\bibinfo  {journal} {Phys. Rev. A}\ }\textbf {\bibinfo {volume}
  {91}},\ \bibinfo {pages} {043637} (\bibinfo {year} {2015})}\BibitemShut
  {NoStop}%
\bibitem [{\citenamefont {Katsimiga}\ \emph
  {et~al.}(2017{\natexlab{c}})\citenamefont {Katsimiga}, \citenamefont
  {Stockhofe}, \citenamefont {Kevrekidis},\ and\ \citenamefont
  {Schmelcher}}]{KatsimigaDarkBrightPairs}%
  \BibitemOpen
  \bibfield  {author} {\bibinfo {author} {\bibfnamefont {G.~C.}\ \bibnamefont
  {Katsimiga}}, \bibinfo {author} {\bibfnamefont {J.}~\bibnamefont
  {Stockhofe}}, \bibinfo {author} {\bibfnamefont {P.~G.}\ \bibnamefont
  {Kevrekidis}}, \ and\ \bibinfo {author} {\bibfnamefont {P.}~\bibnamefont
  {Schmelcher}},\ }\href {http://www.mdpi.com/2076-3417/7/4/388} {\bibfield
  {journal} {\bibinfo  {journal} {Applied Sciences}\ }\textbf {\bibinfo
  {volume} {7}},\ \bibinfo {pages} {388} (\bibinfo {year}
  {2017}{\natexlab{c}})}\BibitemShut {NoStop}%
\bibitem [{\citenamefont {Katsimiga}\ \emph
  {et~al.}(2018{\natexlab{a}})\citenamefont {Katsimiga}, \citenamefont
  {Kevrekidis}, \citenamefont {Prinari}, \citenamefont {Biondini},\ and\
  \citenamefont {Schmelcher}}]{KatsimigaDarkBrightBifurcationCollision}%
  \BibitemOpen
  \bibfield  {author} {\bibinfo {author} {\bibfnamefont {G.~C.}\ \bibnamefont
  {Katsimiga}}, \bibinfo {author} {\bibfnamefont {P.~G.}\ \bibnamefont
  {Kevrekidis}}, \bibinfo {author} {\bibfnamefont {B.}~\bibnamefont {Prinari}},
  \bibinfo {author} {\bibfnamefont {G.}~\bibnamefont {Biondini}}, \ and\
  \bibinfo {author} {\bibfnamefont {P.}~\bibnamefont {Schmelcher}},\ }\href
  {\doibase 10.1103/PhysRevA.97.043623} {\bibfield  {journal} {\bibinfo
  {journal} {Phys. Rev. A}\ }\textbf {\bibinfo {volume} {97}},\ \bibinfo
  {pages} {043623} (\bibinfo {year} {2018}{\natexlab{a}})}\BibitemShut
  {NoStop}%
\bibitem [{\citenamefont {Folman}\ \emph {et~al.}(2000)\citenamefont {Folman},
  \citenamefont {Kr{\"u}ger}, \citenamefont {Cassettari}, \citenamefont
  {Hessmo}, \citenamefont {Maier},\ and\ \citenamefont
  {Schmiedmayer}}]{folman2000controlling}%
  \BibitemOpen
  \bibfield  {author} {\bibinfo {author} {\bibfnamefont {R.}~\bibnamefont
  {Folman}}, \bibinfo {author} {\bibfnamefont {P.}~\bibnamefont {Kr{\"u}ger}},
  \bibinfo {author} {\bibfnamefont {D.}~\bibnamefont {Cassettari}}, \bibinfo
  {author} {\bibfnamefont {B.}~\bibnamefont {Hessmo}}, \bibinfo {author}
  {\bibfnamefont {T.}~\bibnamefont {Maier}}, \ and\ \bibinfo {author}
  {\bibfnamefont {J.}~\bibnamefont {Schmiedmayer}},\ }\href@noop {} {\bibfield
  {journal} {\bibinfo  {journal} {Phys. Rev. Lett.}\ }\textbf {\bibinfo
  {volume} {84}},\ \bibinfo {pages} {4749} (\bibinfo {year}
  {2000})}\BibitemShut {NoStop}%
\bibitem [{\citenamefont {Petrov}\ \emph {et~al.}(2009)\citenamefont {Petrov},
  \citenamefont {Machluf}, \citenamefont {Younis}, \citenamefont {Macaluso},
  \citenamefont {David}, \citenamefont {Hadad}, \citenamefont {Japha},
  \citenamefont {Keil}, \citenamefont {Joselevich},\ and\ \citenamefont
  {Folman}}]{petrov2009trapping}%
  \BibitemOpen
  \bibfield  {author} {\bibinfo {author} {\bibfnamefont {P.}~\bibnamefont
  {Petrov}}, \bibinfo {author} {\bibfnamefont {S.}~\bibnamefont {Machluf}},
  \bibinfo {author} {\bibfnamefont {S.}~\bibnamefont {Younis}}, \bibinfo
  {author} {\bibfnamefont {R.}~\bibnamefont {Macaluso}}, \bibinfo {author}
  {\bibfnamefont {T.}~\bibnamefont {David}}, \bibinfo {author} {\bibfnamefont
  {B.}~\bibnamefont {Hadad}}, \bibinfo {author} {\bibfnamefont
  {Y.}~\bibnamefont {Japha}}, \bibinfo {author} {\bibfnamefont
  {M.}~\bibnamefont {Keil}}, \bibinfo {author} {\bibfnamefont {E.}~\bibnamefont
  {Joselevich}}, \ and\ \bibinfo {author} {\bibfnamefont {R.}~\bibnamefont
  {Folman}},\ }\href@noop {} {\bibfield  {journal} {\bibinfo  {journal} {Phys.
  Rev. A}\ }\textbf {\bibinfo {volume} {79}},\ \bibinfo {pages} {043403}
  (\bibinfo {year} {2009})}\BibitemShut {NoStop}%
\bibitem [{\citenamefont {Rajendran}\ \emph {et~al.}(2011)\citenamefont
  {Rajendran}, \citenamefont {Lakshmanan},\ and\ \citenamefont
  {Muruganandam}}]{rajendran2011matter}%
  \BibitemOpen
  \bibfield  {author} {\bibinfo {author} {\bibfnamefont {S.}~\bibnamefont
  {Rajendran}}, \bibinfo {author} {\bibfnamefont {M.}~\bibnamefont
  {Lakshmanan}}, \ and\ \bibinfo {author} {\bibfnamefont {P.}~\bibnamefont
  {Muruganandam}},\ }\href@noop {} {\bibfield  {journal} {\bibinfo  {journal}
  {J. Math. Phys.}\ }\textbf {\bibinfo {volume} {52}},\ \bibinfo {pages}
  {023515} (\bibinfo {year} {2011})}\BibitemShut {NoStop}%
\bibitem [{\citenamefont {Ho}\ and\ \citenamefont
  {Shenoy}(1996)}]{HoMiscibilityCondition}%
  \BibitemOpen
  \bibfield  {author} {\bibinfo {author} {\bibfnamefont {T.-L.}\ \bibnamefont
  {Ho}}\ and\ \bibinfo {author} {\bibfnamefont {V.~B.}\ \bibnamefont
  {Shenoy}},\ }\href {\doibase 10.1103/PhysRevLett.77.3276} {\bibfield
  {journal} {\bibinfo  {journal} {Phys. Rev. Lett.}\ }\textbf {\bibinfo
  {volume} {77}},\ \bibinfo {pages} {3276} (\bibinfo {year}
  {1996})}\BibitemShut {NoStop}%
\bibitem [{\citenamefont {Timmermans}(1998)}]{TimmermansMiscibilityCondition}%
  \BibitemOpen
  \bibfield  {author} {\bibinfo {author} {\bibfnamefont {E.}~\bibnamefont
  {Timmermans}},\ }\href {\doibase 10.1103/PhysRevLett.81.5718} {\bibfield
  {journal} {\bibinfo  {journal} {Phys. Rev. Lett.}\ }\textbf {\bibinfo
  {volume} {81}},\ \bibinfo {pages} {5718} (\bibinfo {year}
  {1998})}\BibitemShut {NoStop}%
\bibitem [{\citenamefont {Ao}\ and\ \citenamefont
  {Chui}(1998)}]{AoMiscibilityCondition}%
  \BibitemOpen
  \bibfield  {author} {\bibinfo {author} {\bibfnamefont {P.}~\bibnamefont
  {Ao}}\ and\ \bibinfo {author} {\bibfnamefont {S.~T.}\ \bibnamefont {Chui}},\
  }\href {\doibase 10.1103/PhysRevA.58.4836} {\bibfield  {journal} {\bibinfo
  {journal} {Phys. Rev. A}\ }\textbf {\bibinfo {volume} {58}},\ \bibinfo
  {pages} {4836} (\bibinfo {year} {1998})}\BibitemShut {NoStop}%
\bibitem [{\citenamefont {Esry}\ and\ \citenamefont
  {Greene}(1998)}]{esryMisibilityCondition}%
  \BibitemOpen
  \bibfield  {author} {\bibinfo {author} {\bibfnamefont {B.~D.}\ \bibnamefont
  {Esry}}\ and\ \bibinfo {author} {\bibfnamefont {C.~H.}\ \bibnamefont
  {Greene}},\ }\href@noop {} {\bibfield  {journal} {\bibinfo  {journal}
  {Nature}\ }\textbf {\bibinfo {volume} {392}},\ \bibinfo {pages} {434}
  (\bibinfo {year} {1998})}\BibitemShut {NoStop}%
\bibitem [{\citenamefont {Inouye}\ \emph {et~al.}(1998)\citenamefont {Inouye},
  \citenamefont {Andrews}, \citenamefont {Stenger}, \citenamefont {Miesner},
  \citenamefont {Stamper-Kurn},\ and\ \citenamefont
  {Ketterle}}]{InouyeFeshbachResonance}%
  \BibitemOpen
  \bibfield  {author} {\bibinfo {author} {\bibfnamefont {S.}~\bibnamefont
  {Inouye}}, \bibinfo {author} {\bibfnamefont {M.}~\bibnamefont {Andrews}},
  \bibinfo {author} {\bibfnamefont {J.}~\bibnamefont {Stenger}}, \bibinfo
  {author} {\bibfnamefont {H.-J.}\ \bibnamefont {Miesner}}, \bibinfo {author}
  {\bibfnamefont {D.}~\bibnamefont {Stamper-Kurn}}, \ and\ \bibinfo {author}
  {\bibfnamefont {W.}~\bibnamefont {Ketterle}},\ }\href@noop {} {\bibfield
  {journal} {\bibinfo  {journal} {Nature}\ }\textbf {\bibinfo {volume} {392}},\
  \bibinfo {pages} {151} (\bibinfo {year} {1998})}\BibitemShut {NoStop}%
\bibitem [{\citenamefont {Vogels}\ \emph {et~al.}(1997)\citenamefont {Vogels},
  \citenamefont {Tsai}, \citenamefont {Freeland}, \citenamefont {Kokkelmans},
  \citenamefont {Verhaar},\ and\ \citenamefont
  {Heinzen}}]{VogelsFeshbachResonancePrediction}%
  \BibitemOpen
  \bibfield  {author} {\bibinfo {author} {\bibfnamefont {J.~M.}\ \bibnamefont
  {Vogels}}, \bibinfo {author} {\bibfnamefont {C.~C.}\ \bibnamefont {Tsai}},
  \bibinfo {author} {\bibfnamefont {R.~S.}\ \bibnamefont {Freeland}}, \bibinfo
  {author} {\bibfnamefont {S.~J. J. M.~F.}\ \bibnamefont {Kokkelmans}},
  \bibinfo {author} {\bibfnamefont {B.~J.}\ \bibnamefont {Verhaar}}, \ and\
  \bibinfo {author} {\bibfnamefont {D.~J.}\ \bibnamefont {Heinzen}},\ }\href
  {\doibase 10.1103/PhysRevA.56.R1067} {\bibfield  {journal} {\bibinfo
  {journal} {Phys. Rev. A}\ }\textbf {\bibinfo {volume} {56}},\ \bibinfo
  {pages} {R1067} (\bibinfo {year} {1997})}\BibitemShut {NoStop}%
\bibitem [{\citenamefont {Thalhammer}\ \emph {et~al.}(2008)\citenamefont
  {Thalhammer}, \citenamefont {Barontini}, \citenamefont {De~Sarlo},
  \citenamefont {Catani}, \citenamefont {Minardi},\ and\ \citenamefont
  {Inguscio}}]{ThalhammerFeshbachResonance}%
  \BibitemOpen
  \bibfield  {author} {\bibinfo {author} {\bibfnamefont {G.}~\bibnamefont
  {Thalhammer}}, \bibinfo {author} {\bibfnamefont {G.}~\bibnamefont
  {Barontini}}, \bibinfo {author} {\bibfnamefont {L.}~\bibnamefont {De~Sarlo}},
  \bibinfo {author} {\bibfnamefont {J.}~\bibnamefont {Catani}}, \bibinfo
  {author} {\bibfnamefont {F.}~\bibnamefont {Minardi}}, \ and\ \bibinfo
  {author} {\bibfnamefont {M.}~\bibnamefont {Inguscio}},\ }\href {\doibase
  10.1103/PhysRevLett.100.210402} {\bibfield  {journal} {\bibinfo  {journal}
  {Phys. Rev. Lett.}\ }\textbf {\bibinfo {volume} {100}},\ \bibinfo {pages}
  {210402} (\bibinfo {year} {2008})}\BibitemShut {NoStop}%
\bibitem [{\citenamefont {Papp}\ \emph {et~al.}(2008)\citenamefont {Papp},
  \citenamefont {Pino},\ and\ \citenamefont {Wieman}}]{PappTunableMiscibility}%
  \BibitemOpen
  \bibfield  {author} {\bibinfo {author} {\bibfnamefont {S.~B.}\ \bibnamefont
  {Papp}}, \bibinfo {author} {\bibfnamefont {J.~M.}\ \bibnamefont {Pino}}, \
  and\ \bibinfo {author} {\bibfnamefont {C.~E.}\ \bibnamefont {Wieman}},\
  }\href {\doibase 10.1103/PhysRevLett.101.040402} {\bibfield  {journal}
  {\bibinfo  {journal} {Phys. Rev. Lett.}\ }\textbf {\bibinfo {volume} {101}},\
  \bibinfo {pages} {040402} (\bibinfo {year} {2008})}\BibitemShut {NoStop}%
\bibitem [{\citenamefont {Wang}\ \emph {et~al.}(2016)\citenamefont {Wang},
  \citenamefont {Li}, \citenamefont {Xiong},\ and\ \citenamefont
  {Wang}}]{WangTunableMiscibility}%
  \BibitemOpen
  \bibfield  {author} {\bibinfo {author} {\bibfnamefont {F.}~\bibnamefont
  {Wang}}, \bibinfo {author} {\bibfnamefont {X.}~\bibnamefont {Li}}, \bibinfo
  {author} {\bibfnamefont {D.}~\bibnamefont {Xiong}}, \ and\ \bibinfo {author}
  {\bibfnamefont {D.}~\bibnamefont {Wang}},\ }\href
  {http://stacks.iop.org/0953-4075/49/i=1/a=015302} {\bibfield  {journal}
  {\bibinfo  {journal} {J. Phys. B: At. Mol. Opt. Phys.}\ }\textbf {\bibinfo
  {volume} {49}},\ \bibinfo {pages} {015302} (\bibinfo {year}
  {2016})}\BibitemShut {NoStop}%
\bibitem [{\citenamefont {Tojo}\ \emph {et~al.}(2010)\citenamefont {Tojo},
  \citenamefont {Taguchi}, \citenamefont {Masuyama}, \citenamefont {Hayashi},
  \citenamefont {Saito},\ and\ \citenamefont
  {Hirano}}]{TojoControlledPhaseSeparation}%
  \BibitemOpen
  \bibfield  {author} {\bibinfo {author} {\bibfnamefont {S.}~\bibnamefont
  {Tojo}}, \bibinfo {author} {\bibfnamefont {Y.}~\bibnamefont {Taguchi}},
  \bibinfo {author} {\bibfnamefont {Y.}~\bibnamefont {Masuyama}}, \bibinfo
  {author} {\bibfnamefont {T.}~\bibnamefont {Hayashi}}, \bibinfo {author}
  {\bibfnamefont {H.}~\bibnamefont {Saito}}, \ and\ \bibinfo {author}
  {\bibfnamefont {T.}~\bibnamefont {Hirano}},\ }\href {\doibase
  10.1103/PhysRevA.82.033609} {\bibfield  {journal} {\bibinfo  {journal} {Phys.
  Rev. A}\ }\textbf {\bibinfo {volume} {82}},\ \bibinfo {pages} {033609}
  (\bibinfo {year} {2010})}\BibitemShut {NoStop}%
\bibitem [{\citenamefont {Kanna}\ \emph {et~al.}(2013)\citenamefont {Kanna},
  \citenamefont {Mareeswaran}, \citenamefont {Tsitoura}, \citenamefont
  {Nistazakis},\ and\ \citenamefont {Frantzeskakis}}]{kanna2013non}%
  \BibitemOpen
  \bibfield  {author} {\bibinfo {author} {\bibfnamefont {T.}~\bibnamefont
  {Kanna}}, \bibinfo {author} {\bibfnamefont {R.~B.}\ \bibnamefont
  {Mareeswaran}}, \bibinfo {author} {\bibfnamefont {F.}~\bibnamefont
  {Tsitoura}}, \bibinfo {author} {\bibfnamefont {H.}~\bibnamefont
  {Nistazakis}}, \ and\ \bibinfo {author} {\bibfnamefont {D.}~\bibnamefont
  {Frantzeskakis}},\ }\href@noop {} {\bibfield  {journal} {\bibinfo  {journal}
  {J. Phys. A: Math. Theor.}\ }\textbf {\bibinfo {volume} {46}},\ \bibinfo
  {pages} {475201} (\bibinfo {year} {2013})}\BibitemShut {NoStop}%
\bibitem [{\citenamefont {Manikandan}\ \emph {et~al.}(2016)\citenamefont
  {Manikandan}, \citenamefont {Muruganandam}, \citenamefont {Senthilvelan},\
  and\ \citenamefont {Lakshmanan}}]{manikandan2016manipulating}%
  \BibitemOpen
  \bibfield  {author} {\bibinfo {author} {\bibfnamefont {K.}~\bibnamefont
  {Manikandan}}, \bibinfo {author} {\bibfnamefont {P.}~\bibnamefont
  {Muruganandam}}, \bibinfo {author} {\bibfnamefont {M.}~\bibnamefont
  {Senthilvelan}}, \ and\ \bibinfo {author} {\bibfnamefont {M.}~\bibnamefont
  {Lakshmanan}},\ }\href@noop {} {\bibfield  {journal} {\bibinfo  {journal}
  {Phys. Rev. E}\ }\textbf {\bibinfo {volume} {93}},\ \bibinfo {pages} {032212}
  (\bibinfo {year} {2016})}\BibitemShut {NoStop}%
\bibitem [{\citenamefont {Yu}(2016)}]{yu2016nonautonomous}%
  \BibitemOpen
  \bibfield  {author} {\bibinfo {author} {\bibfnamefont {F.}~\bibnamefont
  {Yu}},\ }\href@noop {} {\bibfield  {journal} {\bibinfo  {journal} {Nonl.
  Dyn.}\ }\textbf {\bibinfo {volume} {85}},\ \bibinfo {pages} {1203} (\bibinfo
  {year} {2016})}\BibitemShut {NoStop}%
\bibitem [{\citenamefont {Cheng}(2009)}]{cheng2009effective}%
  \BibitemOpen
  \bibfield  {author} {\bibinfo {author} {\bibfnamefont {Y.}~\bibnamefont
  {Cheng}},\ }\href@noop {} {\bibfield  {journal} {\bibinfo  {journal} {J.
  Phys. B: At. Mol. Opt. Phys.}\ }\textbf {\bibinfo {volume} {42}},\ \bibinfo
  {pages} {205005} (\bibinfo {year} {2009})}\BibitemShut {NoStop}%
\bibitem [{\citenamefont {Belmonte-Beitia}\ \emph {et~al.}(2011)\citenamefont
  {Belmonte-Beitia}, \citenamefont {P{\'e}rez-Garc{\'\i}a},\ and\ \citenamefont
  {Brazhnyi}}]{belmonte2011solitary}%
  \BibitemOpen
  \bibfield  {author} {\bibinfo {author} {\bibfnamefont {J.}~\bibnamefont
  {Belmonte-Beitia}}, \bibinfo {author} {\bibfnamefont {V.~M.}\ \bibnamefont
  {P{\'e}rez-Garc{\'\i}a}}, \ and\ \bibinfo {author} {\bibfnamefont
  {V.}~\bibnamefont {Brazhnyi}},\ }\href@noop {} {\bibfield  {journal}
  {\bibinfo  {journal} {Commun. Nonl. Sc. and Num. Sim.}\ }\textbf {\bibinfo
  {volume} {16}},\ \bibinfo {pages} {158} (\bibinfo {year} {2011})}\BibitemShut
  {NoStop}%
\bibitem [{\citenamefont {Mareeswaran}\ \emph {et~al.}(2014)\citenamefont
  {Mareeswaran}, \citenamefont {Charalampidis}, \citenamefont {Kanna},
  \citenamefont {Kevrekidis},\ and\ \citenamefont
  {Frantzeskakis}}]{mareeswaran2014vector}%
  \BibitemOpen
  \bibfield  {author} {\bibinfo {author} {\bibfnamefont {R.~B.}\ \bibnamefont
  {Mareeswaran}}, \bibinfo {author} {\bibfnamefont {E.}~\bibnamefont
  {Charalampidis}}, \bibinfo {author} {\bibfnamefont {T.}~\bibnamefont
  {Kanna}}, \bibinfo {author} {\bibfnamefont {P.}~\bibnamefont {Kevrekidis}}, \
  and\ \bibinfo {author} {\bibfnamefont {D.}~\bibnamefont {Frantzeskakis}},\
  }\href@noop {} {\bibfield  {journal} {\bibinfo  {journal} {Phys. Rev. E}\
  }\textbf {\bibinfo {volume} {90}},\ \bibinfo {pages} {042912} (\bibinfo
  {year} {2014})}\BibitemShut {NoStop}%
\bibitem [{\citenamefont {Eto}\ \emph {et~al.}(2016)\citenamefont {Eto},
  \citenamefont {Takahashi}, \citenamefont {Kunimi}, \citenamefont {Saito},\
  and\ \citenamefont {Hirano}}]{EtoPhaseSeparation}%
  \BibitemOpen
  \bibfield  {author} {\bibinfo {author} {\bibfnamefont {Y.}~\bibnamefont
  {Eto}}, \bibinfo {author} {\bibfnamefont {M.}~\bibnamefont {Takahashi}},
  \bibinfo {author} {\bibfnamefont {M.}~\bibnamefont {Kunimi}}, \bibinfo
  {author} {\bibfnamefont {H.}~\bibnamefont {Saito}}, \ and\ \bibinfo {author}
  {\bibfnamefont {T.}~\bibnamefont {Hirano}},\ }\href
  {http://stacks.iop.org/1367-2630/18/i=7/a=073029} {\bibfield  {journal}
  {\bibinfo  {journal} {New J. Phys.}\ }\textbf {\bibinfo {volume} {18}},\
  \bibinfo {pages} {073029} (\bibinfo {year} {2016})}\BibitemShut {NoStop}%
\bibitem [{\citenamefont {Manikandan}\ \emph {et~al.}(2014)\citenamefont
  {Manikandan}, \citenamefont {Radhakrishnan},\ and\ \citenamefont
  {Aravinthan}}]{manikandan2014generalized}%
  \BibitemOpen
  \bibfield  {author} {\bibinfo {author} {\bibfnamefont {N.}~\bibnamefont
  {Manikandan}}, \bibinfo {author} {\bibfnamefont {R.}~\bibnamefont
  {Radhakrishnan}}, \ and\ \bibinfo {author} {\bibfnamefont {K.}~\bibnamefont
  {Aravinthan}},\ }\href@noop {} {\bibfield  {journal} {\bibinfo  {journal}
  {Phys. Rev. E}\ }\textbf {\bibinfo {volume} {90}},\ \bibinfo {pages} {022902}
  (\bibinfo {year} {2014})}\BibitemShut {NoStop}%
\bibitem [{\citenamefont {Sartori}\ and\ \citenamefont
  {Recati}(2013)}]{sartori2013dynamics}%
  \BibitemOpen
  \bibfield  {author} {\bibinfo {author} {\bibfnamefont {A.}~\bibnamefont
  {Sartori}}\ and\ \bibinfo {author} {\bibfnamefont {A.}~\bibnamefont
  {Recati}},\ }\href@noop {} {\bibfield  {journal} {\bibinfo  {journal} {Eur.
  Phys. J. D}\ }\textbf {\bibinfo {volume} {67}},\ \bibinfo {pages} {260}
  (\bibinfo {year} {2013})}\BibitemShut {NoStop}%
\bibitem [{\citenamefont {Pu}\ and\ \citenamefont
  {Bigelow}(1998)}]{BigelowBallShell}%
  \BibitemOpen
  \bibfield  {author} {\bibinfo {author} {\bibfnamefont {H.}~\bibnamefont
  {Pu}}\ and\ \bibinfo {author} {\bibfnamefont {N.~P.}\ \bibnamefont
  {Bigelow}},\ }\href {\doibase 10.1103/PhysRevLett.80.1130} {\bibfield
  {journal} {\bibinfo  {journal} {Phys. Rev. Lett.}\ }\textbf {\bibinfo
  {volume} {80}},\ \bibinfo {pages} {1130} (\bibinfo {year}
  {1998})}\BibitemShut {NoStop}%
\bibitem [{\citenamefont {Mertes}\ \emph {et~al.}(2007)\citenamefont {Mertes},
  \citenamefont {Merrill}, \citenamefont {Carretero-Gonz\'alez}, \citenamefont
  {Frantzeskakis}, \citenamefont {Kevrekidis},\ and\ \citenamefont
  {Hall}}]{MertesBallShell}%
  \BibitemOpen
  \bibfield  {author} {\bibinfo {author} {\bibfnamefont {K.~M.}\ \bibnamefont
  {Mertes}}, \bibinfo {author} {\bibfnamefont {J.~W.}\ \bibnamefont {Merrill}},
  \bibinfo {author} {\bibfnamefont {R.}~\bibnamefont {Carretero-Gonz\'alez}},
  \bibinfo {author} {\bibfnamefont {D.~J.}\ \bibnamefont {Frantzeskakis}},
  \bibinfo {author} {\bibfnamefont {P.~G.}\ \bibnamefont {Kevrekidis}}, \ and\
  \bibinfo {author} {\bibfnamefont {D.~S.}\ \bibnamefont {Hall}},\ }\href
  {\doibase 10.1103/PhysRevLett.99.190402} {\bibfield  {journal} {\bibinfo
  {journal} {Phys. Rev. Lett.}\ }\textbf {\bibinfo {volume} {99}},\ \bibinfo
  {pages} {190402} (\bibinfo {year} {2007})}\BibitemShut {NoStop}%
\bibitem [{\citenamefont {Andrews}\ \emph {et~al.}(1997)\citenamefont
  {Andrews}, \citenamefont {Townsend}, \citenamefont {Miesner}, \citenamefont
  {Durfee}, \citenamefont {Kurn},\ and\ \citenamefont
  {Ketterle}}]{AndrewsBECInterference}%
  \BibitemOpen
  \bibfield  {author} {\bibinfo {author} {\bibfnamefont {M.~R.}\ \bibnamefont
  {Andrews}}, \bibinfo {author} {\bibfnamefont {C.~G.}\ \bibnamefont
  {Townsend}}, \bibinfo {author} {\bibfnamefont {H.-J.}\ \bibnamefont
  {Miesner}}, \bibinfo {author} {\bibfnamefont {D.~S.}\ \bibnamefont {Durfee}},
  \bibinfo {author} {\bibfnamefont {D.~M.}\ \bibnamefont {Kurn}}, \ and\
  \bibinfo {author} {\bibfnamefont {W.}~\bibnamefont {Ketterle}},\ }\href
  {\doibase 10.1126/science.275.5300.637} {\ \textbf {\bibinfo {volume}
  {275}},\ \bibinfo {pages} {637} (\bibinfo {year} {1997})}\BibitemShut
  {NoStop}%
\bibitem [{\citenamefont {Naraschewski}\ \emph {et~al.}(1996)\citenamefont
  {Naraschewski}, \citenamefont {Wallis}, \citenamefont {Schenzle},
  \citenamefont {Cirac},\ and\ \citenamefont
  {Zoller}}]{NaraschewskiBECInterference}%
  \BibitemOpen
  \bibfield  {author} {\bibinfo {author} {\bibfnamefont {M.}~\bibnamefont
  {Naraschewski}}, \bibinfo {author} {\bibfnamefont {H.}~\bibnamefont
  {Wallis}}, \bibinfo {author} {\bibfnamefont {A.}~\bibnamefont {Schenzle}},
  \bibinfo {author} {\bibfnamefont {J.~I.}\ \bibnamefont {Cirac}}, \ and\
  \bibinfo {author} {\bibfnamefont {P.}~\bibnamefont {Zoller}},\ }\href
  {\doibase 10.1103/PhysRevA.54.2185} {\bibfield  {journal} {\bibinfo
  {journal} {Phys. Rev. A}\ }\textbf {\bibinfo {volume} {54}},\ \bibinfo
  {pages} {2185} (\bibinfo {year} {1996})}\BibitemShut {NoStop}%
\bibitem [{\citenamefont {Hoston}\ and\ \citenamefont
  {You}(1996)}]{HostonBECInterference}%
  \BibitemOpen
  \bibfield  {author} {\bibinfo {author} {\bibfnamefont {W.}~\bibnamefont
  {Hoston}}\ and\ \bibinfo {author} {\bibfnamefont {L.}~\bibnamefont {You}},\
  }\href {\doibase 10.1103/PhysRevA.53.4254} {\bibfield  {journal} {\bibinfo
  {journal} {Phys. Rev. A}\ }\textbf {\bibinfo {volume} {53}},\ \bibinfo
  {pages} {4254} (\bibinfo {year} {1996})}\BibitemShut {NoStop}%
\bibitem [{\citenamefont {Egorov}\ \emph {et~al.}(2013)\citenamefont {Egorov},
  \citenamefont {Opanchuk}, \citenamefont {Drummond}, \citenamefont {Hall},
  \citenamefont {Hannaford},\ and\ \citenamefont
  {Sidorov}}]{EgorovRbScatteringLengths}%
  \BibitemOpen
  \bibfield  {author} {\bibinfo {author} {\bibfnamefont {M.}~\bibnamefont
  {Egorov}}, \bibinfo {author} {\bibfnamefont {B.}~\bibnamefont {Opanchuk}},
  \bibinfo {author} {\bibfnamefont {P.}~\bibnamefont {Drummond}}, \bibinfo
  {author} {\bibfnamefont {B.~V.}\ \bibnamefont {Hall}}, \bibinfo {author}
  {\bibfnamefont {P.}~\bibnamefont {Hannaford}}, \ and\ \bibinfo {author}
  {\bibfnamefont {A.~I.}\ \bibnamefont {Sidorov}},\ }\href {\doibase
  10.1103/PhysRevA.87.053614} {\bibfield  {journal} {\bibinfo  {journal} {Phys.
  Rev. A}\ }\textbf {\bibinfo {volume} {87}},\ \bibinfo {pages} {053614}
  (\bibinfo {year} {2013})}\BibitemShut {NoStop}%
\bibitem [{\citenamefont {K{\"o}hler}\ \emph {et~al.}(2006)\citenamefont
  {K{\"o}hler}, \citenamefont {G{\'o}ral},\ and\ \citenamefont
  {Julienne}}]{kohler2006production}%
  \BibitemOpen
  \bibfield  {author} {\bibinfo {author} {\bibfnamefont {T.}~\bibnamefont
  {K{\"o}hler}}, \bibinfo {author} {\bibfnamefont {K.}~\bibnamefont
  {G{\'o}ral}}, \ and\ \bibinfo {author} {\bibfnamefont {P.~S.}\ \bibnamefont
  {Julienne}},\ }\href@noop {} {\bibfield  {journal} {\bibinfo  {journal} {Rev.
  Mod. Phys.}\ }\textbf {\bibinfo {volume} {78}},\ \bibinfo {pages} {1311}
  (\bibinfo {year} {2006})}\BibitemShut {NoStop}%
\bibitem [{\citenamefont {Chin}\ \emph {et~al.}(2010)\citenamefont {Chin},
  \citenamefont {Grimm}, \citenamefont {Julienne},\ and\ \citenamefont
  {Tiesinga}}]{chin2010feshbach}%
  \BibitemOpen
  \bibfield  {author} {\bibinfo {author} {\bibfnamefont {C.}~\bibnamefont
  {Chin}}, \bibinfo {author} {\bibfnamefont {R.}~\bibnamefont {Grimm}},
  \bibinfo {author} {\bibfnamefont {P.}~\bibnamefont {Julienne}}, \ and\
  \bibinfo {author} {\bibfnamefont {E.}~\bibnamefont {Tiesinga}},\ }\href@noop
  {} {\bibfield  {journal} {\bibinfo  {journal} {Rev. Mod. Phys.}\ }\textbf
  {\bibinfo {volume} {82}},\ \bibinfo {pages} {1225} (\bibinfo {year}
  {2010})}\BibitemShut {NoStop}%
\bibitem [{\citenamefont {Olshanii}(1998)}]{olshanii1998atomic}%
  \BibitemOpen
  \bibfield  {author} {\bibinfo {author} {\bibfnamefont {M.}~\bibnamefont
  {Olshanii}},\ }\href@noop {} {\bibfield  {journal} {\bibinfo  {journal}
  {Phys. Rev. Lett.}\ }\textbf {\bibinfo {volume} {81}},\ \bibinfo {pages}
  {938} (\bibinfo {year} {1998})}\BibitemShut {NoStop}%
\bibitem [{\citenamefont {Kim}\ \emph {et~al.}(2006)\citenamefont {Kim},
  \citenamefont {Melezhik},\ and\ \citenamefont
  {Schmelcher}}]{kim2006suppression}%
  \BibitemOpen
  \bibfield  {author} {\bibinfo {author} {\bibfnamefont {J.~I.}\ \bibnamefont
  {Kim}}, \bibinfo {author} {\bibfnamefont {V.~S.}\ \bibnamefont {Melezhik}}, \
  and\ \bibinfo {author} {\bibfnamefont {P.}~\bibnamefont {Schmelcher}},\
  }\href@noop {} {\bibfield  {journal} {\bibinfo  {journal} {Phys. Rev. Lett.}\
  }\textbf {\bibinfo {volume} {97}},\ \bibinfo {pages} {193203} (\bibinfo
  {year} {2006})}\BibitemShut {NoStop}%
\bibitem [{\citenamefont {Kelley}(2003)}]{KelleyNewtonKrylov}%
  \BibitemOpen
  \bibfield  {author} {\bibinfo {author} {\bibfnamefont {C.~T.}\ \bibnamefont
  {Kelley}},\ }\href {https://archive.siam.org/books/fa01/} {\emph {\bibinfo
  {title} {Solving Nonlinear Equations with Newton's Method}}}\ (\bibinfo
  {publisher} {Society for Industrial and Applied Mathematics},\ \bibinfo
  {year} {2003})\BibitemShut {NoStop}%
\bibitem [{\citenamefont {Navarro}\ \emph {et~al.}(2009)\citenamefont
  {Navarro}, \citenamefont {Carretero-Gonz\'alez},\ and\ \citenamefont
  {Kevrekidis}}]{NavarroModifiedMiscibilityCondition}%
  \BibitemOpen
  \bibfield  {author} {\bibinfo {author} {\bibfnamefont {R.}~\bibnamefont
  {Navarro}}, \bibinfo {author} {\bibfnamefont {R.}~\bibnamefont
  {Carretero-Gonz\'alez}}, \ and\ \bibinfo {author} {\bibfnamefont {P.~G.}\
  \bibnamefont {Kevrekidis}},\ }\href {\doibase 10.1103/PhysRevA.80.023613}
  {\bibfield  {journal} {\bibinfo  {journal} {Phys. Rev. A}\ }\textbf {\bibinfo
  {volume} {80}},\ \bibinfo {pages} {023613} (\bibinfo {year}
  {2009})}\BibitemShut {NoStop}%
\bibitem [{\citenamefont {Mistakidis}\ \emph {et~al.}(2019)\citenamefont
  {Mistakidis}, \citenamefont {Katsimiga}, \citenamefont {Koutentakis},
  \citenamefont {Busch},\ and\ \citenamefont
  {Schmelcher}}]{mistakidis2019quench}%
  \BibitemOpen
  \bibfield  {author} {\bibinfo {author} {\bibfnamefont {S.~I.}\ \bibnamefont
  {Mistakidis}}, \bibinfo {author} {\bibfnamefont {G.~C.}\ \bibnamefont
  {Katsimiga}}, \bibinfo {author} {\bibfnamefont {G.~M.}\ \bibnamefont
  {Koutentakis}}, \bibinfo {author} {\bibfnamefont {T.}~\bibnamefont {Busch}},
  \ and\ \bibinfo {author} {\bibfnamefont {P.}~\bibnamefont {Schmelcher}},\
  }\href@noop {} {\bibfield  {journal} {\bibinfo  {journal} {Phys. Rev. Lett.}\
  }\textbf {\bibinfo {volume} {122}},\ \bibinfo {pages} {183001} (\bibinfo
  {year} {2019})}\BibitemShut {NoStop}%
\bibitem [{\citenamefont {Ferrier-Barbut}\ \emph {et~al.}(2014)\citenamefont
  {Ferrier-Barbut}, \citenamefont {Delehaye}, \citenamefont {Laurent},
  \citenamefont {Grier}, \citenamefont {Pierce}, \citenamefont {Rem},
  \citenamefont {Chevy},\ and\ \citenamefont {Salomon}}]{ferrier2014mixture}%
  \BibitemOpen
  \bibfield  {author} {\bibinfo {author} {\bibfnamefont {I.}~\bibnamefont
  {Ferrier-Barbut}}, \bibinfo {author} {\bibfnamefont {M.}~\bibnamefont
  {Delehaye}}, \bibinfo {author} {\bibfnamefont {S.}~\bibnamefont {Laurent}},
  \bibinfo {author} {\bibfnamefont {A.~T.}\ \bibnamefont {Grier}}, \bibinfo
  {author} {\bibfnamefont {M.}~\bibnamefont {Pierce}}, \bibinfo {author}
  {\bibfnamefont {B.~S.}\ \bibnamefont {Rem}}, \bibinfo {author} {\bibfnamefont
  {F.}~\bibnamefont {Chevy}}, \ and\ \bibinfo {author} {\bibfnamefont
  {C.}~\bibnamefont {Salomon}},\ }\href@noop {} {\bibfield  {journal} {\bibinfo
   {journal} {Science}\ }\textbf {\bibinfo {volume} {345}},\ \bibinfo {pages}
  {1035} (\bibinfo {year} {2014})}\BibitemShut {NoStop}%
\bibitem [{\citenamefont {Tommasini}\ \emph {et~al.}(2003)\citenamefont
  {Tommasini}, \citenamefont {de~Passos}, \citenamefont {de~Toledo~Piza},
  \citenamefont {Hussein},\ and\ \citenamefont
  {Timmermans}}]{TommasiniBogoliubov}%
  \BibitemOpen
  \bibfield  {author} {\bibinfo {author} {\bibfnamefont {P.}~\bibnamefont
  {Tommasini}}, \bibinfo {author} {\bibfnamefont {E.~J.~V.}\ \bibnamefont
  {de~Passos}}, \bibinfo {author} {\bibfnamefont {A.~F.~R.}\ \bibnamefont
  {de~Toledo~Piza}}, \bibinfo {author} {\bibfnamefont {M.~S.}\ \bibnamefont
  {Hussein}}, \ and\ \bibinfo {author} {\bibfnamefont {E.}~\bibnamefont
  {Timmermans}},\ }\href {\doibase 10.1103/PhysRevA.67.023606} {\bibfield
  {journal} {\bibinfo  {journal} {Phys. Rev. A}\ }\textbf {\bibinfo {volume}
  {67}},\ \bibinfo {pages} {023606} (\bibinfo {year} {2003})}\BibitemShut
  {NoStop}%
\bibitem [{Note1()}]{Note1}%
  \BibitemOpen
  \bibinfo {note} {For the numerical evaluation of $N_b^{num}$ for a selected
  DB pair we used $N_b^{num}=(1/N_B)\DOTSI \intop \ilimits@ _{c_1}^{c_2} dx
  \left | \Psi _B \right |^2$. Here, $N_B$ is the total number of particles in
  the $B$-species and the integration interval $[ c_1$, $c_2 ]$ refers to the
  spatial area around the center of the selected bright soliton. The observed
  deviation between $N_b^{num}$ and $N_b$ is smaller than $1\%$.}\BibitemShut
  {Stop}%
\bibitem [{Note2()}]{Note2}%
  \BibitemOpen
  \bibinfo {note} {Note that these equalities are exact at the integrable
  limit, but approximately hold also here with a deviation of about
  $2\%$}\BibitemShut {NoStop}%
\bibitem [{\citenamefont {Bloch}\ and\ \citenamefont
  {Zwerger}(2008)}]{bloch2008many}%
  \BibitemOpen
  \bibfield  {author} {\bibinfo {author} {\bibfnamefont {J.}~\bibnamefont
  {Bloch}, \bibfnamefont {I.~Dalibard}}\ and\ \bibinfo {author} {\bibfnamefont
  {W.}~\bibnamefont {Zwerger}},\ }\href@noop {} {\bibfield  {journal} {\bibinfo
   {journal} {Rev. Mod. Phys.}\ }\textbf {\bibinfo {volume} {80}},\ \bibinfo
  {pages} {885} (\bibinfo {year} {2008})}\BibitemShut {NoStop}%
\bibitem [{\citenamefont {Katsimiga}\ \emph
  {et~al.}(2017{\natexlab{d}})\citenamefont {Katsimiga}, \citenamefont
  {Mistakidis}, \citenamefont {Koutentakis}, \citenamefont {Kevrekidis},\ and\
  \citenamefont {Schmelcher}}]{katsimiga2017many}%
  \BibitemOpen
  \bibfield  {author} {\bibinfo {author} {\bibfnamefont {G.~C.}\ \bibnamefont
  {Katsimiga}}, \bibinfo {author} {\bibfnamefont {S.~I.}\ \bibnamefont
  {Mistakidis}}, \bibinfo {author} {\bibfnamefont {G.~M.}\ \bibnamefont
  {Koutentakis}}, \bibinfo {author} {\bibfnamefont {P.~G.}\ \bibnamefont
  {Kevrekidis}}, \ and\ \bibinfo {author} {\bibfnamefont {P.}~\bibnamefont
  {Schmelcher}},\ }\href@noop {} {\bibfield  {journal} {\bibinfo  {journal}
  {New J. Phys.}\ }\textbf {\bibinfo {volume} {19}},\ \bibinfo {pages} {123012}
  (\bibinfo {year} {2017}{\natexlab{d}})}\BibitemShut {NoStop}%
\bibitem [{\citenamefont {Zakharov}\ and\ \citenamefont
  {Shabat}(1973)}]{Zakharov1973}%
  \BibitemOpen
  \bibfield  {author} {\bibinfo {author} {\bibfnamefont {V.~E.}\ \bibnamefont
  {Zakharov}}\ and\ \bibinfo {author} {\bibfnamefont {A.~B.}\ \bibnamefont
  {Shabat}},\ }\href@noop {} {\bibfield  {journal} {\bibinfo  {journal} {Sov.
  J. Exp. Theor. Phys.}\ }\textbf {\bibinfo {volume} {37}},\ \bibinfo {pages}
  {823} (\bibinfo {year} {1973})}\BibitemShut {NoStop}%
\bibitem [{\citenamefont {Scott}\ \emph {et~al.}(1998)\citenamefont {Scott},
  \citenamefont {Ballagh},\ and\ \citenamefont {Burnett}}]{Scott1998}%
  \BibitemOpen
  \bibfield  {author} {\bibinfo {author} {\bibfnamefont {T.~F.}\ \bibnamefont
  {Scott}}, \bibinfo {author} {\bibfnamefont {R.~J.}\ \bibnamefont {Ballagh}},
  \ and\ \bibinfo {author} {\bibfnamefont {K.}~\bibnamefont {Burnett}},\ }\href
  {\doibase 10.1088/0953-4075/31/8/001} {\bibfield  {journal} {\bibinfo
  {journal} {J. Phys. B: At. Mol. Opt. Phys.}\ }\textbf {\bibinfo {volume}
  {31}},\ \bibinfo {pages} {329} (\bibinfo {year} {1998})}\BibitemShut
  {NoStop}%
\bibitem [{\citenamefont {Law}\ \emph {et~al.}(2010)\citenamefont {Law},
  \citenamefont {Kevrekidis},\ and\ \citenamefont
  {Tuckerman}}]{LawVortexBright}%
  \BibitemOpen
  \bibfield  {author} {\bibinfo {author} {\bibfnamefont {K.~J.~H.}\
  \bibnamefont {Law}}, \bibinfo {author} {\bibfnamefont {P.~G.}\ \bibnamefont
  {Kevrekidis}}, \ and\ \bibinfo {author} {\bibfnamefont {L.~S.}\ \bibnamefont
  {Tuckerman}},\ }\href {\doibase 10.1103/PhysRevLett.105.160405} {\bibfield
  {journal} {\bibinfo  {journal} {Phys. Rev. Lett.}\ }\textbf {\bibinfo
  {volume} {105}},\ \bibinfo {pages} {160405} (\bibinfo {year}
  {2010})}\BibitemShut {NoStop}%
\bibitem [{\citenamefont {Pola}\ \emph {et~al.}(2012)\citenamefont {Pola},
  \citenamefont {Stockhofe}, \citenamefont {Schmelcher},\ and\ \citenamefont
  {Kevrekidis}}]{PolaVortexBright}%
  \BibitemOpen
  \bibfield  {author} {\bibinfo {author} {\bibfnamefont {M.}~\bibnamefont
  {Pola}}, \bibinfo {author} {\bibfnamefont {J.}~\bibnamefont {Stockhofe}},
  \bibinfo {author} {\bibfnamefont {P.}~\bibnamefont {Schmelcher}}, \ and\
  \bibinfo {author} {\bibfnamefont {P.~G.}\ \bibnamefont {Kevrekidis}},\ }\href
  {\doibase 10.1103/PhysRevA.86.053601} {\bibfield  {journal} {\bibinfo
  {journal} {Phys. Rev. A}\ }\textbf {\bibinfo {volume} {86}},\ \bibinfo
  {pages} {053601} (\bibinfo {year} {2012})}\BibitemShut {NoStop}%
\bibitem [{\citenamefont {Mukherjee}\ \emph {et~al.}(2019)\citenamefont
  {Mukherjee}, \citenamefont {Mistakidis}, \citenamefont {Kevrekidis},\ and\
  \citenamefont {Schmelcher}}]{mukherjee2019quench}%
  \BibitemOpen
  \bibfield  {author} {\bibinfo {author} {\bibfnamefont {K.}~\bibnamefont
  {Mukherjee}}, \bibinfo {author} {\bibfnamefont {S.~I.}\ \bibnamefont
  {Mistakidis}}, \bibinfo {author} {\bibfnamefont {P.~G.}\ \bibnamefont
  {Kevrekidis}}, \ and\ \bibinfo {author} {\bibfnamefont {P.}~\bibnamefont
  {Schmelcher}},\ }\href@noop {} {\bibfield  {journal} {\bibinfo  {journal}
  {arXiv:\textbf{1904.06208}}\ } (\bibinfo {year} {2019})}\BibitemShut
  {NoStop}%
\bibitem [{\citenamefont {Charalampidis}\ \emph {et~al.}(2016)\citenamefont
  {Charalampidis}, \citenamefont {Wang}, \citenamefont {Kevrekidis},
  \citenamefont {Frantzeskakis},\ and\ \citenamefont
  {Cuevas-Maraver}}]{charalampidis2016so}%
  \BibitemOpen
  \bibfield  {author} {\bibinfo {author} {\bibfnamefont {E.~G.}\ \bibnamefont
  {Charalampidis}}, \bibinfo {author} {\bibfnamefont {W.}~\bibnamefont {Wang}},
  \bibinfo {author} {\bibfnamefont {P.~G.}\ \bibnamefont {Kevrekidis}},
  \bibinfo {author} {\bibfnamefont {D.~J.}\ \bibnamefont {Frantzeskakis}}, \
  and\ \bibinfo {author} {\bibfnamefont {J.}~\bibnamefont {Cuevas-Maraver}},\
  }\href@noop {} {\bibfield  {journal} {\bibinfo  {journal} {Phys. Rev. A}\
  }\textbf {\bibinfo {volume} {93}},\ \bibinfo {pages} {063623} (\bibinfo
  {year} {2016})}\BibitemShut {NoStop}%
\bibitem [{\citenamefont {Bersano}\ \emph {et~al.}(2018)\citenamefont
  {Bersano}, \citenamefont {Gokhroo}, \citenamefont {Khamehchi}, \citenamefont
  {D’~Ambroise}, \citenamefont {Frantzeskakis}, \citenamefont {Engels},\ and\
  \citenamefont {Kevrekidis}}]{bersano2018three}%
  \BibitemOpen
  \bibfield  {author} {\bibinfo {author} {\bibfnamefont {T.}~\bibnamefont
  {Bersano}}, \bibinfo {author} {\bibfnamefont {V.}~\bibnamefont {Gokhroo}},
  \bibinfo {author} {\bibfnamefont {M.}~\bibnamefont {Khamehchi}}, \bibinfo
  {author} {\bibfnamefont {J.}~\bibnamefont {D’~Ambroise}}, \bibinfo {author}
  {\bibfnamefont {D.}~\bibnamefont {Frantzeskakis}}, \bibinfo {author}
  {\bibfnamefont {P.}~\bibnamefont {Engels}}, \ and\ \bibinfo {author}
  {\bibfnamefont {P.}~\bibnamefont {Kevrekidis}},\ }\href@noop {} {\bibfield
  {journal} {\bibinfo  {journal} {Phys. Rev. Lett.}\ }\textbf {\bibinfo
  {volume} {120}},\ \bibinfo {pages} {063202} (\bibinfo {year}
  {2018})}\BibitemShut {NoStop}%
\bibitem [{\citenamefont {Nistazakis}\ \emph
  {et~al.}(2008{\natexlab{b}})\citenamefont {Nistazakis}, \citenamefont
  {Frantzeskakis}, \citenamefont {Kevrekidis}, \citenamefont {Malomed},\ and\
  \citenamefont {Carretero-Gonz{\'a}lez}}]{nistazakis2008bright}%
  \BibitemOpen
  \bibfield  {author} {\bibinfo {author} {\bibfnamefont {H.}~\bibnamefont
  {Nistazakis}}, \bibinfo {author} {\bibfnamefont {D.}~\bibnamefont
  {Frantzeskakis}}, \bibinfo {author} {\bibfnamefont {P.}~\bibnamefont
  {Kevrekidis}}, \bibinfo {author} {\bibfnamefont {B.}~\bibnamefont {Malomed}},
  \ and\ \bibinfo {author} {\bibfnamefont {R.}~\bibnamefont
  {Carretero-Gonz{\'a}lez}},\ }\href@noop {} {\bibfield  {journal} {\bibinfo
  {journal} {Phys. Rev. A}\ }\textbf {\bibinfo {volume} {77}},\ \bibinfo
  {pages} {033612} (\bibinfo {year} {2008}{\natexlab{b}})}\BibitemShut
  {NoStop}%
\bibitem [{\citenamefont {Katsimiga}\ \emph
  {et~al.}(2018{\natexlab{b}})\citenamefont {Katsimiga}, \citenamefont
  {Mistakidis}, \citenamefont {Koutentakis}, \citenamefont {Kevrekidis},\ and\
  \citenamefont {Schmelcher}}]{katsimiga2018many}%
  \BibitemOpen
  \bibfield  {author} {\bibinfo {author} {\bibfnamefont {G.~C.}\ \bibnamefont
  {Katsimiga}}, \bibinfo {author} {\bibfnamefont {S.~I.}\ \bibnamefont
  {Mistakidis}}, \bibinfo {author} {\bibfnamefont {G.~M.}\ \bibnamefont
  {Koutentakis}}, \bibinfo {author} {\bibfnamefont {P.~G.}\ \bibnamefont
  {Kevrekidis}}, \ and\ \bibinfo {author} {\bibfnamefont {P.}~\bibnamefont
  {Schmelcher}},\ }\href {\doibase 10.1103/PhysRevA.98.013632} {\bibfield
  {journal} {\bibinfo  {journal} {Phys. Rev. A}\ }\textbf {\bibinfo {volume}
  {98}},\ \bibinfo {pages} {013632} (\bibinfo {year}
  {2018}{\natexlab{b}})}\BibitemShut {NoStop}%
\end{thebibliography}%

\end{document}